\documentclass[twocolumn]{aastex631}
\pdfoutput=1 %for arXiv submission

\usepackage{newtxtext,newtxmath}
\usepackage[T1]{fontenc}

\usepackage{nicefrac}
\usepackage{latexsym}
\usepackage{amsmath, mathrsfs, bm}
\usepackage{eqnarray}
\usepackage{url}
\usepackage{cases}
\usepackage{epsfig}
\usepackage{color}
\usepackage{graphicx}
\usepackage{grffile}
\usepackage{float}
\usepackage{soul}
\usepackage{enumerate}
\usepackage{multirow}
\usepackage{xspace}
\usepackage{bm}
\usepackage{times}

\usepackage{etoolbox}
\preto\align{\par\nobreak\noindent}
\expandafter\preto\csname align*\endcsname{\par\nobreak\noindent}
\preto\multline{\par\nobreak\noindent}
\expandafter\preto\csname align*\endcsname{\par\nobreak\noindent}
\preto\flalign{\par\nobreak\noindent}
\expandafter\preto\csname align*\endcsname{\par\nobreak\noindent}
\preto\eqnarray{\par\nobreak\noindent}
\expandafter\preto\csname align*\endcsname{\par\nobreak\noindent}

\newcommand{\reffg}[1]{Figure~\ref{#1}}
\newcommand{\reftb}[1]{Table~\ref{#1}}
\newcommand{\refeq}[1]{Equation~(\ref{#1})}
\newcommand{\refsc}[1]{Section~\ref{#1}}
\newcommand*{\msk}{\\[0.25cm]} % mathlineskip
\newcommand*{\nmsk}{\notag\msk} % mathlineskip with no label
\definecolor{indiagreen}{rgb}{0.07, 0.53, 0.03}

\def\T{{\rm T}}
\def\hi{\textsc{Hi}\xspace}
\def\hiim{\textsc{Hi\xspace IM}\xspace}

\def\dd{{\rm d}\xspace}

\def\apjl{Astrophysical Journal Letters}
\def\aap{Astronomy and Astrophysics}

\def\apjs{The Astrophysical Journal Supplement Series}

\begin{document}

\title{FAST drift scan survey for \hi intensity mapping: I. preliminary data analysis}

%% The \author command is the same as before except it now takes an optional
%% argument which is the 16 digit ORCID. The syntax is:
%% \author[xxxx-xxxx-xxxx-xxxx]{Author Name}

%\correspondingauthor{August Muench}
%\email{greg.schwarz@aas.org, gus.muench@aas.org}

\author[0000-0003-1962-2013]{Yichao Li}
\affiliation{Key Laboratory of Cosmology and Astrophysics (Liaoning) \& College of Sciences, Northeastern University, Shenyang 110819, China}

\correspondingauthor{Yougang Wang}
\author[0000-0003-0631-568X]{Yougang Wang}
\email{wangyg@bao.ac.cn}
\affiliation{National Astronomical Observatories, Chinese Academy of Sciences, Beijing 100101, China}
\affiliation{School of Astronomy and Space Science, University of Chinese Academy of Sciences, Beijing 100049, China}
\affiliation{Key Laboratory of Radio Astronomy and Technology, Chinese Academy of Sciences, A20 Datun Road, Chaoyang District, Beijing 100101, China}

\author[0000-0001-8075-0909]{Furen Deng}
\affiliation{National Astronomical Observatories, Chinese Academy of Sciences, Beijing 100101, China}
\affiliation{School of Astronomy and Space Science, University of Chinese Academy of Sciences, Beijing 100049, China}
\affiliation{Key Laboratory of Radio Astronomy and Technology, Chinese Academy of Sciences, A20 Datun Road, Chaoyang District, Beijing 100101, China}

\author[0009-0006-2521-025X]{Wenxiu Yang}
\affiliation{National Astronomical Observatories, Chinese Academy of Sciences, Beijing 100101, China}
\affiliation{School of Astronomy and Space Science, University of Chinese Academy of Sciences, Beijing 100049, China}
\affiliation{Key Laboratory of Radio Astronomy and Technology, Chinese Academy of Sciences, A20 Datun Road, Chaoyang District, Beijing 100101, China}

\author[0000-0002-3108-5591]{Wenkai Hu}
\affiliation{Aix Marseille Univ, CNRS, CNES, LAM, Marseille, France}
\affiliation{ARC Centre of Excellence for All Sky Astrophysics in 3 Dimensions (ASTRO 3D), Australia}

\author[0009-0000-6895-9136]{Diyang Liu}
\affiliation{Key Laboratory of Cosmology and Astrophysics (Liaoning) \& College of Sciences, Northeastern University, Shenyang 110819, China}

\author[0009-0008-2564-9398 ]{Xinyang Zhao}
\affiliation{Key Laboratory of Cosmology and Astrophysics (Liaoning) \& College of Sciences, Northeastern University, Shenyang 110819, China}

\author[0000-0003-3858-6361]{Shifan Zuo}
\affiliation{National Astronomical Observatories, Chinese Academy of Sciences, Beijing 100101, China}
\affiliation{Key Laboratory of Radio Astronomy and Technology, Chinese Academy of Sciences, A20 Datun Road, Chaoyang District, Beijing 100101, China}

\author[0009-0004-8919-7088]{Shuanghao Shu}
\affiliation{National Astronomical Observatories, Chinese Academy of Sciences, Beijing 100101, China}
\affiliation{School of Astronomy and Space Science, University of Chinese Academy of Sciences, Beijing 100049, China}

\author[0000-0001-9652-1377]{Jixia Li}
\affiliation{National Astronomical Observatories, Chinese Academy of Sciences, Beijing 100101, China}
\affiliation{Key Laboratory of Radio Astronomy and Technology, Chinese Academy of Sciences, A20 Datun Road, Chaoyang District, Beijing 100101, China}

\author[0000-0003-0325-1633]{Peter Timbie}
\affiliation{Department of Physics, University of Wisconsin -- Madison, Madison, Wisconsin 53706, USA}

\author{R\'eza Ansari}
\affiliation{Université Paris-Saclay, Université Paris Cité, CEA, CNRS, AIM, 91191, Gif-sur-Yvette, France}

\author{Olivier Perdereau}
\affiliation{Universit\'e Paris-Saclay, CNRS/IN2P3, IJCLab, 91405 Orsay, France}

\author[0000-0002-3807-7252]{Albert Stebbins}
\affiliation{Fermi National Accelerator Laboratory, P.O. Box 500, Batavia IL 60510, USA}

\author[0000-0003-3334-3037]{Laura Wolz}
\affiliation{Jodrell Bank Centre for Astrophysics, Department of Physics \& Astronomy, The University of Manchester, Manchester M13 9PL, UK}

\author[0000-0002-6174-8640]{Fengquan Wu}
\affiliation{National Astronomical Observatories, Chinese Academy of Sciences, Beijing 100101, China}
\affiliation{School of Astronomy and Space Science, University of Chinese Academy of Sciences, Beijing 100049, China}
\affiliation{Key Laboratory of Radio Astronomy and Technology, Chinese Academy of Sciences, A20 Datun Road, Chaoyang District, Beijing 100101, China}

\correspondingauthor{Xin Zhang}
\author[0000-0002-6029-1933]{Xin Zhang}
\email{zhangxin@mail.neu.edu.cn}
\affiliation{Key Laboratory of Cosmology and Astrophysics (Liaoning) \& College of Sciences, Northeastern University, Shenyang 110819, China}
\affiliation{National Frontiers Science Center for Industrial Intelligence and Systems Optimization, Northeastern University, Shenyang 110819, China}
\affiliation{Key Laboratory of Data Analytics and Optimization for Smart Industry (Ministry of Education), Northeastern University, Shenyang 110819, China}

\correspondingauthor{Xuelei Chen}
\author[0000-0001-6475-8863]{Xuelei Chen}
\email{xuelei@cosmology.bao.ac.cn}
\affiliation{National Astronomical Observatories, Chinese Academy of Sciences, Beijing 100101, China}
\affiliation{Key Laboratory of Cosmology and Astrophysics (Liaoning) \& College of Sciences, Northeastern University, Shenyang 110819, China}
\affiliation{Key Laboratory of Radio Astronomy and Technology, Chinese Academy of Sciences, A20 Datun Road, Chaoyang District, Beijing 100101, China}
\affiliation{School of Astronomy and Space Science, University of Chinese Academy of Sciences, Beijing 100049, China}

%% automatically takes care of all commas and "and"s between authors' names.


\begin{abstract}
This work presents the initial results of the drift-scan observation for the neutral 
hydrogen (\hi) intensity mapping survey with the Five-hundred-meter Aperture Spherical radio Telescope (FAST). 
The data analyzed in this work were collected in night observations from 2019 through 2021. 
The primary findings are based on 28 hours of drift-scan observation carried out over seven nights in 2021, 
which covers $60\,{\rm deg}^2$ sky area. Our main findings are: 
(i) Our calibration strategy can successfully correct both the temporal and bandpass gain variation over the $4$-hour drift-scan observation.
(ii) The continuum maps of the surveyed region are made with frequency resolution of $28$ kHz 
and pixel area of $2.95\,{\rm arcmin}^2$. The pixel noise levels of the continuum maps are slightly higher than the forecast assuming $T_{\rm sys}=20\,{\rm K}$, which are $36.0$ mK (for 10.0 s integration time) at the $1050$--$1150$ MHz band, and $25.9$ mK (for 16.7 s integration time) at the $1323$--$1450$ MHz band, respectively.  
(iii) The flux-weighted differential number count is consistent with the NRAO-VLA Sky Survey (NVSS) catalog down to the confusion limit $\sim7\,{\rm mJy}/{\rm beam}^{-1}$.
(iv) The continuum flux measurements of the sources are consistent with that found in the literature. The difference in the flux measurement of $81$ isolated NVSS sources is about $6.3\%$. 
Our research offers a systematic analysis for the FAST \hi intensity mapping drift-scan survey and serves as a helpful resource for further cosmology and associated galaxies sciences with the FAST drift-scan survey.
\end{abstract}

%% You will be asked to select these concepts during the submission process


\section{Introduction}
 
Measurements of the cosmological large-scale structure (LSS) play an important role 
in studying the evolution of the Universe. 
In the past decades, the LSS fluctuations have been explored by 
observing the galaxy distribution in the Universe with wide-field
spectroscopic and photometric surveys 
\citep{2005MNRAS.362..505C,2005ApJ...633..560E,
2014MNRAS.441...24A,2017MNRAS.464.4807H,2020arXiv200708991E}.
Recently, it has been proposed another cosmological probe of 
the LSS by observing the neutral hydrogen (\hi) in the galaxies 
via its 21 cm emission line of hyperfine spin-flip transition
\citep[e.g.][]{2004MNRAS.355.1339B,2006ApJ...653..815M,
2012RPPh...75h6901P}. 

A number of \hi galaxy surveys have been carried out, 
e.g. the 64 m Parkes telescope in Australia with the HI Parkes All-Sky Survey
(HIPASS; \citealt{2001MNRAS.322..486B,2004MNRAS.350.1195M,2004MNRAS.350.1210Z}),
the 76 m Lovell Telescope at Jodrell Bank with the HI Jodrell All-Sky Survey 
(HIJASS;\citealt{2003MNRAS.342..738L}), the Arecibo Legacy Fast ALFA (ALFALFA) survey
\citep{2005AJ....130.2598G,2007AJ....133.2569G,
2007AJ....133.2087S} and Jansky Very Large Array (JVLA) $10\,{\rm deg}^2$ deep survey
\citep{2014arXiv1401.4018J}. 
However, limited by the sensitivity and the angular resolution of
the radio telescopes, the redshift range of these surveys is much smaller than the current optical surveys.
To resolve the \hi emission line from individual distant galaxies at 
centimeter wavelength requires a large radio interferometer and it is time-consuming. 
Instead, a technique known as \hi intensity mapping (\hiim), which 
is to measure the total \hi intensity of many galaxies 
within large voxels \citep{2008PhRvL.100i1303C,2008PhRvL.100p1301L,
2008PhRvD..78b3529M,2008PhRvD..78j3511P,2008MNRAS.383..606W,2008MNRAS.383.1195W,
2009astro2010S.234P,2010MNRAS.407..567B,2010ApJ...721..164S,2011ApJ...741...70L,
2012A&A...540A.129A,2013MNRAS.434.1239B},
can be quickly carried out and extended to very large survey volume
and is ideal for cosmological surveys
\citep{2015ApJ...798...40X,Zhang:2021yof,Jin:2021pcv,Wu:2021vfz,Wu:2022dgy,Wu:2022jkf,Zhang:2023gaz}.

\begin{table*}
\caption{Observation mode. 
Column (1): The observation date. 
Column (2): The sample rate for recording the data. 
Column (3): The level of noise diode.
Column (4): The rotated angle of the feed array.}\label{table:obs}
\begin{center}
\hspace*{-1.5cm}
\begin{tabular}{ccccccc}\hline\hline
Field center    &     Date    & Frequency resolution   & Integration time & Noise diode level  & Rotation angle \\
                &             &  $[{\rm kHz}]$                  & $[{\rm s}]$        &                    & $[^\circ]$     \\\hline    
HIMGS 1100+2539 & 2019-05-27  &  0.5                   & 1                & high               & 0             \\
HIMGS 1100+2554 & 2019-05-28  &  0.5                   & 0.1              & high               & 0             \\
HIMGS 1100+2609 & 2019-05-29  &  7.6                   & 0.1              & low                & 0             \\
HIMGS 1100+2554 & 2019-05-30  &  7.6                   & 1                & low                & 0             \\
HIMGS 1100+2639 & 2019-05-31  &  7.6                   & 1                & low                & 23.4          \\
HIMGS 1100+2639 & 2020-05-08  &  7.6                   & 1                & low                & 23.4          \\\hline
HIMGS 1100+2600 & 2021-03-02  &  7.6                   & 1                & low                & 23.4          \\
HIMGS 1100+2632 & 2021-03-05  &  7.6                   & 1                & low                & 23.4          \\
HIMGS 1100+2643 & 2021-03-06  &  7.6                   & 1                & low                & 23.4          \\
HIMGS 1100+2654 & 2021-03-07  &  7.6                   & 1                & low                & 23.4          \\
HIMGS 1100+2610 & 2021-03-09  &  7.6                   & 1                & low                & 23.4          \\
HIMGS 1100+2621 & 2021-03-13  &  7.6                   & 1                & low                & 23.4          \\
HIMGS 1100+2610 & 2021-03-14  &  7.6                   & 1                & low                & 23.4          \\
\hline\hline
\end{tabular}
\end{center}
\end{table*}

\begin{figure}
\centering
\hspace*{-0.3cm}
\includegraphics[width=0.5\textwidth]{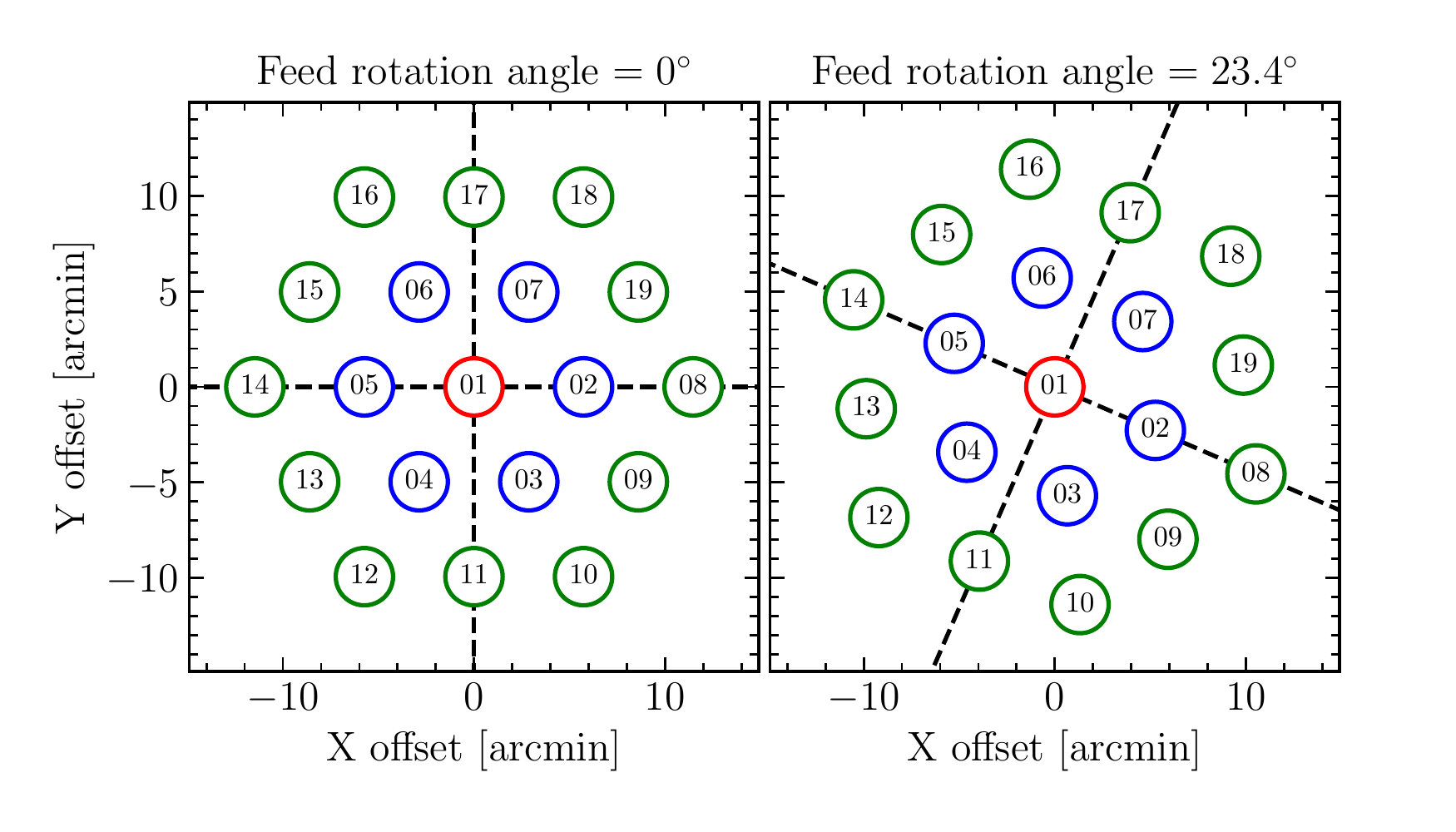}
\caption{ The position of the feed in the FAST L-band 19-feed array. 
The feed array without rotation is shown in the left panel and
the feed array rotated by $23\overset{\circ}{.}4$ is shown in the right panel. 
The circles represent the $3\,{\rm arcmin}$ beam size at frequency of $1420\,{\rm MHz}$.
The central, inner-circle, and outer-circle feeds are shown in red, blue, and green,
respectively.
}\label{fig:feedpos}
\end{figure}

\begin{figure*}
\centering
\includegraphics[width=\textwidth]{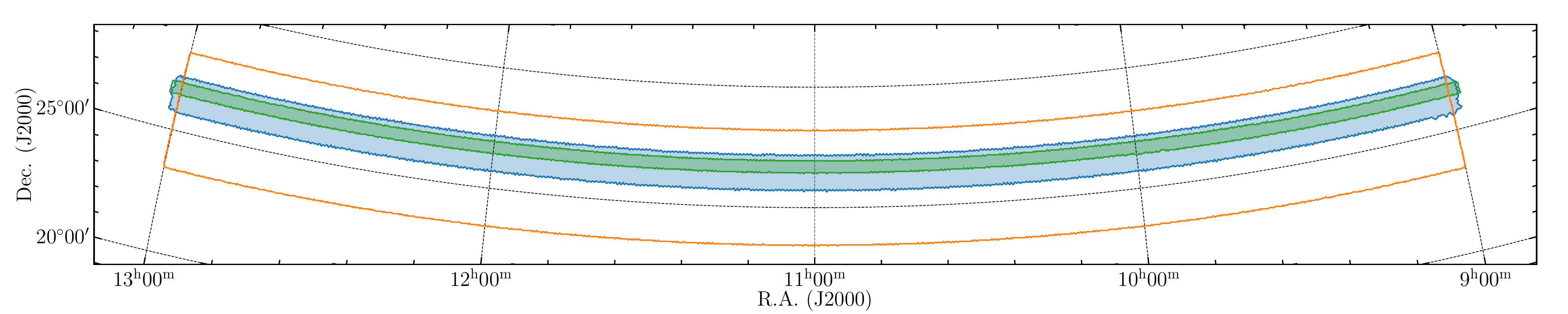}
\caption{
The footprint of the FAST \hiim Pilot Survey observation using Zenith equal area projection (ZEA).
The filled blue region shows the sky area of the major data observed in 2021;
The green stripe indicates the test observation carried out on May 31st, 2019
and May 8th, 2020; and the orange region shows the full target area for 
FAST \hiim Pilot Survey. %\red{what is the projection used in this map?}
%\green{using Zenith equal area projection (ZEA)}
}\label{fig:footprint}
\end{figure*}

\begin{figure*}
\centering
\includegraphics[width=\textwidth]{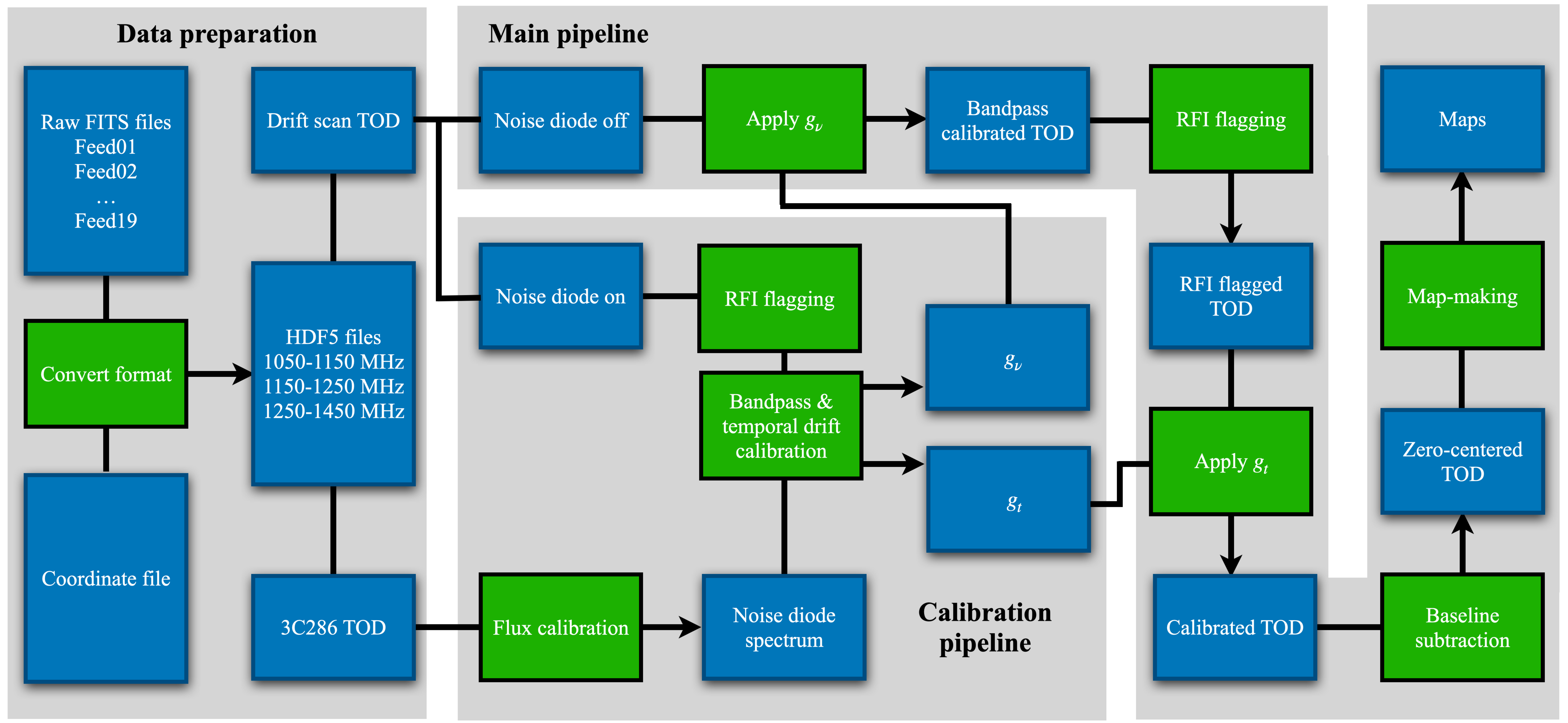}
\caption{
TOD analysis pipeline. The blue rectangular indicates the 
input/output data and the green rectangular indicates the operation.
}\label{fig:pipeline}
\end{figure*}

The \hiim technique was explored by measuring the cross-correlation function 
between an \hiim survey carried out with Green Bank Telescope (GBT) and an
optical galaxy survey \citep{2010Natur.466..463C}. 
Later, a few detections of the cross-correlation power spectrum 
between an \hiim survey and an optical galaxy survey were reported
with GBT and Parkes telescopes
\citep{2013ApJ...763L..20M,2018MNRAS.476.3382A,2017MNRAS.464.4938W,2022MNRAS.510.3495W,
2022arXiv220201242C}.
There are several ongoing \hiim experiments focusing on the
post-reionization epoch, such as the Tianlai project \citep{2012IJMPS..12..256C,
2020SCPMA..6329862L,2021MNRAS.506.3455W,2022MNRAS.517.4637P,2022RAA....22f5020S}, the
Canadian Hydrogen Intensity Mapping Experiment ~\citep[CHIME,][]{2014SPIE.9145E..22B}.
A couple of \hiim experiments are under construction, such as 
the Baryonic Acoustic Oscillations from Integrated Neutral Gas Observations
~\citep[BINGO,][]{2013MNRAS.434.1239B} and the
Hydrogen Intensity and Real-Time Analysis experiment ~\citep[HIRAX,][]{2016SPIE.9906E..5XN}.
The \hiim technique is also proposed as the major cosmology project with 
the Square Kilometre Array (SKA)\footnote{\url{https://www.skao.int}} \citep{2015aska.confE..19S,2020PASA...37....7S}
and MeerKAT \citep{2015ApJ...803...21B, 2017arXiv170906099S,2021MNRAS.501.4344L,2021MNRAS.505.3698W,
2021MNRAS.505.2039P, 2023arXiv230211504C}.
Recently, the MeerKAT \hiim survey 
reported the cross-correlation power spectrum detection with the
optical galaxy survey \citep{2022arXiv220601579C}. 
Meanwhile, using the MeerKAT interferometric observations, 
\citet{2023arXiv230111943P} reports the \hiim auto power spectrum detection 
on Mpc scales. The \hiim auto power spectrum on large scales 
remains undetected \citep{2013MNRAS.434L..46S}. 

The Five-hundred-meter Aperture Spherical radio Telescope
\citep[FAST,][]{2011IJMPD..20..989N,2016RaSc...51.1060L}
was recently built and operating for observations
\citep{2020raa....20...64j}.
FAST is located in Dawodang karst depression, a natural basin in Guizhou province, China
$({\rm E}\,106\overset{\circ}{.}86,\,{\rm N}\,25\overset{\circ}{.}65)$.
The FAST can reach all parts of the sky within $40^{\circ}$ from the zenith, 
corresponding to $\sim 25000\, \rm{deg}^2$ sky area. 
With the $300$ m effective aperture diameter, FAST becomes the most sensitive radio 
telescope in the world. 
In the meanwhile, the L-band 19-feed receiver,
working at frequencies between $1.05$ GHz and $1.45$ GHz,
increases the field of view, which is ideal for the large-area survey.
Additionally, with $\sim 3\,{\rm arcmin}$ angular resolution, 
the large area survey with FAST can potentially resolve the \hi environment in
a large number of galaxy clusters, groups, filaments, and voids.
 
Forecasts and simulations show that the large area \hiim survey with the 
FAST is ideal for cosmological studies \citep{2017PhRvD..96f3525L,2020MNRAS.493.5854H}.
Before committing to a large survey, we proposed a pilot survey.
With the pilot survey, we aim to find out the relevant characteristics of
FAST receiver system. The systematic 1/f-type gain variation is studied 
in \citet{2021MNRAS.508.2897H}.
In this work, we address the time-ordered data (TOD) analysis pipeline for 
the \hiim with FAST drift-scan observation. This paper is organized as follows:
We describe the data collected for \hiim pilot survey in \refsc{sec:data} and 
the TOD analysis method in \refsc{sec:pipeline}. 
The results and implications of the TOD analysis are discussed 
in \refsc{sec:disc}. In \refsc{sec:summary} a summary of this work is presented.

\section{Observational Data}\label{sec:data}

The data analyzed in this work were collected in night sessions spanning in 
$2019$, $2020$ and $2021$. During the observation of each night, the telescope 
was pointed at a fixed altitude angle. We chose an area close to the 
zenith, corresponding to ${\rm Dec.} \sim 26^\circ$ at the telescope site.
The resulting Dec. in the J2000 equatorial coordinate varies slightly 
(at a magnitude of $\sim 3$ arcmin) during the $4$ hours drift-scan. 

During the observations in 2019 and 2020, we investigated different configurations of the 
observation parameters, e.g the frequency resolution, sampling rate, feed array rotation angle 
and the level of the noise diode power, to test the impacts on the observation results. 
The finer resolution in frequency and time is beneficial for identifying narrow radio frequency
interference (RFI) and resolving the spectroscopic profile of \hi galaxy.
However, it produces a significant amount of data and takes a long time to process.
Our testing results show that, with the exception of a few badly contaminated frequency ranges, 
RFI contamination can be properly detected utilizing the frequency resolution of $7.6$ kHz. 
In the meanwhile, the spectral features of interest can be resolved using this frequency
resolution.
Data with finer frequency resolution and higher sampling rates are used for 
measuring the systematic 1/f noise at different temporal and spectroscopic scales.
Our analysis \citep{2021MNRAS.508.2897H} show that the systematic 1/f noise is negligible 
within a few hundred seconds time scale. Thus, we adopt $7.6$ kHz frequency resolution 
and $1\,{\rm s}$ integration time for the survey observations.

We use the FAST L-band 19-feed receiver. 
The position of the feed in the 19-feed array is shown in \reffg{fig:feedpos}
The feed array is rotated by $23\overset{\circ}{.}4$ 
to obtain the maximum span of Dec. coverage during the drift scans \citep{2018IMMag..19..112L}.
We also have a couple of observations without rotating the feed array.
During such drift-scan observations, the same sky stripe is repeatedly scanned by 
different feeds, which is ideal for systematic checking via cross-correlating the 
observation of different feeds.

A noise diode is built into the receiver system and its output can be injected as a real-time 
calibrator during the observation. In every $8\,{\rm s}$, the noise diode was fired for 
$0.9\,{\rm s}$, which is slightly shorter than the integration time to avoid power 
leakage to the nearby time stamps. 
In addition, the noise diode spectrum is calibrated by observing the celestial 
point source calibrator 3C286. 
The noise diode calibrator can be fired at either the high-power or the low-power level.  
With our test data, we found that the low-power level is sufficient for our calibration.

The observation of the $7$ nights in 2021, that adopt the optimized observation parameter
configuration, are the major survey data used in these analyses below.
The pointing direction shifts by $10.835\,{\rm arcmin}$ per night in Dec.
direction. With the 7 nights drift-scan observations, a range of $\sim 1^\circ$ across the Dec. 
direction is covered. 
With $4$ hours drift scan, the observation covers the right ascension (R.A.) range 
from $9$ hr to $13$ hr, which overlaps with the
Northern Galactic Cap (NGP) area of the Sloan Digital Sky Survey (SDSS; \citealt{2016MNRAS.455.1553R}). 
The detailed observation information is summarized in \reftb{table:obs}.
The observation footprints of different data sets are shown in \reffg{fig:footprint}.

\section{Time-ordered data analysis}\label{sec:pipeline}

The raw data of the 19 feeds are dumped into files individually using FITS 
\footnote{\url{https://fits.gsfc.nasa.gov/}} (Flexible Image Transport System) format.
Each FITS file contains a chunk of TOD with all frequency channels. The telescope-pointing 
direction data are recorded separately. 
In order to simplify data analysis, we convert the initial data format by
combining the 19-feeds data as an extra axis and splitting the full frequency 
band into three sub-bands, i.e. the low-frequency band $1050$--$1150$ MHz, 
mid-frequency band $1150$--$1250$ MHz and high-frequency band $1250$--$1450$ MHz.
The telescope pointing directions of the 19 feeds at each time stamp is calculated
and written into the same data file. 

The observation data include both the drift-scan data and the flux calibration data.
The noise diode signal is used for real-time relative calibration of the gain, which is discussed in \refsc{sec:bandpasscal} and \refsc{sec:driftcal}.
The noise diode flux spectrum is calibrated by observing a standard source, as discussed in \refsc{sec:abscal}. 
The RFI flagging is applied to the calibrated data. The details of the RFI flagging are described in \refsc{sec:rfi}.
The data are then zero-centered by removing the temporal baseline variation
before being used for map-making. The TOD data analysis pipeline is illustrated in 
\reffg{fig:pipeline}, where the blue rectangular indicates the input/output data
and the green rectangular indicates the operation.

\subsection{Bandpass calibration}\label{sec:bandpasscal}

\begin{figure*}
\centering
\includegraphics[width=0.49\textwidth]{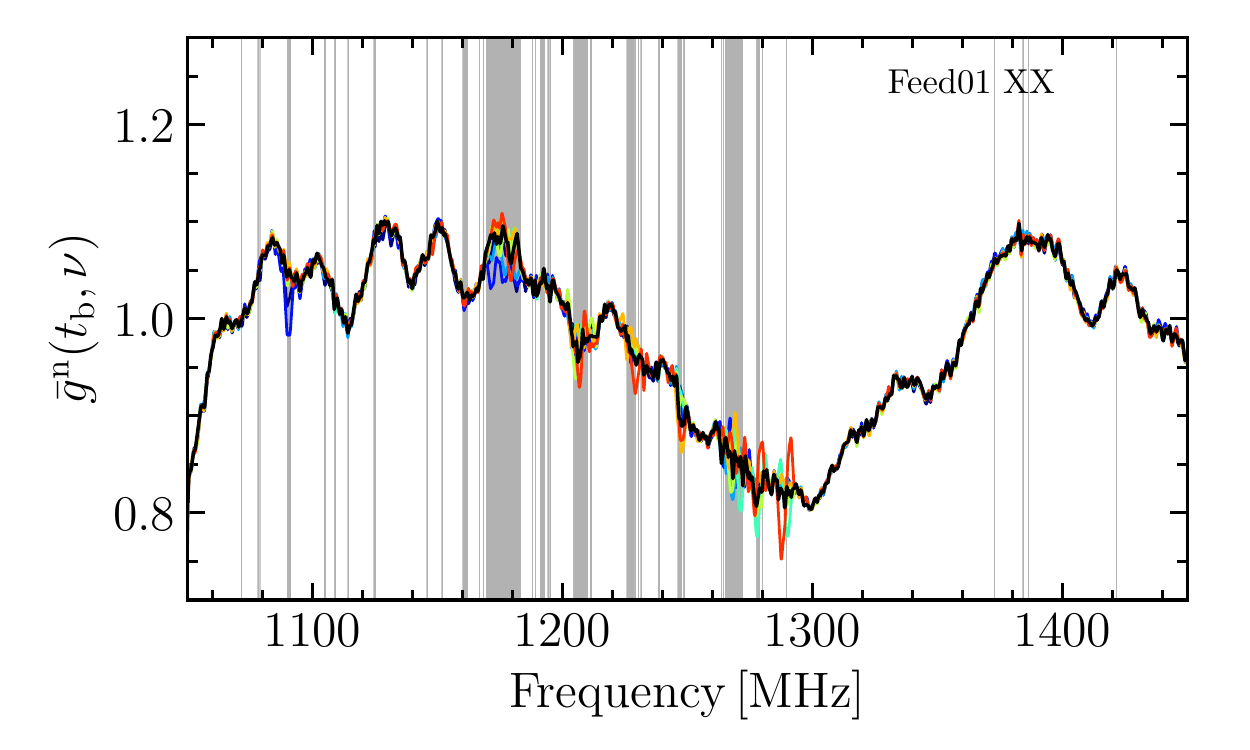}
\includegraphics[width=0.49\textwidth]{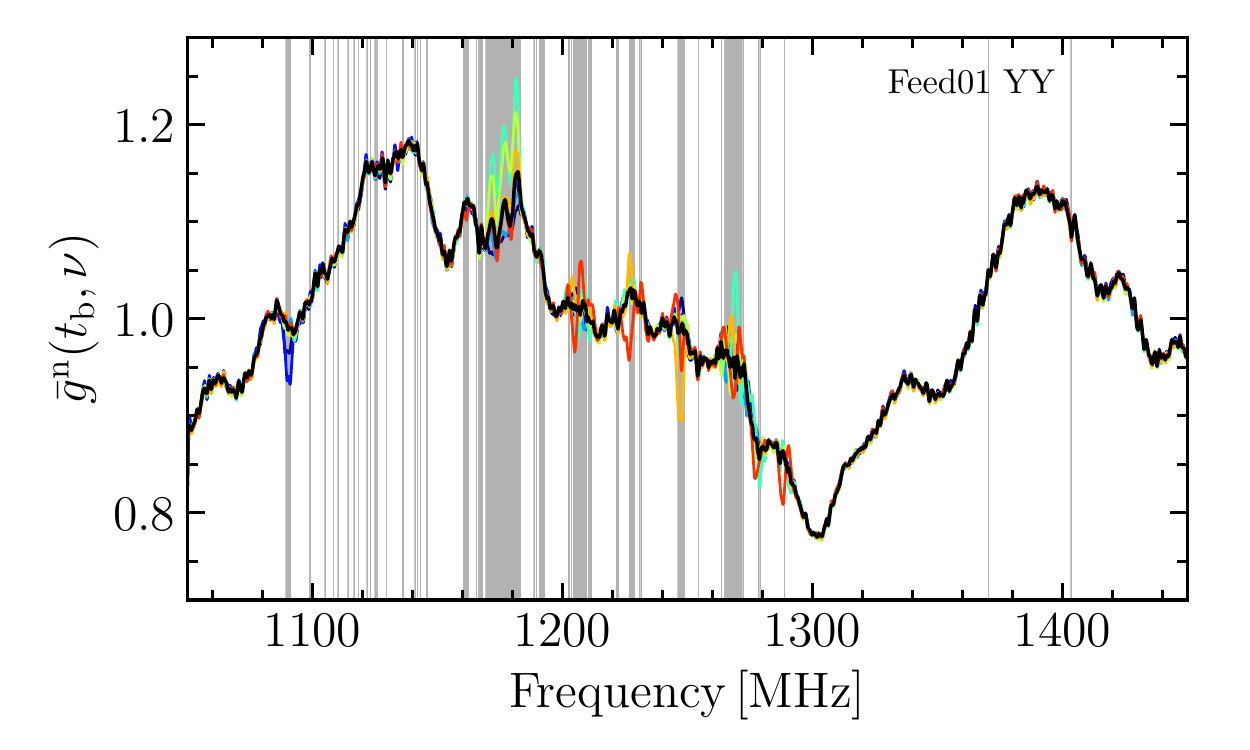}
\includegraphics[width=0.49\textwidth]{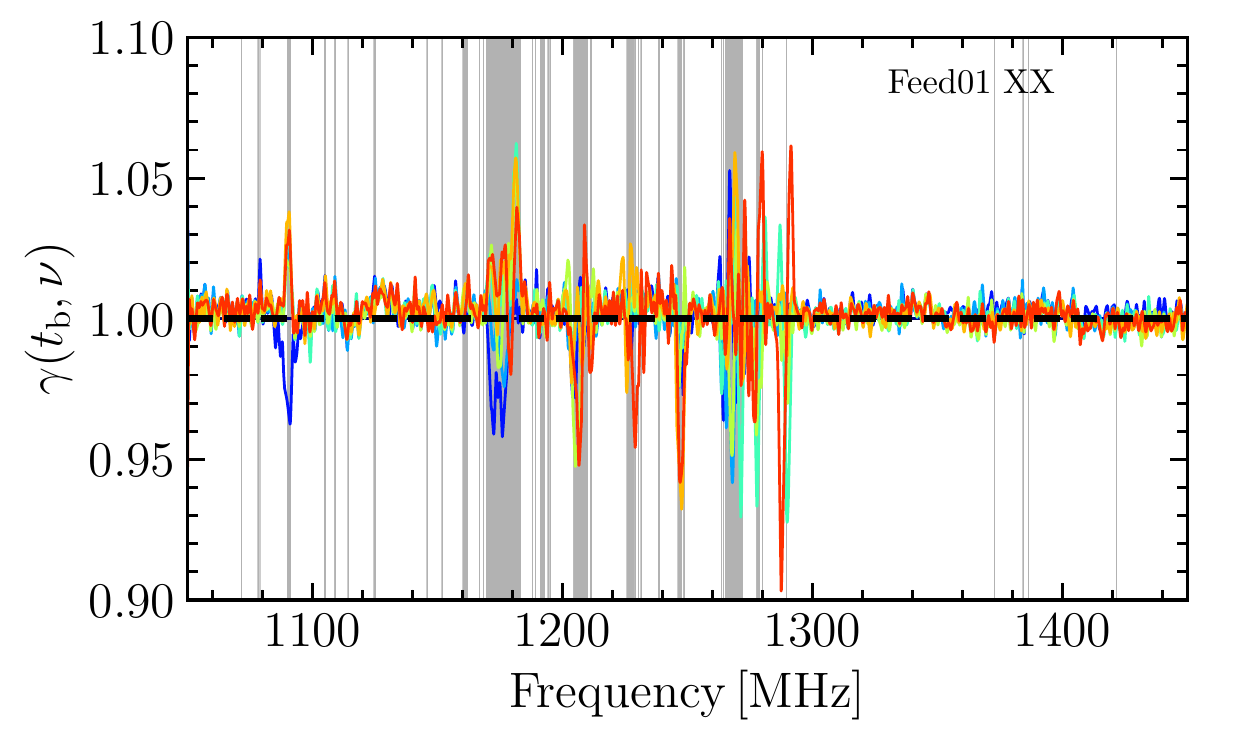}
\includegraphics[width=0.49\textwidth]{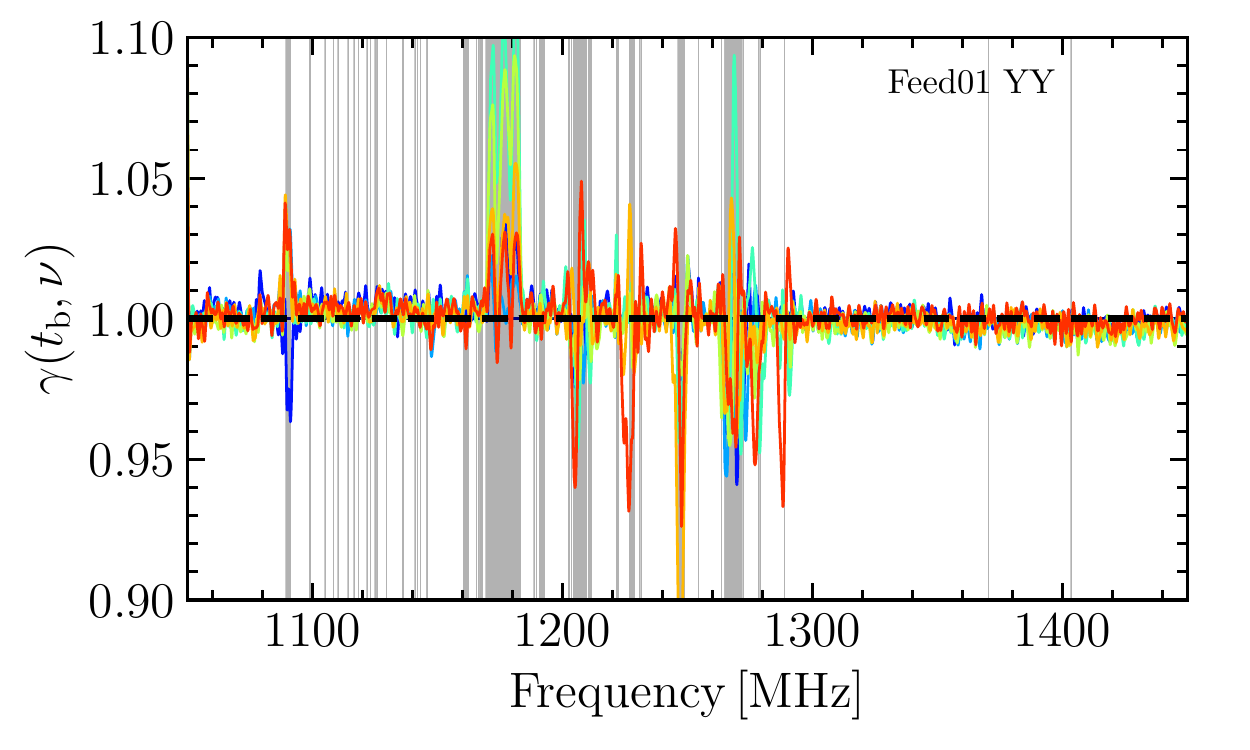}
\caption{ The smoothed bandpass of the XX (left) and YY (right) polarizations for the 20210302 data (i.e. those taken on March 2nd, 2021). The raw frequency resolution is $7.6$ kHz, 
Top: The block averaged bandpass,  each block is shown in a different color, and the black curve shows the averaged
bandpass across all blocks. 
Bottom: the bandpass relative variation with respect to the first block. 
The gray areas mark the frequency channels with over $20\%$ time stamps flagged.
}\label{fig:bandpass}
\end{figure*}

\begin{figure*}
\centering
\includegraphics[width=0.49\textwidth]{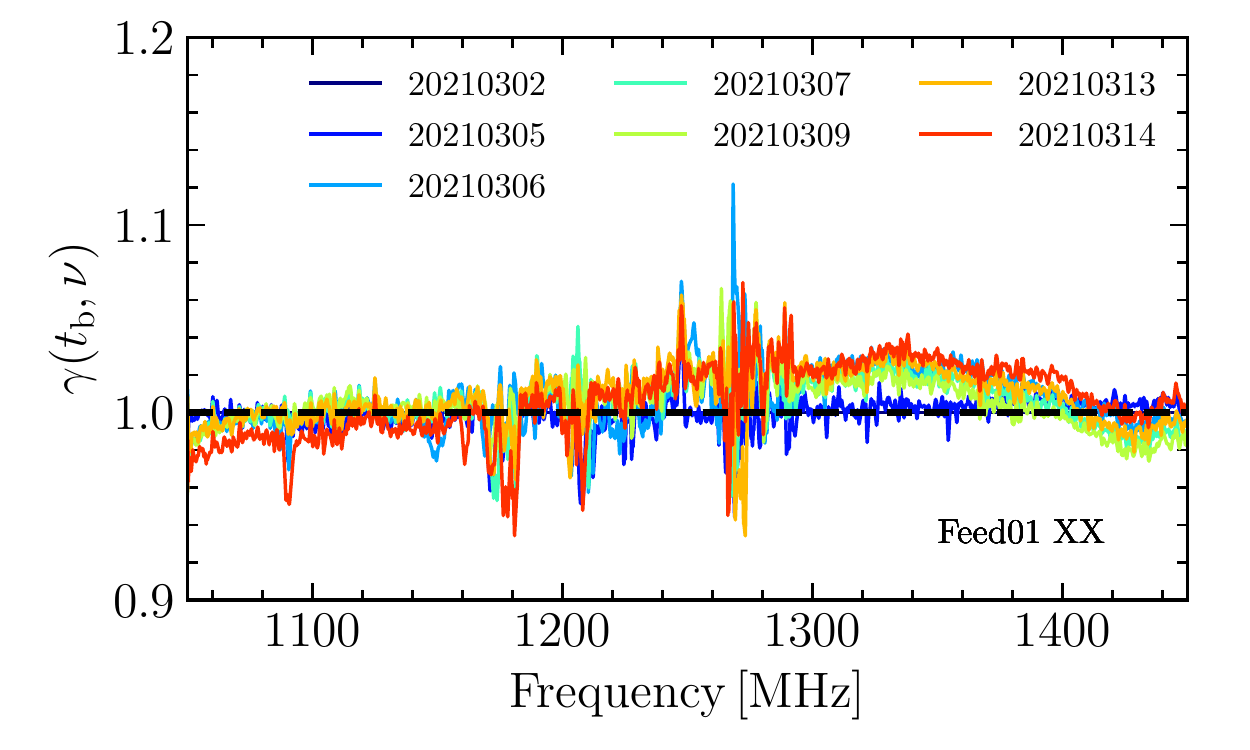}
\includegraphics[width=0.49\textwidth]{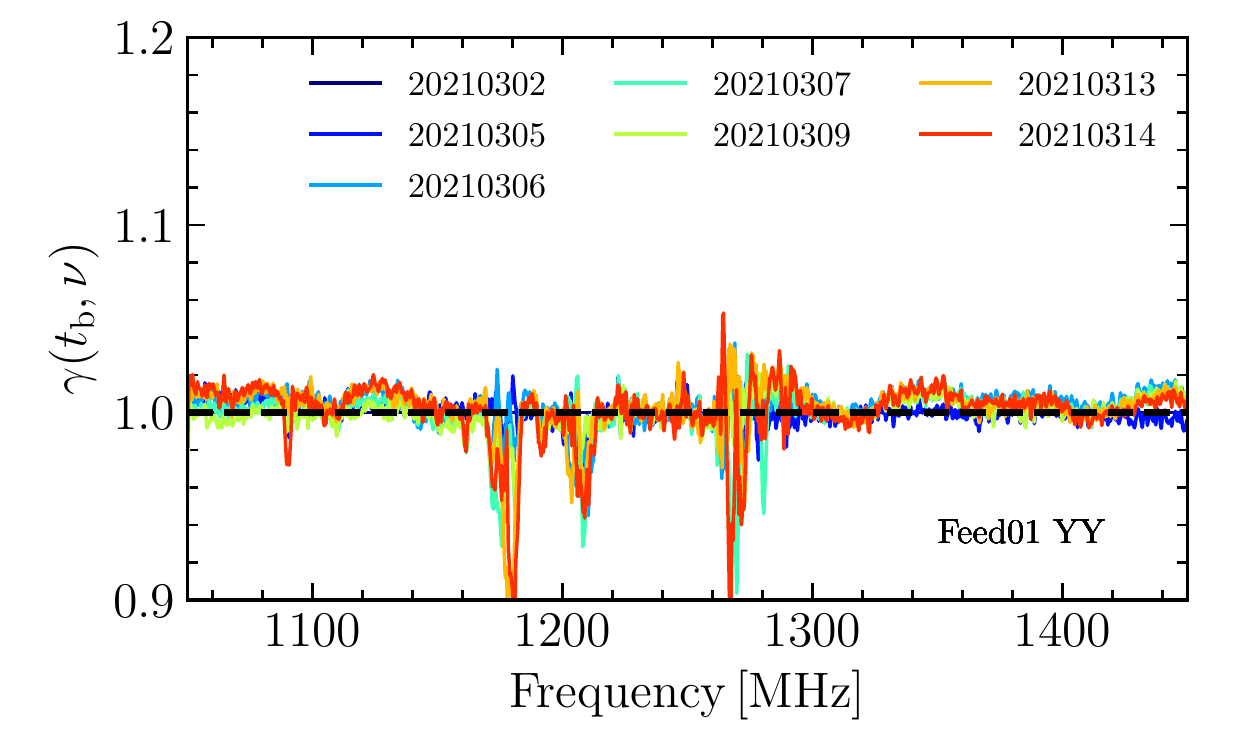}
\caption{
The bandpass relative variation between different nights with respect to 
the first night observation (20210302).
}\label{fig:bandpassdays}
\end{figure*}

The observed data value, $V(t, \nu)$, is the system gain multiplied with the
combination of the input signal and noise,
\begin{align}
    V(t, \nu) = g(t, \nu)\left(T(t, \nu) + n(t, \nu)\right) ,
\end{align}
where $T(t, \nu)$ is the antenna temperature corresponding to the total power collected by
the telescope, $n(t, \nu)$ is the noise with $\langle n(t, \nu)\rangle_t = 0$, and 
$g(t, \nu)$ is the system gain.
We assume that the system gain can be decomposed into a time-dependent component and a
frequency-dependent  component: 
\begin{align}
g(t, \nu) = g_t(t) g_{\nu}(\nu), 
\label{eq:g_decompose}
\end{align}
where $g_{\nu}(\nu)$ is the 
bandpass gain factor and $g_t(t)$ is the temporal drift factor of the gain.
In our analysis, $g_{\nu}(\nu)$ and $g_t(t)$ are calibrated using the
noise diode, which is fired for $0.9\,{\rm s}$ in every $8\,{\rm s}$ as a
relative flux calibrator. We assume that the noise diode
temperature and spectrum are both stable during the observation. In every $8\,{\rm s}$, 
we pick up the power value when
the noise diode fired on, $V_{\rm ND on}(t, \nu)$, subtract the average
power value at the two nearby time stamps, $V_{\rm ND off}(t, \nu)$,
\begin{align}\label{eq:nd}
    V_{\rm ND on}(t, \nu) - V_{\rm ND off}(t, \nu) = 
    g_t(t) g_{\nu}(\nu)\left(T_{\rm ND}(\nu) + n(t, \nu)\right),
\end{align}
in which, we assume both $g_t$ and the background emission are constant 
during the short time interval.

To check if the gain variation can be decomposed into factors of time variation and 
constant spectral shape as in Eq.(\ref{eq:g_decompose}), 
we break the full drift scan into a few of $\sim 30\,{\rm min}$ time blocks ($t_{\rm b}$) 
and evaluate the block-averaged bandpass gain, 
\begin{align}\label{eq:gnu}
\bar{g}(t_{\rm b}, \nu)=\frac{\langle V_{\rm ND on}(t, \nu) - V_{\rm ND off}(t, \nu)\rangle_{t_{\rm b}}}{T_{\rm ND}(\nu)},
\end{align}
where $\langle \cdots \rangle_{t_{\rm b}}$ represent the averaging across the time block and 
$T_{\rm ND}(\nu)$ is the noise diode spectrum.

The block-averaged bandpass is contaminated by the RFI.
To remove the RFI contamination, we first smooth the data by applying a median filter across the frequency channels to 
obtain an estimation of the bandpass, then estimate the root mean square (rms) of the residual of 
the data after subtracting the smooth bandpass. 
The frequency channels with values greater than $3$ times the rms are flagged as RFI contaminated. 
We iterate the flagging until there are no extra masked channels.
Note that the RFI flagging processing applied to the bandpass determination procedure is much 
more strict than that applied to the survey data. 
Some of the flagged channels here may actually not be real RFIs. 
However, it does not hurt to take a more strict criterion in the bandpass determination.
The RFI flagging is applied at each time stamp and the block-averaged bandpass is 
determined by taking the median value across the time block. 
A few of the frequency channels that are badly contaminated by RFI are fully flagged.
The bandpass values at the fully flagged frequency channels are interpolated from the smoothed bandpass.
Finally, in order to eliminate the noise, the bandpass is further smoothed with a 
3rd-order Butterworth low-pass filter with a critical delay frequency of $\tau=0.7\,{\rm \mu s}$, 
which corresponds to a window size of $200$ frequency bins.
The choice of the window function size is further discussed in \refsc{sec:ripple}.

The overall gain drift of the block-averaged bandpass,
as well as the noise diode spectrum is removed by normalizing $\bar{g}(t_{\rm b}, \nu)$ 
with its mean and produce the normalized bandpass,
\begin{align}
    \bar{g}^{\rm n}(t_{\rm b}, \nu) = \bar{g}(t_{\rm b}, \nu) / \langle \bar{g}(t_{\rm b}, \nu) \rangle_{\nu},
\end{align}
where $\langle \cdots \rangle_{\nu}$ represent the averaging across the frequencies.
In order to visualize the bandpass shape evolution, 
we check the normalized bandpass ratio with respect to the first block,
\begin{align}
\gamma(t_{\rm b}, \nu) = g^{\rm n}(t_{\rm b}, \nu)/g^{\rm n}(t_0, \nu),
\end{align}
where $t_0$ denotes the first time block.

An example of the results of Feed 01 for observation 20210302 (i.e. those taken on March 2nd, 2021) is shown in \reffg{fig:bandpass}, and different time blocks are represented by different colors. 
The black curve is the mean bandpass across all time blocks.
The gray areas mark the frequency channels with over $20\%$ time stamps flagged.
As shown by this figure, the bandpass has significant variation over frequency 
and also varies with time, but its shape is nearly constant over a few hours,
except for a small fraction of frequency channels, which are badly contaminated by RFI.
The bottom panels show the bandpass ratio $\gamma(t_{\rm b}, \nu)$, which has a variation
less than $1\%$ over $4\,{\rm hr}$ observation for most frequency channels. 

Assuming the stable bandpass shape, \refeq{eq:nd} is further averaged across the 4 hours of 
each night observation,
\begin{align}\label{eq:t_mean}    
\langle V_{\rm ND on}(t, \nu) - V_{\rm ND off}(t, \nu)\rangle_t = 
\bar{g}_t g_\nu(\nu) T_{\rm ND}(\nu),
\end{align}
where $\bar{g}_t = \langle g_t(t) \rangle_t$.
The bandpass relative variation between the seven nights observations in 2021 is shown
in \reffg{fig:bandpassdays}. As an example,  we show the two polarizations of the center feed in
the left and right panels. The bandpass shape varies significantly between different days.
Therefore, we emphasize that the bandpass shape needs to be determined for each night observation.

The bandpass-calibrated data is given by 
\begin{align}
    V_1 \equiv \frac{V(t, \nu)}{\langle V_{\rm ND on}(t, \nu) - V_{\rm ND off}(t, \nu)\rangle_t}
    = \frac{g_t(t)}{\bar{g}_t}\frac{T(t, \nu) + n(t, \nu)}{T_{\rm ND}(\nu)}.
\end{align}

\subsection{Temporal drift calibration}\label{sec:driftcal}

\begin{figure*}
\centering
\includegraphics[width=\textwidth]{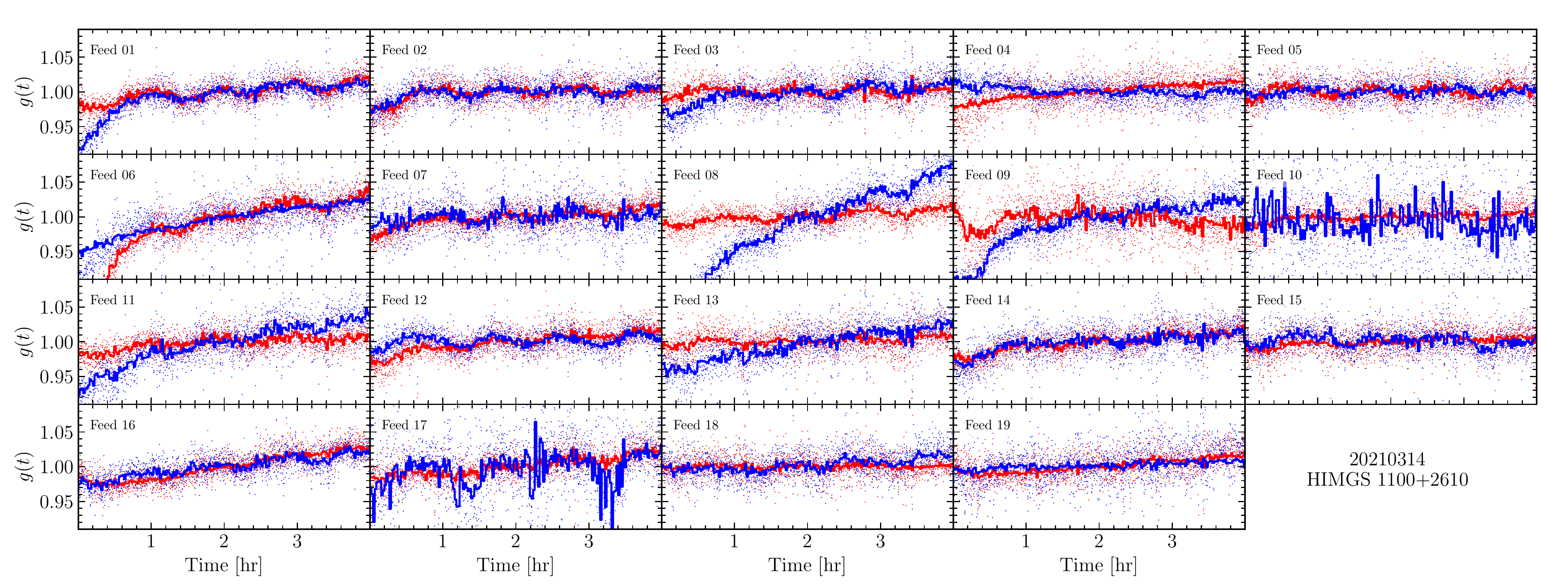}
\caption{The gain temporal variations of $4$ hours observation. 
Each sub-panel shows the gain of one feed, the two polarizations 
are shown in red and blue colors. The dot markers represent the measurements with the noise diode;
and the thick curves are the reconstructed gain variation.
}\label{fig:gt}
\end{figure*}

\begin{figure*}
\centering
\includegraphics[width=\textwidth]{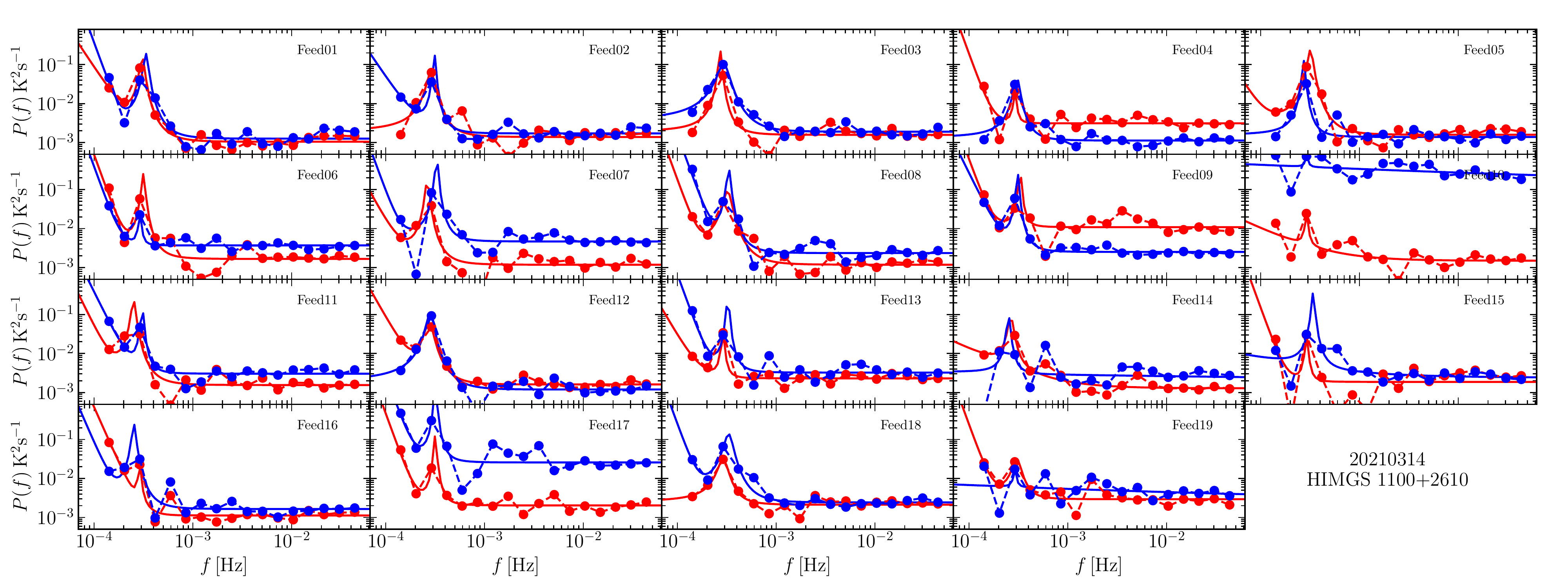}
\caption{The  temporal power spectra of the time-ordered data. 
Each sub-panel shows the gain of one feed, the two polarizations are shown in red and blue colors. The dashed lines with markers show the measured power spectra and the 
solid lines show the best-fit model of \refeq{eq:pf}.
}\label{fig:gtpow}
\end{figure*}

The temporal drift is calibrated with the noise diode as well. We average \refeq{eq:nd} across frequencies,
\begin{align}
    \langle V_{\rm ND on}(t, \nu) - V_{\rm ND off}(t, \nu) \rangle_\nu =
    \bar{g}_\nu g_t(t) \left(\langle T_{\rm ND}(\nu) \rangle_\nu + n(t)\right),
\end{align}
where $\bar{g}_\nu = \langle g(\nu) \rangle_\nu$ and 
$n(t) = \langle n(t, \nu) \rangle_\nu$.
If we further normalize with the time mean, the temporal drift is,
\begin{align}
    V_t &= \frac{\langle V_{\rm ND on}(t, \nu) - V_{\rm ND off}(t, \nu) \rangle_\nu}
    {\langle V_{\rm ND on}(t, \nu) - V_{\rm ND off}(t, \nu) \rangle_{\nu, t}}  \nmsk
    & = \frac{g_t(t)}{\bar{g}_t} + {n}^\prime(t),
    \label{eq:vt} 
\end{align}
where we assume that both $\bar{g}_\nu$ and $T_{\rm ND}$ are constant over time
and ${n}^\prime(t) = \left( g_t(t)/\bar{g}_tT_{\rm ND}\right) n(t)$.
The first term of \refeq{eq:vt}, i.e. the normalized gain, represents the
drifting of the actual gain, while the second term represents the variation 
caused by the measurement error in the calibration. 
$V_t$ represents the measurements of the gain value at each firing of the 
noise diode. Written in discrete form, we denote the gain measurements $V_t$ as vector 
${\bf g}_{\rm m}$ and \refeq{eq:vt} is expressed as
\begin{align}
{\bf g}_{\rm m} = {\bf g}_t + {\bf n}
\end{align}
We split the full-time stream into $\alpha$ short time blocks, $\Delta_\alpha$.
The gain is assumed to be constant within each short block and varying between 
different blocks. 
Using a set of the base function 
${\bf F} = \left\{ F_1(t), F_2(t), \cdots, F_\alpha(t) \right\}$, where
\begin{align}
F_\alpha(t) = 
\begin{cases}
1 & t \in \Delta_{\alpha},\\
0 & {\rm otherwise},
\end{cases}
\end{align}
the drifting of the gain is expressed as
${\bf g}_t = {\bf F} {\bf g}$, 
where ${\bf g}^{\rm T} = \left\{ g_1, g_2, \cdots, g_\alpha \right\}$ is the parameter
sets that need to be determined.
In our analysis, we use a short block length of $\Delta_\alpha = 20\, {\rm s}$
to avoid overfitting the temporal variation of the gain. 

With the amplitude vector ${\bf g}$ as the parameter, and the measured gain values 
${\bf g}_{\rm m} $, the likelihood is 
\begin{align}
P({\bf g}_{\rm m}) &= P({\bf g}_{\rm m} | {\bf g}) P({\bf g})\nonumber\\
&\propto \exp\left( -\frac{1}{2}{\bf n}^{\rm T} {\bf N}^{-1}{\bf n}\right)
\exp\left(-\frac{1}{2}{\bf g}^{\rm T}{\bf C}_g^{-1}{\bf g}\right),
\end{align}
where ${\bf N}$ is the measurement noise covariance matrix, and ${\bf C}_g = \langle {\bf g} {\bf g}^\T \rangle$ is the covariance matrix of the gain amplitude vector. The temporal variation can be modeled as follows:
%then ${\bf C}_g$ can be further expressed as,
\begin{align}
{\bf C}_g = \left({\bf F}^\T{\bf F}\right)^{-1} {\bf F}^\T {\bf C}_N {\bf F} \left({\bf F}^\T{\bf F}\right)^{-1},
\end{align}
where the covariance matrix ${\bf C}_N$ is related to the noise power spectrum $P(f)$ as,
\begin{align}
C_N(\delta t) &= \int P(f) e^{2\pi i f \delta t} \dd f.
\end{align}
Note that, $P(f)$ represents only the power spectrum of the correlated  noise (1/f noise), the total noise power spectrum is the combination of the white and correlated noise power spectrum, i.e. $P_{\rm total}(f) = \frac{\sigma^2}{\delta \nu}\left(1 + P(f)\right)$, where $\sigma$ is the systematic rms and $\delta \nu$ is the frequency resolution \citep{2018MNRAS.478.2416H,2021MNRAS.501.4344L}.

The amplitude vector ${\bf g}$ can be solved by the maximum likelihood method as
\begin{align}
\hat{{\bf g}} = 
\left( {\bf F}^\T {\bf N}^{-1} {\bf F} + {\bf C}_g^{-1} \right)^{-1}
{\bf F}^\T {\bf N}^{-1} {\bf g}_m,
\end{align}
which is equivalent to Wiener filtering.

The measurements of $g_m$ for observation on 20210314, as an example,
are shown in  \reffg{fig:gt} with the dot markers, %thin curves,
the frequency range between  $1150\,{\rm MHz}$ to $1250\,{\rm MHz}$ is 
ignored due to the serious RFI contamination. The averaging bandwidth of two 
separated sub-bands is $\sim 300\,{\rm MHz}$ in total.
The measurements of different feeds are shown in different panels 
and the two polarizations are shown in red and blue colors, respectively. 
The corresponding temporal power spectrum is shown with the dashed lines
in \reffg{fig:gtpow}. 
There is a peak in the power spectrum at $f\sim3\times10^{-4}\,{\rm Hz}$., 
corresponding to an oscillation in $V_t$ with period $\sim 1\,{\rm hour}$.
\footnote{The cause of the $\sim 1\,{\rm hour}$ period oscillation is unknown. However,\\
such oscillation is only observed in the 2021 data and disappeared\\
in the later observations.} 
Except for this, the power spectrum has the 1/f-type shape, which has higher
power at the lower end of the $f$-axis. The 1/f-type shape power spectrum is 
due to the overall drift of the $V_t$ across time.
The 1/f noise power spectrum is finally modeled as the combination 
of the 1/f-type power spectrum and a Lorenz profile,
\begin{align}\label{eq:pf}
P_{\rm total}(f) &= \frac{\sigma^2}{\delta \nu} \left(1 + P(f) \right)\nmsk 
&= \frac{\sigma^2}{\delta \nu} \left(1 + \left(\frac{f}{f_k}\right)^{\alpha} 
      + \frac{B}{1 + \left(\left(f - f_{\rm peak}\right)/w\right)^2}\right),
\end{align}
where, $\alpha$, $f_k$, $B$, $f_{\rm peak}$, and $w$ are the parameters that need to be fitted with the measured power spectrum. The solid lines in 
\reffg{fig:gtpow} show the best-fit power spectrum.

With the best-fit noise power spectrum, we can estimate ${\bf g}$ and 
the temporal gain variation can be reconstructed with 
\begin{align}\label{eq:gt}
{\bf g}_t = {\bf F}{\bf g}.
\end{align}
The reconstructed temporal gains for observation 20210314 are shown 
with thick curves in \reffg{fig:gt}.
The temporal gain variation is finally calibrated via
\begin{align}
    V_2 = V_1 / g_t.
\end{align}

\subsection{Absolute flux calibration}\label{sec:abscal}

\begin{figure*}
\centering
\includegraphics[width=\textwidth]{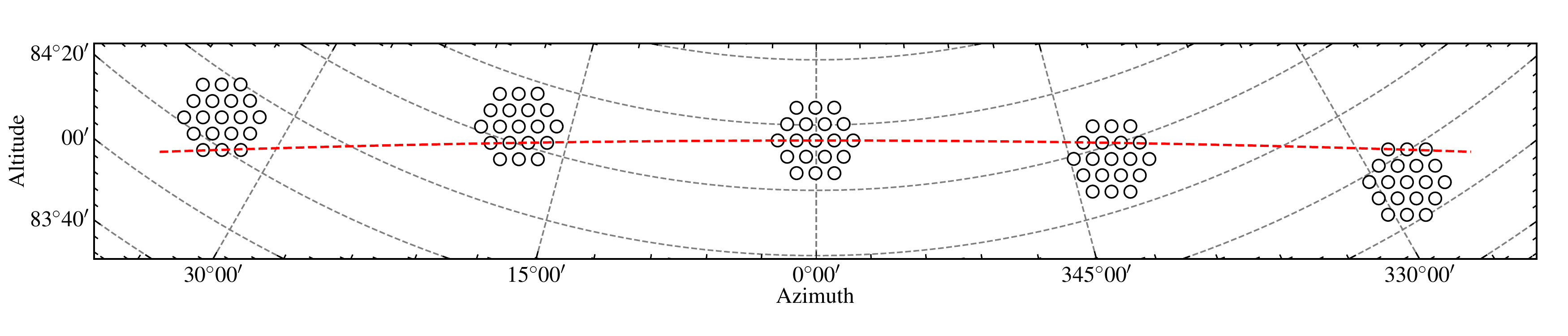}
\caption{
The footprint of the calibration observation.
The red dashed line represents 3C286's drift route as it crosses the meridian 
in the local alt-az coordinate at FAST site.
The calibrator drifts through each line of feeds with five 
different pointing directions. 
The circles packed in hexagons illustrate the position of the 19 feeds in each pointing direction. 
}\label{fig:calscans}
\end{figure*}

\begin{figure*}
\centering
\includegraphics[width=0.9\textwidth]{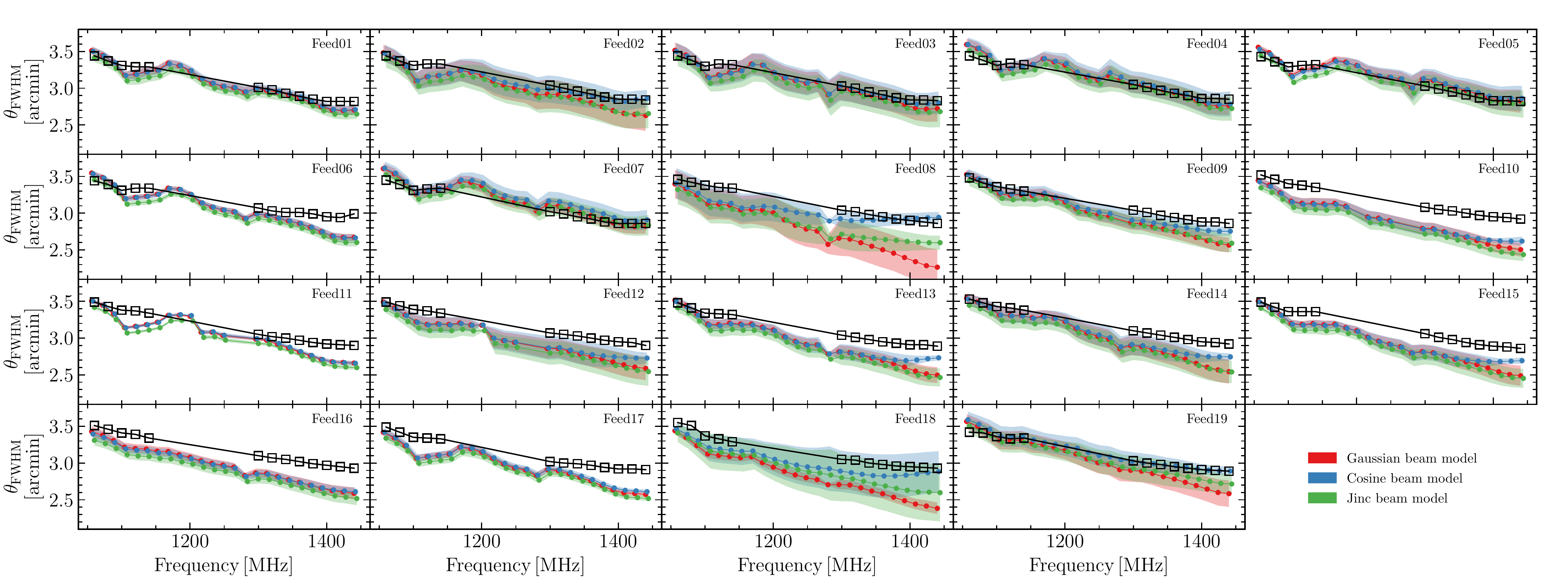}
\caption{
$\theta_{\rm FWHM}$ measurements using 3C286 drift scan observation. 
The circle markers show the mean of the best-fit $\theta_{\rm FWHM}$
across the measurements of different days and the filled region 
indicates the corresponding rms.
The results of different beam models are shown in different colors. 
The black squares show the measurements from \citet{2020raa....20...64j}.
}\label{fig:fwhm}
\end{figure*}

The absolute flux calibration is done by multiplying the temporal gain calibrated data $V_2$ with the noise diode spectrum,
\begin{align}
    T = V_2 \times T_{\rm ND},
\end{align}
where $T$ is referred below as the calibrated data, and $T_{\rm ND}$ as
the temperature of the noise diode.
The noise diode temperature is measured via a series of hot load measurements
\citep{2020raa....20...64j} and it is assumed to be stable during the observations.
During our observations, we performed several absolute flux
calibrations using known celestial calibrators.

The absolute flux calibrations were made in drift scan mode.
The 19 feeds were grouped into $5$ east-west lines. With $5$ different 
pointings, the calibrator drifted across each feed in the same
east-west line. To minimize systematic differences compared to 
the target field observation, a calibrator with its Dec.
close to the target field is required. We chose 3C286 as our flux calibrator
and performed the calibration observation after the target observation of each day.
The calibration pointing direction is shown in \reffg{fig:calscans}.

The observation time is long enough to have the calibrator fully 
transits across the beam. The great-circle distance between the pointing
direction and the calibrator, $\theta$, is calculated with,
\begin{align}
\cos(\theta) = \sin(\delta_{\rm p}) \sin(\delta_{\rm cal}) 
        + \cos(\delta_{\rm p}) \cos(\delta_{\rm cal}) 
          \cos(\alpha_{p} - \alpha_{\rm cal}),
\end{align}
where $(\alpha_{\rm p}, \delta_{\rm p})$ and $(\alpha_{\rm cal}, \delta_{\rm cal})$
are the R.A. and Dec. of the pointing direction and
the calibrator, respectively. We use the data within the time range
with $\theta<3\,{\rm arcmin}$ as the source-on power $V_{\rm on}(\theta)$, and average 
across the time range with $\theta>12\,{\rm arcmin}$ as source-off power $V_{\rm off}$.

During the calibration, the noise diode is also fired in the same way as
the target field observation. The data are firstly calibrated 
against the noise diode power, $V_{\rm ND}$, to cancel the bandpass gain.
The corresponding main beam brightness temperature of the calibrator is
\begin{align}
T_{\rm cal}B(\theta) = \frac{V_{\rm on}(\theta) - V_{\rm off}}{V_{\rm ND}} T_{\rm ND},
\end{align}
where $B(\theta) = \exp\left[-\frac{\theta^2}{2\sigma^2}\right]$ is the normalized 
beam pattern with $\sigma=\theta_{\rm FWHM}/ (2\sqrt{2\ln 2})$.
The antenna temperature is converted from the source flux density via
\begin{align}\label{eq:Ta}
T_{\rm cal} = \frac{\eta_{\rm M} \lambda^2 }{2 k_{\rm B} \Omega_{\rm MB}} S,
\end{align}
where $k_{\rm B}$ is the Boltzmann constant, $\lambda$ is the wavelength,
$\Omega_{\rm MB}$ is the main beam solid angle and $\eta_{\rm M}$ is the main beam
efficiency. Assuming a symmetric Gaussian beam with half power beam width 
$\theta_{\rm FWHM}$, the main beam
solid angle is given by,
\begin{align}
\Omega_{\rm MB} = 2 \pi \left(\frac{\theta_{\rm FWHM}}{2\sqrt{2\ln 2}}\right)^2
\approx 1.133 \theta_{\rm FWHM}^2,
\end{align}
The spectrum flux density of 3C286 can be modeled as \citet{2017ApJS..230....7P},
\begin{align}
\log \left(\frac{S}{{\rm Jy}}\right) = 
1.2481 - 0.4507 x  - 0.1798 x^2 + 0.0357 x^3,
\end{align}
in which, $x=\log \left(\frac{\nu}{\rm GHz}\right)$. 
The noise diode spectrum is evaluated by minimizing the following 
residual function for each feed, frequency, and polarization,
\begin{align}
\chi^2 = \sum_{\theta<3\,{\rm arcmin}} \left| \frac{\lambda^2 B(\theta) S}{2k_{\rm B}\Omega_{\rm MB}}
- \frac{V_{\rm on}(\theta) - V_{\rm off}}{V_{\rm ND}} \frac{T_{\rm ND}}{\eta_{\rm M}} \right|^2.
\end{align}
where $T_{\rm ND}$ is the parameter to be determined.

We can also leave the half power beam width, $\theta_{\rm FWHM}$, as another free parameter
fit with the observation data. 
In order to model the sidelobes, we use the Jinc function beam model, 
\begin{equation}
B_{\rm Jinc} = 4 \left(\frac{J_1\left(\pi \theta / \theta_{\rm FWHM}\right)}
{\pi \theta / \theta_{\rm FWHM}}\right)^2,
\label{eq:jinc}
\end{equation}
where $J_1(x)$ is the Bessel Function of the First Kind;
and the cosine beam model,
\begin{equation}
B_{\rm cos} = \left( \frac{\cos \left( 1.189 \pi \theta / \theta_{\rm FWHM} \right) }
{1 - 4 (1.189\theta/\theta_{\rm FWHM})^2} \right)^2,
\label{eq:cosine}
\end{equation}
which is known to have lower sidelobes compared to the Jinc function \citep{2020arXiv201110815M}.
We fit $\theta_{\rm FWHM}$ at each frequency with initial frequency resolution 
of $7.6$ kHz. 
In order to reduce the variance, the best-fit values are then averaged in 
each $16$ MHz frequency bin.
The best-fit $\theta_{\rm FWHM}$ of the XX polarization from different beam models 
are shown in \reffg{fig:fwhm} with different colors.
The circle markers show the mean $\theta_{\rm FWHM}$ across the measurements in different 
days and the filled region indicates the corresponding rms.
The beam width reported in \cite{2020raa....20...64j} is shown
with the black square markers. 
The best-fit $\theta_{\rm FWHM}$ of Feed 01, which is in the center of the FAST 19-feed array shows consistent 
results across different days and beam models. 
Meanwhile, it is also consistent with $\theta_{\rm FWHM}$ reported in \cite{2020raa....20...64j}. 
However, the best-fit results of the other feeds reveal significant scattering
between various days, for example, Feed 02, or when using a different beam model,
for example, Feed 08 and Feed18. Some of the results, for example, Feed 10, show deviation 
from the results of \cite{2020raa....20...64j}. 

A possible reason for this is that here we assumed a symmetric beam profile,
but in reality, the beams are asymmetric. 
With a single transit observation, we can only measure the beam profile
across one direction for both the XX and YY polarizations.
As shown in \cite{2020raa....20...64j}, the full beam shape is significantly asymmetric and can be well fit 
using a 'skew Gaussian' profile, which takes into account the ellipticity.
A complete analysis needs more observation and we will improve the measurements 
in further work.
In the rest of the analysis, we interpolate the beam width using the 
results reported in \cite{2020raa....20...64j}. 

\begin{figure*}
\centering
\includegraphics[width=\textwidth]{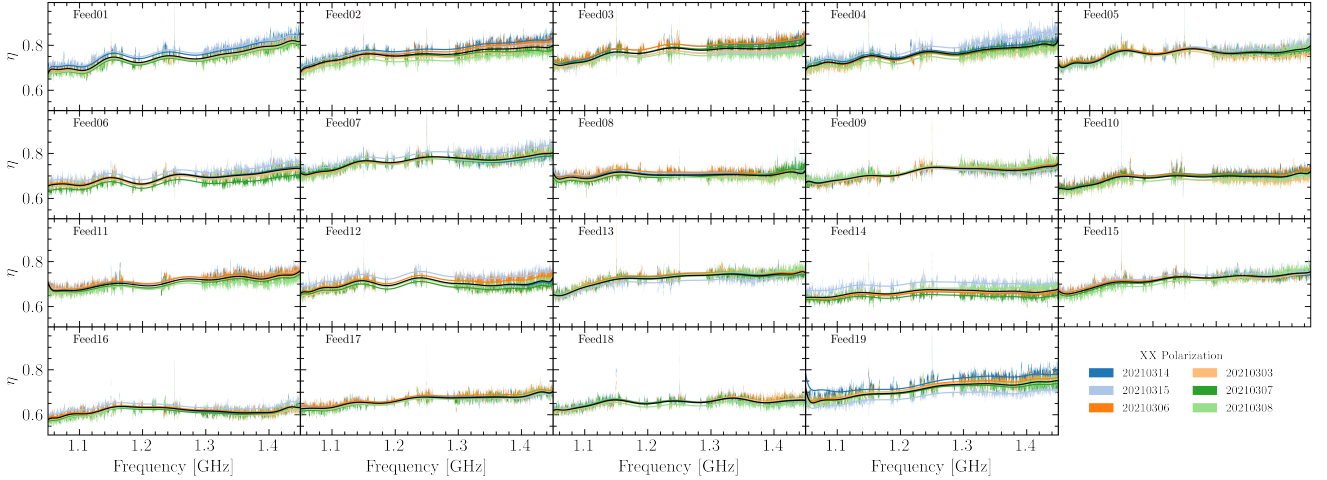}
\caption{
The best-fit $\eta$ for different measurements;
}\label{fig:eta1}
\end{figure*}

Because the noise diode signal is injected into the receiver system
between the feed and low-noise-amplifier (LNA) \citep{2020raa....20...64j}, its sky-source-calibrated spectrum is slightly different from the noise diode spectrum model, which is measured using the hot-load. The difference is parameterized as,
\begin{align}
\epsilon = \frac{T_{\rm ND}^{\rm model}}{T_{\rm ND}},
\end{align}
where $T_{\rm ND}^{\rm model}$ is the noise diode spectrum model and 
$T_{\rm ND}$ represents the noise diode spectrum determined using the celestial calibrator.
In fact, $\epsilon$ degenerates with the main beam efficiency of FAST, $\eta_{\rm M}$.
Thus $\epsilon$ and $\eta_{\rm M}$ are combined as the total aperture efficiency 
$\eta = \eta_{\rm M} \epsilon$ and determined using the celestial calibrator for each 
frequency, polarization, and feed.

It is known the aperture efficiency of FAST is weakly dependent on the Zenith Angle (ZA) with
${\rm ZA}<26\overset{\circ}{.}4$; and decrease quickly with ZA beyond $26\overset{\circ}{.}4$
\citep{2020raa....20...64j}. Because our calibration observations were always 
carried out when 3C286 is near its transit time, the pointing directions
of the same feed are relatively consistent between different days and 
the ZA are all within $26\overset{\circ}{.}4$. Thus $\eta$ is assumed to be relatively
stable between days for our calibration observations.
The measured $\eta$ on different days are shown  in \reffg{fig:eta1}. 
Generally, the measured $\eta$ varies between different feeds but keeps a similar shape
between different days for the same feed.
The measured $\eta$ on each day is contaminated by RFI. In order to fill in the RFI gaps, 
we produce a template $\bar{\eta}$ for each feed by taking the median values of $\eta$ across the measurements 
on different days and fitting with a $15th$-order polynomial function.
The $\eta$ template is shown with the black solid line in \reffg{fig:eta1}. 
To recover the variations of $\eta$ between different days, 
the $\eta$ measurement on each day is fitted to the template via,
\begin{align}
{\eta} = (p_0 + p_1 \nu + p_2 \nu^2 + p_3 \nu^3) \bar{\eta},
\end{align}
where $\{p_0, p_1, p_2, p_3\}$ are the parameters.
The best-fit ${\eta}$ for different measurements is shown in \reffg{fig:eta1} with
solid curves in the same colors as the corresponding measurements.
The sky absolute flux density is finally obtained as
\begin{align}
T = V_2 \times T_{\rm ND}^{\rm model} / {\eta}.
\end{align}

\subsection{Temporal baseline subtraction}\label{sec:baseline}

\begin{figure*}
\centering
\includegraphics[width=0.9\textwidth]{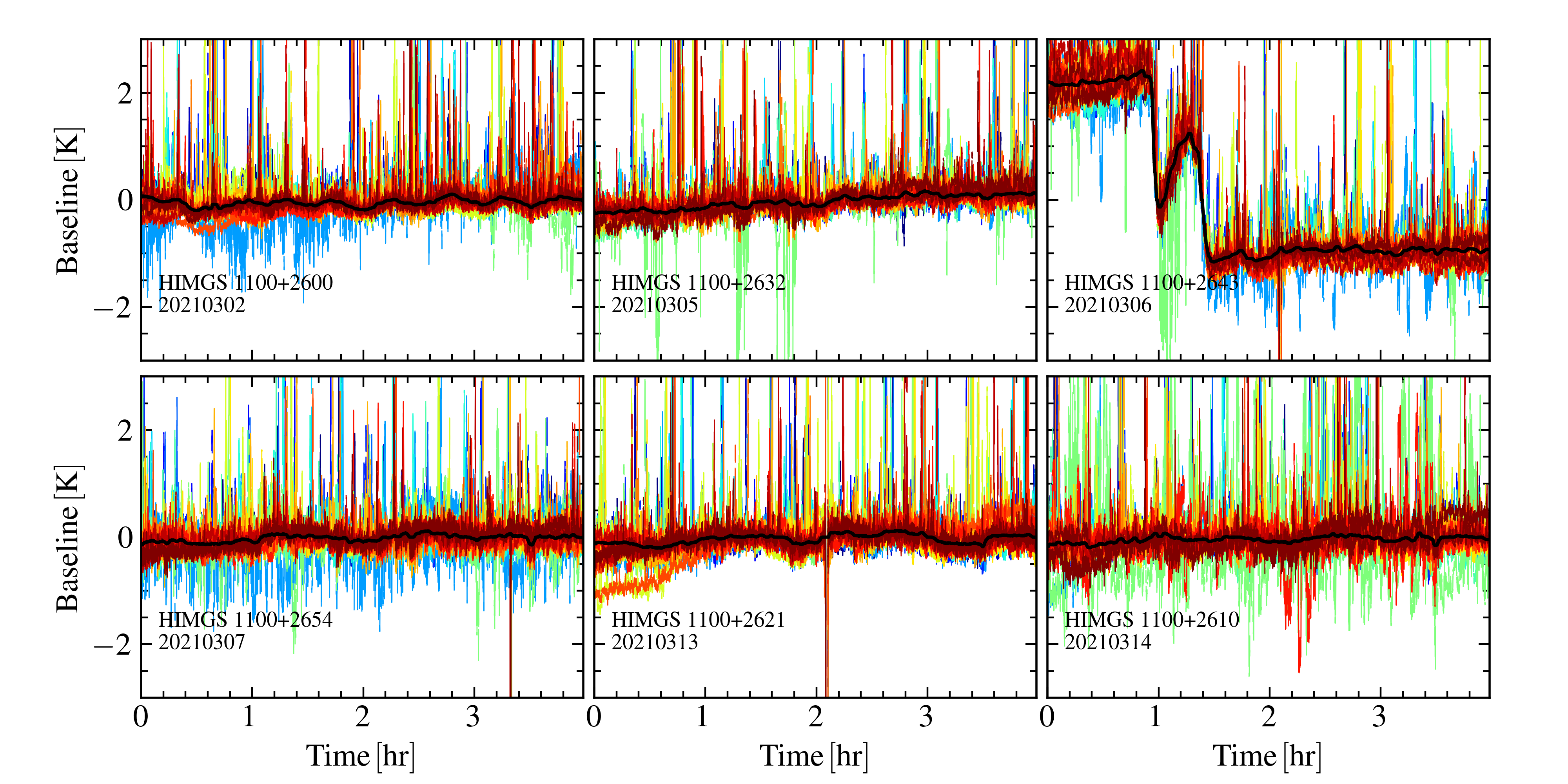}
\caption{
The temporal variation of baselines for the 6 days of observation. 
The different color shows different feeds. 
The temporal baselines for each feed are centered by subtracting the mean 
across the full observation, individually. The positive peaks are due to the
bright continuum sources, and the few negative spikes are bad data due to the 
strong RFI contamination. 
}\label{fig:bsl}
\end{figure*}

We shall call the average of the calibrated data across the frequency band as the {\it baseline} of the data.
The baselines for $6$ days observation are shown in \reffg{fig:bsl} with different colors. 
The baselines are centered by subtracting the mean across the full observation time. 
The positive peaks in this data are due to bright continuum sources. 
Due to unknown reasons, some feeds occasionally perform badly during the observation, 
producing significantly larger fluctuations. 
Because different feeds point to different sky positions, we average the baselines 
across feeds to eliminate the flux variation from the sky.
Such averaging across different feeds also reduces the baseline variance.
Some of the data are contaminated by very strong RFI from satellites, which are fully 
flagged across the full frequency band. 
However, the data adjacent to these bad times are not flagged across the full frequency range.
As the baseline is estimated by taking the median value across the frequency band, those partially flagged time stamps may have significantly lower median values than the rest,
and result in negative spikes after subtracting the temporal mean.

In most cases, the baselines are steady, though the 20210306 data show some significant sharp variations, 
for which the reason is unknown.  The shape of the baselines for different feeds is generally consistent 
as expected, even for the 20210306 observation. This might indicate the baseline variation is due to a systematic 
background noise level variation during the observation time.
We take the median across the $19$ baselines to get rid of 
the spikes and further smoothed across along the time with a median
value filter. The smoothed mean baseline is shown with the thick black curve. 
The smoothed baseline is fit to the calibrated TOD at each frequency, 
\begin{align}
{\bf A} = \left({\bf b}^\T{\bf b}\right)^{-1} {\bf b}^\T {{\bf T}},
\end{align}
where ${\bf b}$ represents the $n_{\rm t} \times 1$ vector of baseline template, 
$\bf T$ represents the $n_{\rm t}\times n_{\nu}$ matrix of TOD and 
${\bf A}$ is the $1\times n_{\nu}$ vector of the fitting parameter.
The baseline is subtracted via
\begin{align}
{\bf T}^{\rm c}_\nu = {\bf T}_\nu - {\bf b}{\bf A} ,
\end{align}
where ${T}^{\rm c}$ represent the baseline centred TOD.

\begin{figure*}
\centering
\includegraphics[width=0.95\textwidth]{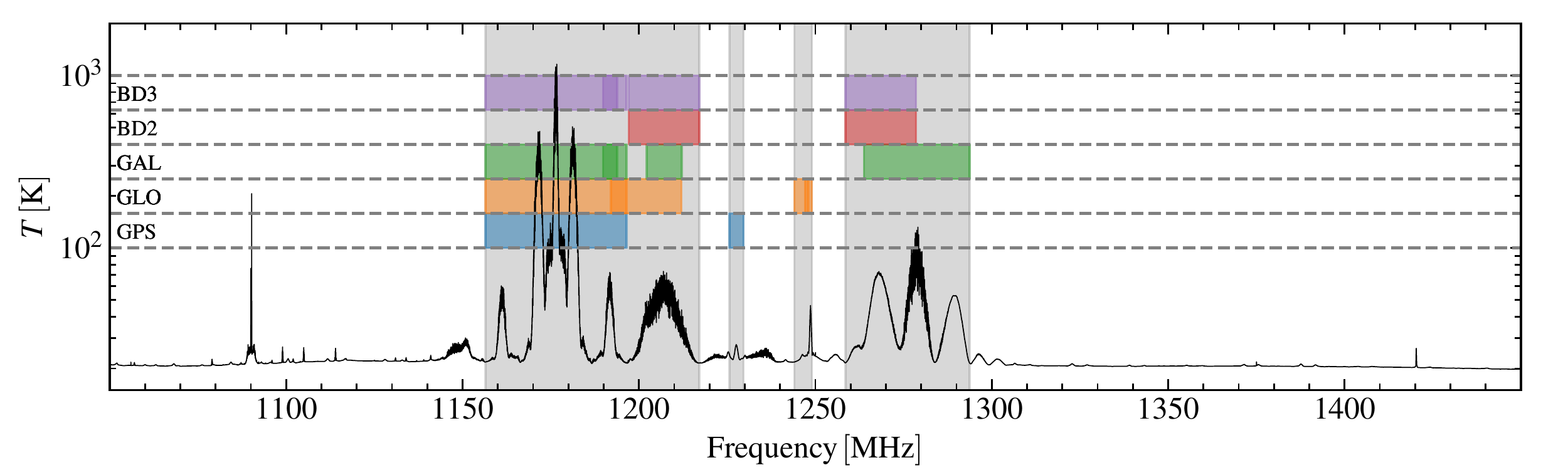}
\caption{
The spectrum from a feed averaged over $\sim 30$ min. The channels contaminated by RFI from GNSS are marked by the gray region. 
The allocated frequency channels used for the GPS, Galileo (GAL), Glonass (GLO), and Beidou (BD2/BD3) are marked in different colors \citep{handbook.gnss}.
}\label{fig:rfi}
\end{figure*}

\begin{figure*}
\centering
\includegraphics[width=\textwidth]{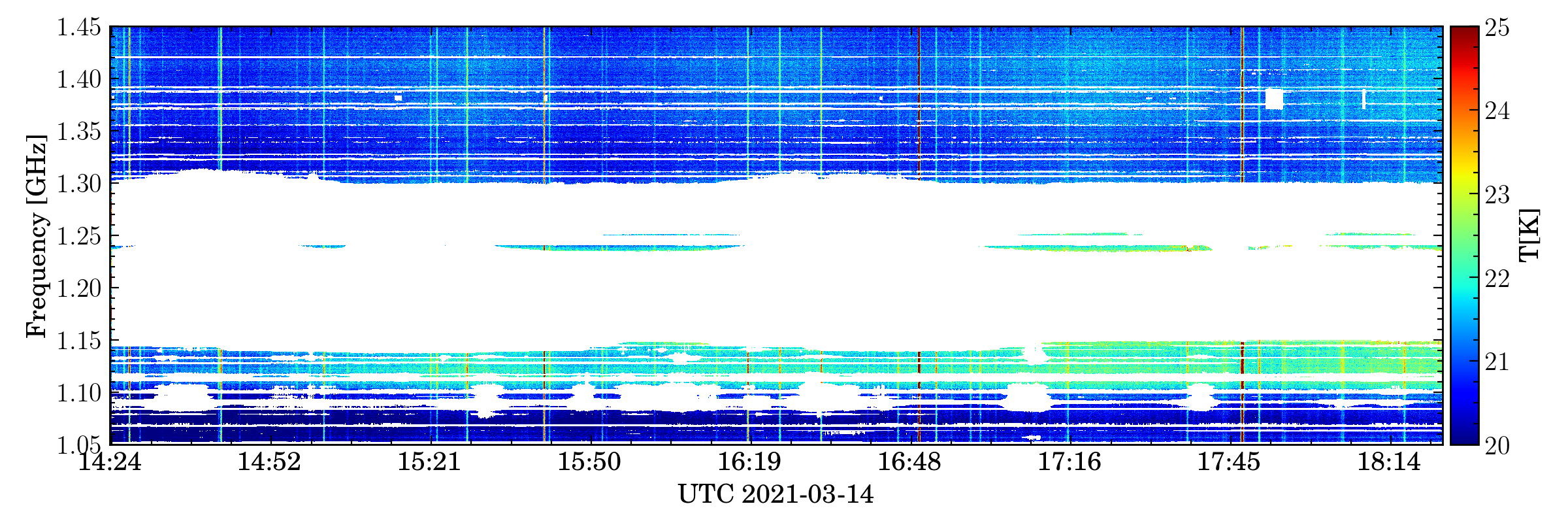}
\caption{
Waterfall plot of $\sim4\,{\rm hr}$ TOD observed with Feed01 XX polarization.
The blank region is the RFI-flagged data.
}\label{fig:rfitod}
\end{figure*}

\subsection{RFI flagging}\label{sec:rfi}

In \reffg{fig:rfi}, the black curve shows the spectrum from one feed averaged over 
$\sim 30\, {\rm min}$. 
There is strong RFI contamination in the frequency band between $1150\,{\rm MHz}$ to 
$1300\,{\rm MHz}$, which are produced by the Global Navigation Satellite Systems (GNSS), 
including the GPS, Galileo, Glonass, and Beidou \citep{handbook.gnss}.
The frequency channels allocated for GNSS are marked in different colors.
The contamination of such GNSS bands can leak into the neighboring channels
due to the extended GNSS signal spectrum profile.
We first remove the channels badly contaminated by GNSS, 
including those allocated frequency channels, as well as the neighboring 
channels within $1\sim20\,{\rm MHz}$.
The pre-removed frequency channels are shown with the gray area in
\reffg{fig:rfi}.

\begin{figure}
\centering
\includegraphics[width=0.47\textwidth]{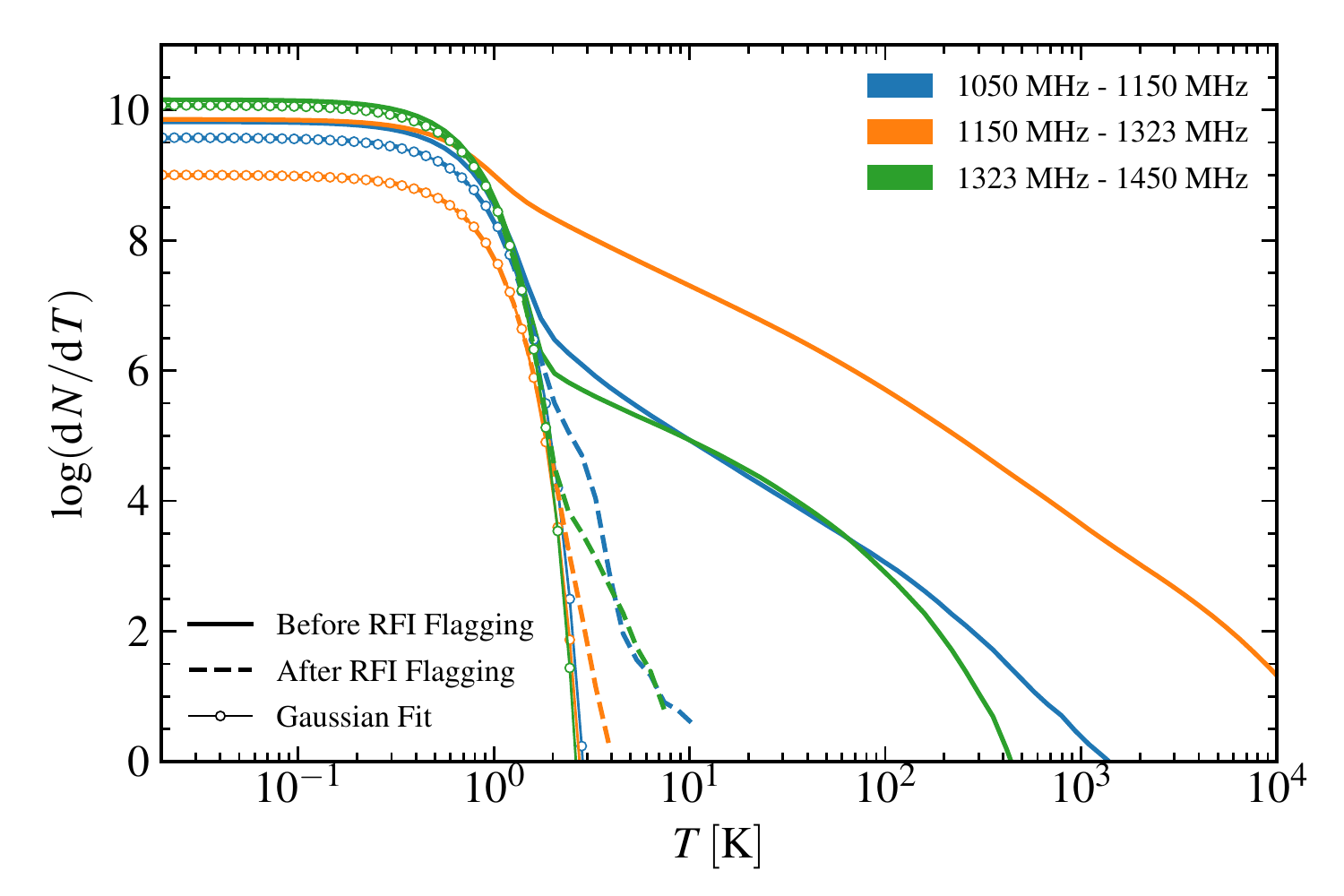}
\caption{
Histogram statistic of the temperature difference before and after
the RFI flagging.
}\label{fig:rfihist}
\end{figure}

We then apply the {\tt SumThreshold} and {\tt SIR} (Scale-Invariant Rank) RFI flagging program
\citep{2010MNRAS.405..155O,2012A&A...539A..95O,2021A&C....3400439Z} 
to the bandpass calibrated data. The {\tt SumThreshold} algorithm searches for 
consecutive points of different numbers (increase from 1 to $2^n$) in TOD as the potential RFI 
contaminated points which have value above certain preset thresholds, 
i.e. $\chi_n = \chi_1/1.5^{\log_2 n}$, where $\chi_1 = 10$ is the initial threshold
and $n$ is the number of data sample considered.
The thresholds are varying according to the number of data samples considered.
In order to find the extra weak contamination near the flagged high values, 
the {\tt SIR} RFI flagging is then applied.
The {\tt SIR} RFI flagging method uses the one-dimensional mathematical morphology technique to find 
the neighbored intervals in the time or frequency domain that are likely to be affected by RFI.

Compared to other radio telescopes, the FAST is much more sensitive, with many genuine celestial radio sources, 
and even the \hi emissions from nearby galaxies could be detected with a high signal-to-noise ratio (SNR) 
in the raw data, therefore, care must be taken to avoid removing them by mistake. 
We stack the same time outputs of the different feeds, which would lower the 
celestial source signal as the different feeds are pointed at slightly different sky directions at any given time
while enhancing the RFIs which enter through the beam side lobe and are simultaneous on all feeds.  
Also, the {\tt SumThreshold} algorithm is only applied along the frequency axis.

As an example, the RFI flagged 20210314 data for Feed01 XX polarization is shown in a waterfall plot \reffg{fig:rfitod}.
The blank regions are flagged data.
Most frequency bands between $1150\,{\rm MHz}$ and  $1323\,{\rm MHz}$ are
flagged as they are badly contaminated by the GNSS. 
Another severely RFI contaminate part is  
around $1090\,{\rm MHz}$, which is the band allocated to aircraft  
Automatic Dependent Surveillance-Broadcast (ADS-B) data 
communication, where the RFI occurs frequently during the 4-hour observation.
The frequency range beyond $1323\,{\rm MHz}$ is relatively free of RFI.

To check the RFI residual after flagging, we show a histogram of the 
data. The contribution of natural continuum emission can be removed by taking the difference 
between the two neighboring frequency channels. 
The histograms are produced using data in three different frequency ranges, i.e.
$1050-1150~{\rm MHz}$, $1150-1323~{\rm MHz}$ and $1323-1450~{\rm MHz}$, and 
the results are shown in \reffg{fig:rfihist}. 
The results of the data before RFI flagging are shown with solid lines and those after 
RFI flagging is shown with dashed lines. 
Before RFI flagging, the data shows a combined profile of a Gaussian distribution and 
a high-temperature tail, which indicates a significant RFI contamination. 
After the RFI flagging, the high-temperature tails are greatly
reduced in all three frequency bands.
The histogram statistic also shows that, with the {\tt SumThreshold}
flagging, there are $34.7\%$, $79.2\%$ and $12.2\%$ data flagged
in the three frequency bands, respectively; and with 
the additional SIR flagging, another $6.5\%$, $8.5\%$ and $2.0\%$
data are flagged. For the data on other days, the ratios are more 
or less similar.

The frequency channels within $1150$-$1323~{\rm MHz}$ are badly contaminated by
the strong RFI contamination. Especially, the channels at the lower-end 
of the frequency band, $1250$-$1450$ MHz contain some GNSS signal bands.
We adjust the valid frequency range of the high-frequency band to $1323$-$1450$ MHz.

\subsection{Maps}

\begin{figure*}
\centering
\includegraphics[width=\textwidth]{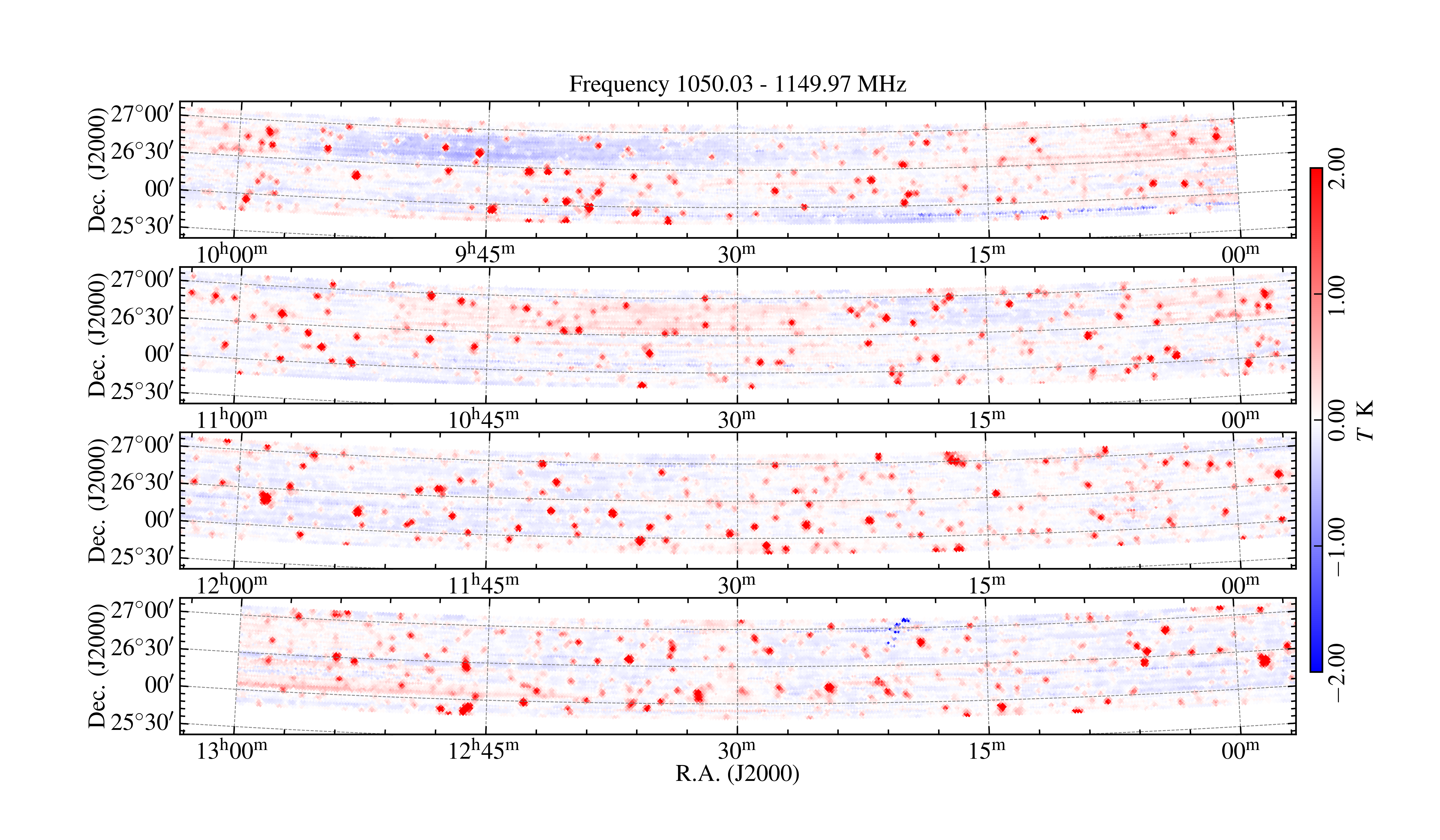}
\includegraphics[width=\textwidth]{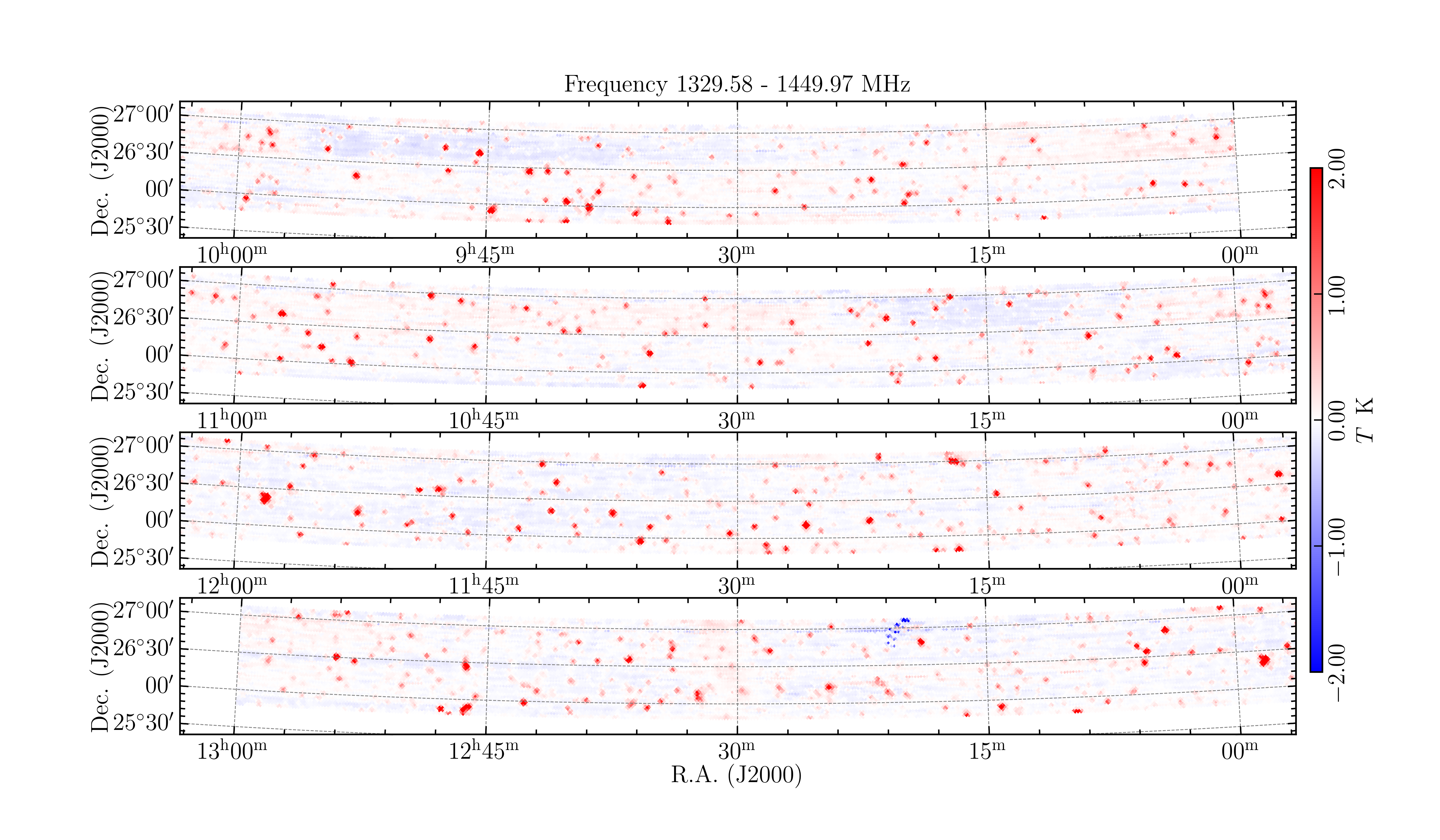}
\caption{Total intensity map of the surveyed regions. The top and bottom panels show the integrated intensity between 1050-1150 MHz and 1323-1450 MHz respectively. 
}\label{fig:map}
\end{figure*}

The calibrated data are zero-centered by subtracting the baseline
and the XX and YY polarization are combined into Stokes I.
The TOD is then projected to the map domain via the standard map-making procedure  \citep{1997ApJ...480L..87T}.
In order to save the computation time for map-making,  we re-binned the data to $28\,{\rm kHz}$ frequency resolution. 
The maps are made for each frequency without considering the correlation between different frequency channels.

We use the variance across time of the $\sim30\,{\rm min}$ data block for each feed and polarization
as the noise variance. The noise is assumed to be uncorrelated between different
time blocks, and its covariance matrix is assumed to be diagonal, i.e. 
${\bf N} = {\rm diag}\{\sigma^2(t)\}$. 
The map is obtained by 
\begin{align}
\hat{{\bf m}} = \left({\bf P}^\T {\bf N}^{-1} {\bf P}\right)^{-1} {\bf P}^\T {\bf N}^{-1}{\bf T},
\end{align}
in which, ${\bf P}$ is the pointing matrix, which relates the time to the map 
coordinate. We use the HEALPix scheme for the sky with ${\rm NSIDE}=2048$,
corresponding to a pixel size of $1.72\,{\rm arcmin}$ (pixel area of $2.95\,{\rm arcmin}^2$). 

\reffg{fig:map} shows the frequency-averaged maps.
The maps made with the data of two RFI-free frequency bands are shown in
the top and bottom panels, respectively. 
As the surveyed sky is a long strip, we divide the full region into several 
pieces in the R.A. direction, each is about $1~{\rm hr}$ (i.e. $15^\circ$) in R.A. 
We can see many point sources on the map clearly, and the point sources in 
the $1050$-$1150$ MHz band has good correspondence with the
point sources in the $1323$-$1450$ MHz band, which is what we would expect for the radio continuum 
sources such as quasars and radio galaxies. 

In order to improve the flux measurements of point sources, we applied an alternative map-making
procedure similar to \citet{2018ApJ...861...49H}
\begin{align}
\hat{m}_p = { \left[ \left({\bf P}{\bf K}\right)^{\rm T} {\bf T} \right]_p} \Big/ { \left[ \left( {\bf P} {\bf K} \right)^{\rm T} {\bf I} \right]_p },
\end{align}
where $p$ indicate the $p$-th pixels of the map, ${\bf I}=\{1, 1,\cdots,1\}^{\rm T}$
is a column vector with all elements equal to $1$ and ${\bf K}$ represents a kernel function 
that paints the antenna temperature to the nearby pixels.
We use a Gaussian kernel function
\begin{align}\label{eq:todkernel}
K_{pq} = \exp\left[-\frac{1}{2}\left(\frac{r_{pq}}{\sigma_K}\right)^2\right],
\end{align}
where $K_{pq}$ the element of the ${\bf K}$ matrix, $r_{pq}$ is the great circle 
distance between the $p$-th and $q$-th pixels of the map and $\sigma_K$ indicates 
the kernel size. We set $\sigma_K = 1.5\,{\rm arcmin}$ 
and use HEALPix scheme with ${\rm NSIDE}=4096$,
corresponding to a pixel size of $0.86\,{\rm arcmin}$.
To make the map-making process easier, the TOD from $1375$ MHz to $1425$ MHz
are averaged into a single frequency channel before the map-making.
Such a map is used for flux comparison with the continuum sources.
We extract the flux of point sources and compare them with the NVSS continuum measurements. 
Finally, $81$ isolated point sources with flux over $14$ mJy are identified within 
the surveyed region. The flux measurements are presented in \reftb{tab:sources} and 
the detailed discussion can be found in \refsc{sec:fluxmap}.

As the observed data is already the convolution of the sky signal and the telescope 
beam pattern, using the kernel function \refeq{eq:todkernel} to create the map is 
equivalent to an additional convolution. 
If we assume a Gaussian beam model with a beam width of $\theta_{\rm FWHM}$,
the final map is then smoothed with a Gaussian function with a kernel size of
$\sigma_{K'} = \sigma_{K} + \theta_{\rm FWHM}/\left(2\sqrt{2\ln 2}\right)$.

\section{Discussion}\label{sec:disc}

%\begin{figure*}
%\includegraphics[width=\textwidth]{plots_ripple_obvious.pdf}
%\includegraphics[width=\textwidth]{plots_ripple.pdf}
%\caption{
%The measured delay spectrum of the data observed using the central feed. 
%The delay spectra with and without obvious peak of the standing wave,
%i.e. using data 20210313 and 20210302, are shown in the top and bottom panels, respectively.
%The delay spectra of the uncalibrated data, the bandpass, and the bandpass calibrated 
%data are shown in the left, middle, and right panels, respectively.
%The blue curves in the middle and right panels show the results using the bandpass 
%smoothed with a Butterworth filter, while the orange curves show the results using the Hanning filter instead.
%The dashed line indicates the expected ripple frequency of $0.9\,{\mu s}$ as reported in the literature. 
%}\label{fig:pdfripple}
%\end{figure*}

\subsection{Bandpass ripple}\label{sec:ripple}

The bandpass calibration changes the shape of the spectrum significantly. To understand the ripple structure in the bandpass, we estimate the delay spectrum  $|\tilde{V}(\tau)|^2 = \langle \delta \tilde{V}(\tau) \delta \tilde{V}^*(\tau) \rangle$,
where 
$$\delta \tilde{V}(\tau) = \int {\rm d}\nu~ \delta V(\nu) e^{-2\pi i \nu \tau}$$ 
is the Fourier transform of the data across the frequency. We make the analysis with the data contrast, i.e.
$\delta V(\nu) = \frac{V(\nu)}{\bar{V}} - 1$, where $V(\nu) = \langle V(t, \nu) \rangle_t$ is the data value at frequency $\nu$ averaged across observation time,
and $\bar{V} = \langle V(\nu) \rangle_{\nu}$ is the mean $V(\nu)$ averaged across the frequency band. The delay spectrum is taken for the data before the bandpass calibration, after bandpass calibration, and also for the measured bandpass itself.
The bandpass ripple structure would show up as a peak in the delay spectrum at a particular delay value. 
 
\begin{figure}
\centering
\includegraphics[width=0.47\textwidth]{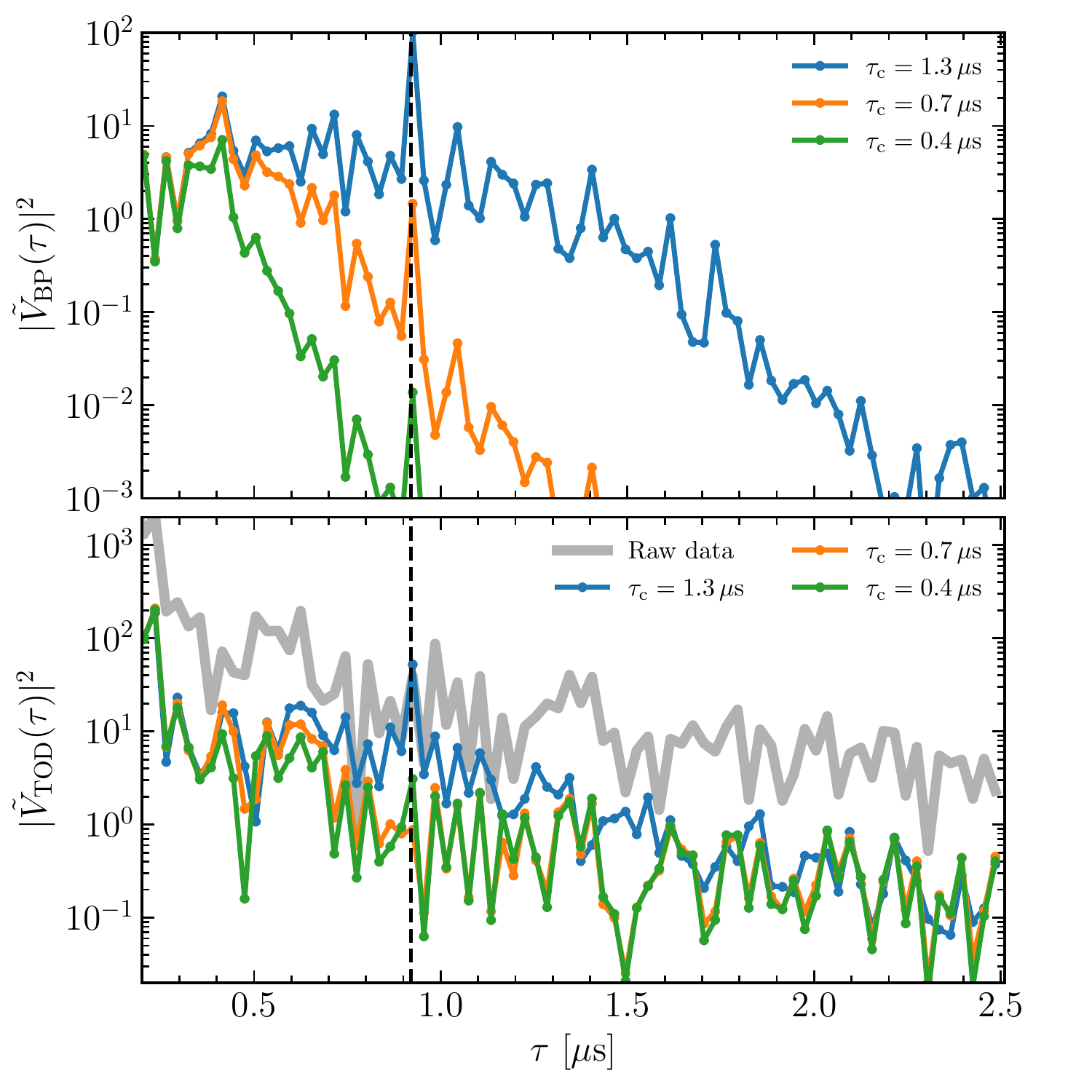}
\caption{
Top panel: the delay spectrum of the bandpass. The delay spectra of
the bandpass smoothed with different window function sizes are 
shown in different colors. Bottom panel: the delay spectrum of the TOD. 
The gray curve shows the delay spectrum of the TOD before bandpass calibration,
while the rest curves show the results of bandpass-calibrated data using
bandpass measurement smoothed with different window function sizes.
All the delay spectra are estimated using the XX polarization of the central beam data
of 20210302. The dashed vertical line marks the delay of $\tau\sim0.92\,{ \rm \mu s}$,
associated with the standing wave between the FAST feed and reflector. 
}\label{fig:pdfbp}
\end{figure}

\begin{figure}
\centering
\includegraphics[width=0.47\textwidth]{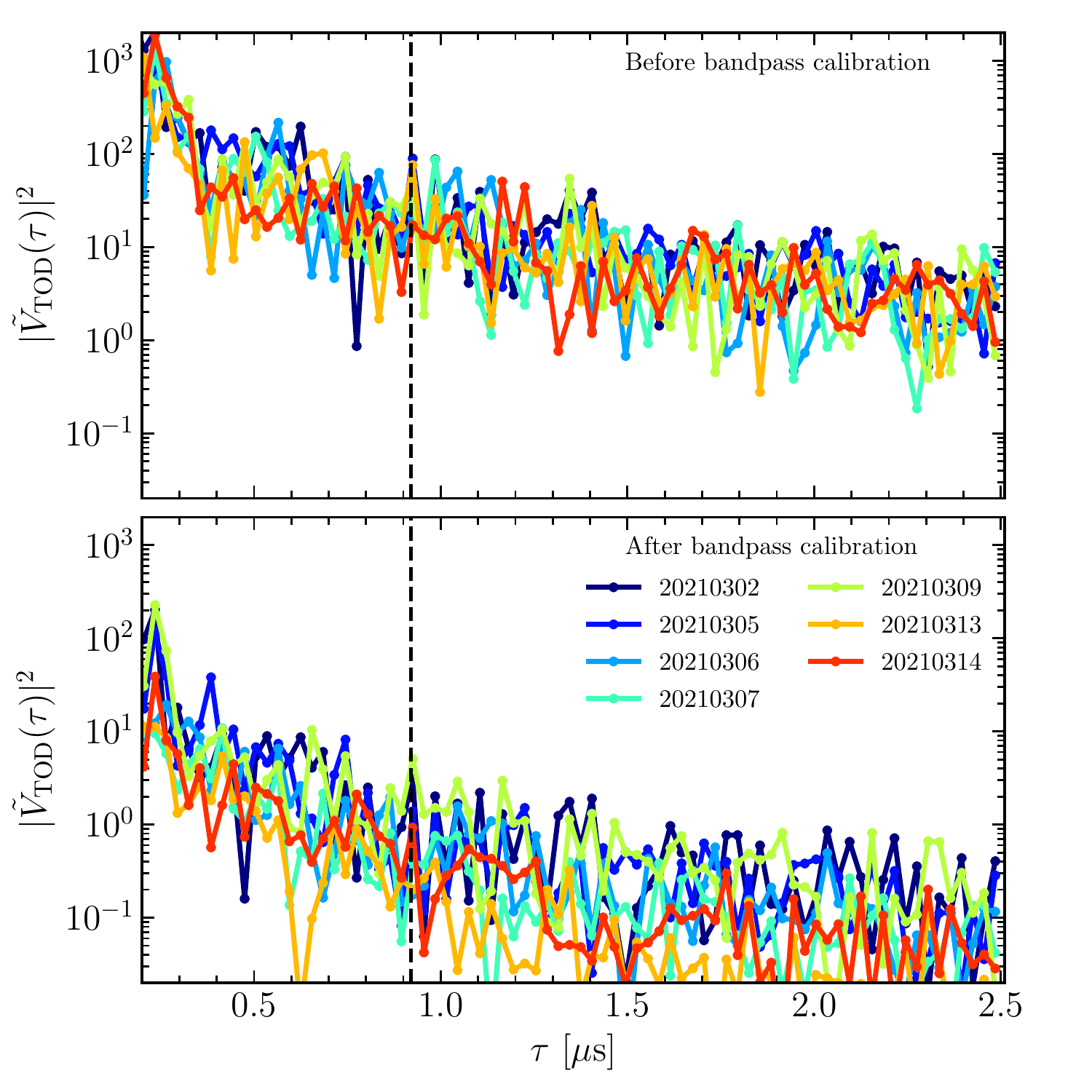}
\caption{
The TOD delay spectrum of different days.
The top panel shows the delay spectra of the TOD before bandpass calibration,
while the bottom panel shows the results after bandpass calibration.
All the delay spectra are estimated using the XX polarization of the central beam data.
The dashed vertical line marks the delay of $\tau\sim0.92\,{ \rm \mu s}$,
associated with the standing wave between the FAST feed and reflector. 
}\label{fig:pdfripple}
\end{figure}

In \reffg{fig:pdfbp} we plot 
the delay spectra of the bandpass (top panel) and the data (bottom panel) of the XX polarization of the center feed during the 20210302 observation. The spectra for other feeds and polarizations are similar. 
For the bandpass (top panel), as mentioned in \refsc{sec:bandpasscal}, we show the results of smoothing with the 3rd-order Butterworth low-pass filter of three different sizes (defined as the 3 dB compression point) $\tau_{\rm c} = 1.3\,{\rm \mu s}$, $0.7\,{\rm \mu s}$ and $0.4\,{\rm \mu s}$, respectively. As expected, the delay spectra are strongly suppressed at the small scale by the smaller-sized filters. For the delay spectra, we show the raw data (i.e. TOD before calibration) and the calibrated data using the bandpass smoothed with the three different filter sizes.  

In the bandpass delay spectrum, we can see an obvious peak which is marked by a dashed vertical line in the figure. This peak is associated with a standing wave between the FAST feed and reflector, with a delay of $\tau \sim 0.92~{\rm \mu s}$, corresponding to a standing wave with a ripple wavelength of $1.087$ MHz in the spectrum. 
However, we do not see a significant standing-wave peak in the raw TOD, the standing-wave signature is more prominent in the bandpass measurements than in the sky observation, probably because this standing wave is induced by the noise diode itself. 
If the data is calibrated with such a bandpass, it would induce the ripple structure which is not present in the raw  data itself. This can be avoided by employing the bandpass smoothed with small-sized filters, e.g. those with $\tau_{\rm c} = 0.7\,{\rm \mu s}$ and $\tau_{\rm c} = 0.4\,{\rm \mu s}$ filters, as shown in the bottom panel of \reffg{fig:pdfbp}. Note the smoothing filter is applied to the bandpass, not the sky spectrum. 

The delay spectra of the TOD before bandpass calibration for all seven nights are shown in the top panel 
of \reffg{fig:pdfripple}. 
Two out of the seven nights' data show weak standing-wave peaks. 
The standing-wave signature is generally consistent across different polarization and beams within the same night's observation. %and is negligible in the raw TOD. 
We use the low-pass filter with the size of $\tau_{\rm c} = 0.7\,{\rm \mu s}$ to suppress both the noise and standing-wave signature during the bandpass determination.
The delay spectra of the bandpass calibrated data are shown in the bottom panel of \reffg{fig:pdfripple}, the standing-wave peak is negligible.

\subsection{Measurement uncertainty}\label{sec:tsys}

\begin{figure*}
\begin{minipage}[t]{0.47\textwidth}
\centering
\includegraphics[width=\textwidth]{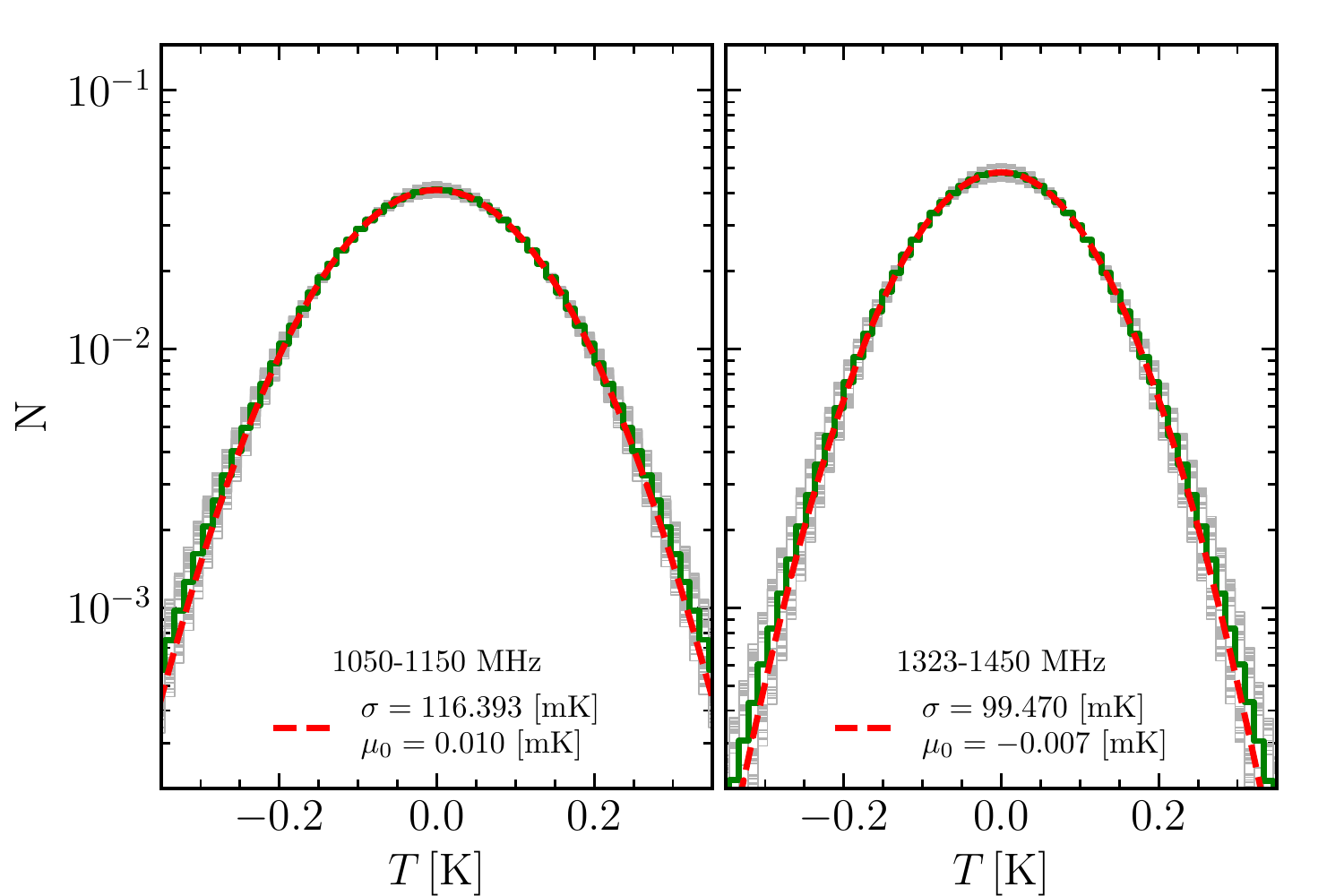}
\caption{
The histogram of the residual TOD.
The solid green curve represents the combined data histogram and
the dashed red line represents the best-fit Gaussian distribution.
The gray thin curves represent the histograms of different data blocks. 
Each curve is normalized by its total number of data points.
}\label{fig:todrmshist}
\end{minipage} \hfill
\begin{minipage}[t]{0.47\textwidth}
\centering
\includegraphics[width=\textwidth]{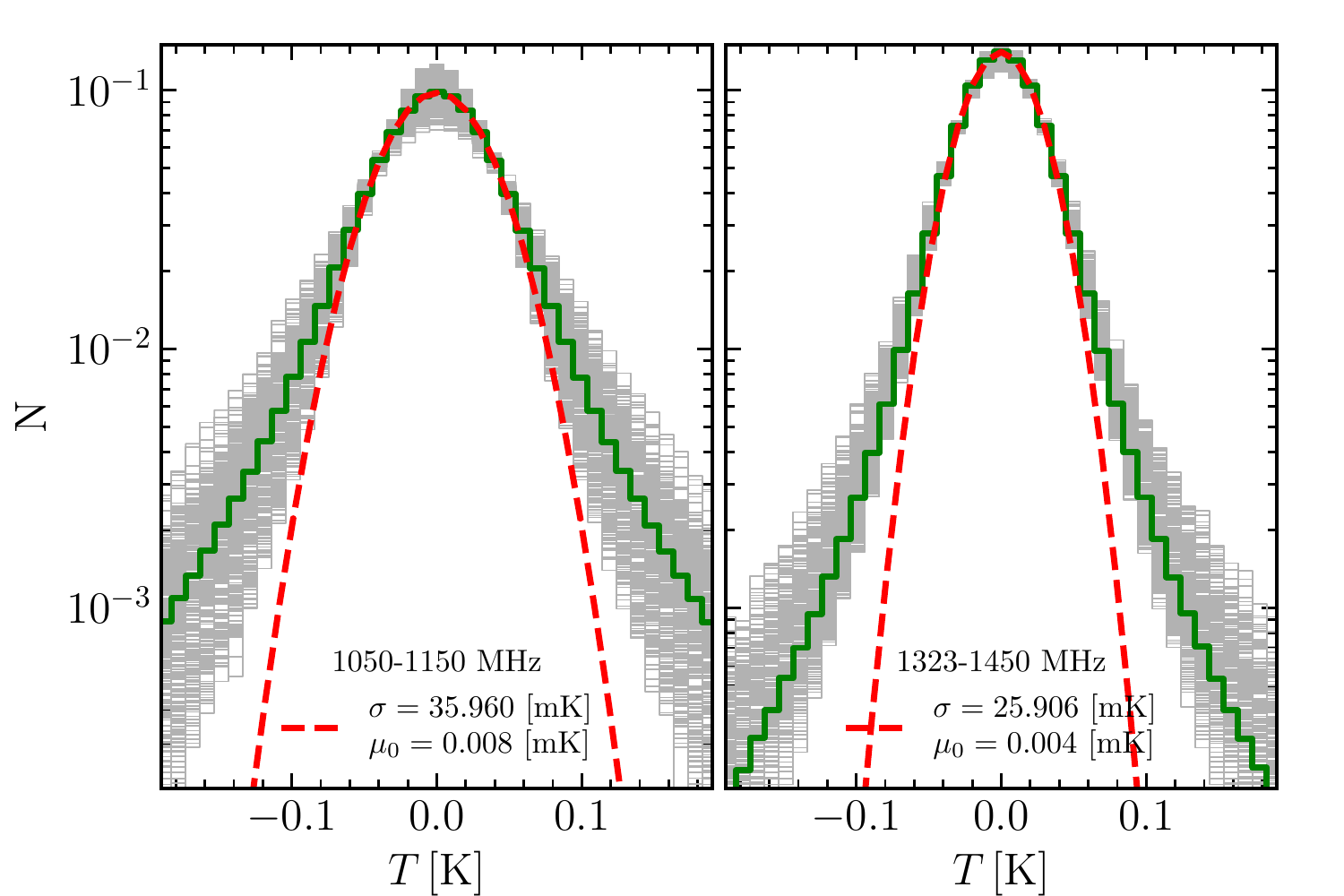}
\caption{
Same as \reffg{fig:todrmshist} but for the residual maps.
The gray thin curves represent the histograms of different frequencies. 
}\label{fig:maprmshist}
\end{minipage}
\end{figure*}

The system temperature $T_{\rm sys}$, is related to the measurement noise level via 
the radiometer equation,
\begin{align}\label{eq:radiometer}
\sigma = \frac{T_{\rm sys}}{\sqrt{N_{\rm pol} N_{\rm feed} \Delta t \Delta \nu}},
\end{align}
where $\sigma$ is the rms of the measurements representing the noise level,
$N_{\rm pol} = 2$ is the number of polarization,
$\Delta t = 1~{\rm s} $ is the integration time, 
$\Delta \nu = 28~{\rm kHz}$ is the frequency resolution and
$N_{\rm feed}$ is the number of the feed.
We estimate the measurement noise level for each of the feeds individually and
adopt $N_{\rm feed}=1$ in the following analysis.

To calculate the rms, we subtract the continuum emission of the point sources.
We revise the rms estimation method introduced in \cite{2021MNRAS.505.3698W},
that uses the difference between four adjacent frequency channels,
\begin{align}\label{eq:abba}
\Delta T(\tilde{\nu}) = \frac{1}{2}\left( T(\nu_1) + T(\nu_3) \right) - 
\frac{1}{2}\left( T(\nu_2) + T(\nu_4) \right),
\end{align}
where $\nu_1<\nu_2<\nu_3<\nu_4$ are the four adjacent frequency channels and
$\tilde{\nu} = (\nu_1 + \nu_2 + \nu_3 + \nu_4)/4$ is the reduced frequency of the 
residual data. This can remove most of the continuum emission from the sky
that is linear across the frequencies.
The rms of the residual data is related to the original data rms via,
\begin{align}
\sigma(\tilde{\nu}) = \sqrt{\frac{1}{4}\left(\sigma^2(\nu_1) + \sigma^2(\nu_3)\right)
+ \frac{1}{4}\left(\sigma^2(\nu_2) + \sigma^2(\nu_4)\right)}.
\end{align}

We calculate the noise level using the calibrated data. 
Each day's data is divided into seven blocks, each $\sim 30$-minute;
and the two polarizations are combined into the total intensity, i.e.
$T=(T_{\rm XX} + T_{\rm YY})/2$.
\reffg{fig:todrmshist} shows the histogram statistics of the residual data value.
For each time block, the histogram statistic includes all the $19$ beams' data and the result 
is shown with the gray curve.
The histogram for all time blocks combined is shown with the green curve.  
All the histograms are normalized with the total number of data samples.
The results of the low-frequency band, i.e. $1050$-$1150$ MHz, are shown
in the left panel, and the high-frequency band, i.e. $1323$-$1450$ MHz,
are shown in the right panel, respectively.

We fit the histogram with a Gaussian function,
\begin{align}\label{eq:histfit}
N(T)= A \exp\left[-\frac{1}{2}\frac{\left(T-\mu_0\right)^2}{\sigma^2}\right].
\end{align}
The best-fit function of the all-time-block combined histogram is
shown with the dashed red curve in \reffg{fig:todrmshist} and the
best-fit $\sigma$ and $\mu_0$ are also shown in the legend.
Both of the two frequency bands' data fit the Gaussian function well, which indicates 
that the residual data are dominated by white noise.
The best-fit noise levels of the data are $116.39$ mK and $99.47$ mK for the
low-frequency band and high-frequency band, respectively.

The system temperature includes several different components and can be expressed as,
\begin{align}
T_{\rm sys} = T_{\rm rec} + T_{\rm sky} + T_{\rm CMB},
\end{align}
where $T_{\rm rec}$ is the receiver temperature; $T_{\rm sky}$ is the
sky temperature; and $T_{\rm CMB}=2.725~{\rm K}$ is the mean brightness temperature of the 
cosmic microwave background (CMB). We ignore the temperature from ground-spill
when the telescope is pointing close to the Zenith.
The major component of the $T_{\rm sky}$ is the diffuse emission of the Galactic synchrotron. 
According to the measurements in the work of \citet{2020raa....20...64j},
the system temperature close to the Zenith is about $20$ K.
Using \refeq{eq:radiometer} and substituting $N_{\rm pol}=2$, 
$N_{\rm feed}=1$, $\Delta t=1~{\rm s}$ and $\Delta \nu = 28~{\rm kHz}$,
we should have $\sigma = 84.5\,{\rm mK} $. 
The slightly higher noise level could be caused by residual continuum emissions.

\begin{figure}
\centering
\includegraphics[width=0.47\textwidth]{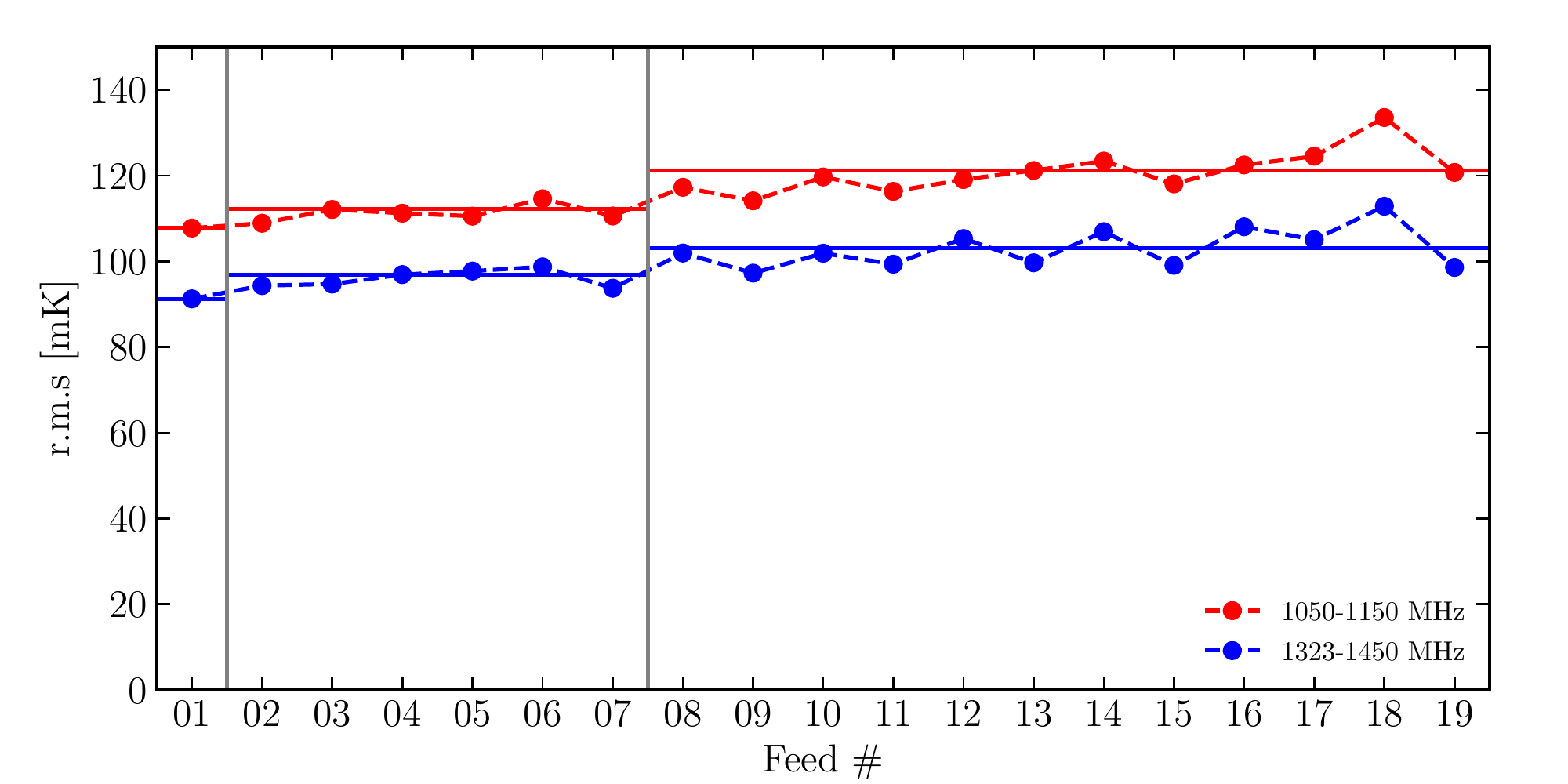}
\caption{The noise level of each feed. The results for lower and higher
frequency bands are shown in red and blue markers.
The feeds are split into three categories, i.e. the feeds in the central (Feed 01), 
the inner circle (Feed 02 - Feed 07), and the 
outer circle (Feed 08 - Feed 19) of the feed array.
The horizontal lines indicate the mean noise level of each feed category.}\label{fig:beamrms}
\end{figure}

The noise level also fluctuates between different feeds. 
As shown in \reffg{fig:beamrms}, the noise level of the lower and higher
frequency bands are shown in red and blue markers.
The feeds are split into three categories, i.e. the feeds in the central (Feed 01), 
the inner circle (Feed 02 - Feed 07) and the 
outer circle (Feed 08 - Feed 19) of the feed array.
The horizontal lines indicate the mean noise level of each feed category.
Although the noise levels of different feeds are varying, there is a trend that the feeds in the 
outer circle of the feed array have higher noise, i.e. about $10\%$ increases in the noise level
with respect to the central feed.

Such system temperature estimation can also be applied in the map domain.
We estimate the pixel noise level of the map at each reduced frequency, $\tilde{\nu}$. 
The histogram of each reduced frequency 
is shown with the gray curve in \reffg{fig:maprmshist}. The green curve 
shows the total histogram using all frequencies. The results of the low/high-frequency band
are shown in the left/right panels. The dashed red curve shows the best-fit
Gaussian function. Only histograms with amplitudes greater than half of their maximum 
are used in the Gaussian function fitting.
The best-fit $\sigma$ of the Gaussian function indicates the pixel noise level.

Clearly, the total histogram profile departs from the Gaussian function at $|T| \gtrsim 0.05 $ K. 
The pixel noise level for the low-frequency and high-frequency bands are 
$36.0~{\rm mK}$ and $25.9~{\rm mK}$, respectively.
Because the fit only uses the histograms with amplitudes greater than 
half of their maximum, the pixel noise level does not take into account the
effect of the large residual values. Nevertheless, the pixel noise level is higher than the forecast. 
Due to the different RFI flagging fractions, the mean integration times for 
the low-frequency and the high-frequency band are $10.0$ s and $16.7$ s, respectively.
Thus, assuming $T_{\rm sys}=20\,{\rm K}$, the pixel noise levels for such two frequency bands 
are $26.8$ mK and $20.7$ mK, respectively.
A couple of reasons could potentially increase the noise.
For example, the weak RFI contamination, which is below the noise level of the original TOD, 
becomes dominant when the pixel noise level is lowered by integrating data via the map-making process;
the residual sky contamination that is not removed with \refeq{eq:abba}; or the 
weakly correlated noise in the original TOD.

\subsection{Spectra of sources}

\begin{figure*}
\centering
\includegraphics[width=\textwidth]{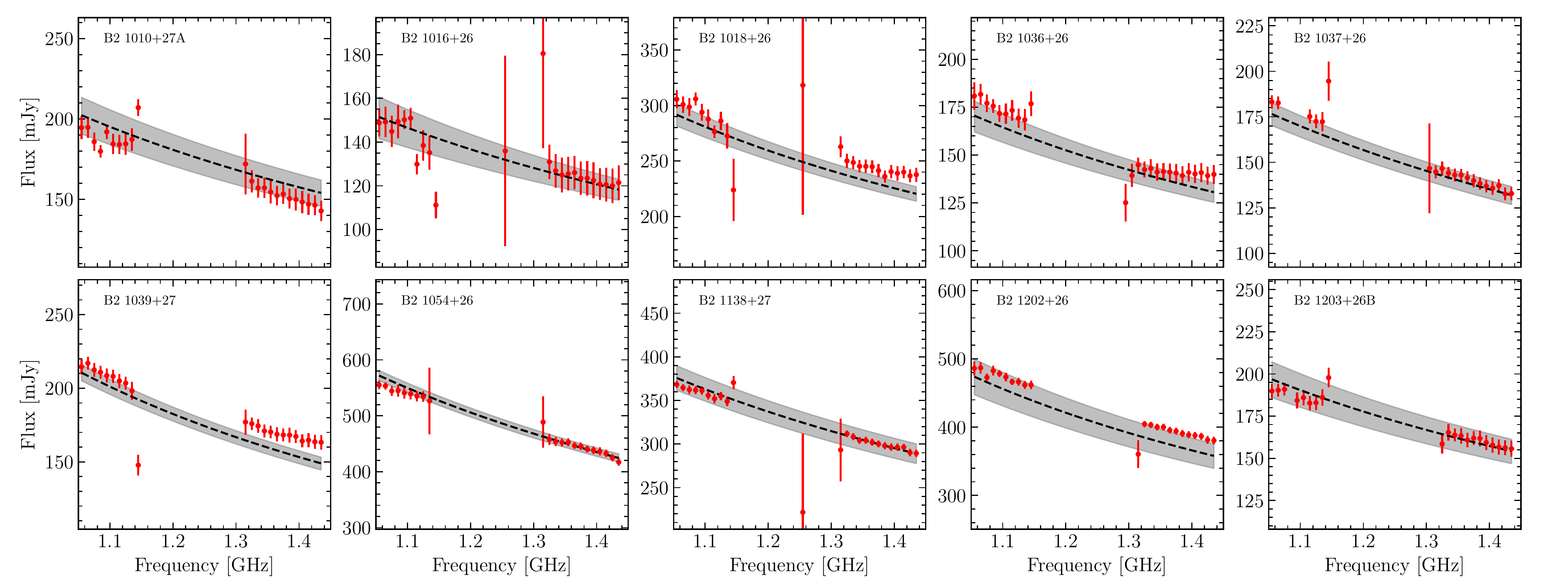}
\caption{The spectra of bright sources in the field. The red markers show the measured source spectrum. 
The flux is averaged within each $10\,{\rm MHz}$ frequency bin and the error bar indicates the 
flux rms in each frequency bin. 
The gap between $1150$ MHz and $1300$ MHz is due to RFI contamination. 
The dashed line indicates the source spectrum model fit using flux measurements in the literature 
and the gray area indicates the spectrum model uncertainty.}\label{fig:spec}
\end{figure*}

We also inspect the bandpass shape by comparing the spectra of bright sources to the
flux measurements in the literature. 
We chose the bright sources that are closely scanned by at least one beam, i.e. the minimal angular 
distance between the source and the beam center is less than $0.5\,{\rm arcmin}$. 
In the meanwhile, the source should have flux measurements at multiple frequency bands.  
In this analysis, we use $10$ radio sources with flux measurements at 
$74\,{\rm MHz}$ \citep{2007AJ....134.1245C}, 
$151\,{\rm MHz}$ \citep{1996MNRAS.282..779W},
$365\,{\rm MHz}$ \citep{1996AJ....111.1945D}, 
$408\,{\rm MHz}$ \citep{1972A&AS....7....1C}, 
$1.4\,{\rm GHz}$ \citep{1998AJ....115.1693C}, and 
$4.85\,{\rm GHz}$ \citep{1991ApJS...75....1B}. The flux of such $10$ sources are listed in
\reftb{tab:b2sources}.
The source spectrum is modeled by fitting a 3rd-order polynomial function to the flux measurements.

The source spectra are extracted from the calibrated data by taking the spectra at the time stamp 
when the source center is mostly close to the pointing direction. The data within the frequency band 
$1150$--$1250\,{\rm MHz}$ is ignored due to serious RFI contamination. 
The spectrum of each source is then averaged in every $10\,{\rm MHz}$ frequency bin.
The measured spectra of the $10$ sources are shown in \reffg{fig:spec}.
The error bar indicates the rms of the flux measurements within each $10$ MHz frequency bin.
The polynomial-fitted source spectrum model is shown with the black dashed line and 
the gray area indicates the model uncertainty, i.e. the upper/lower bound is
estimated by fitting the 3rd-order polynomial function to the upper/lower limit of 
$68\%$ flux measurement confidence interval.
The gap between $1150$ MHz and $1300$ MHz is due to RFI contamination.

Generally, our measurements produce a smooth power-law shape spectrum, which indicates that
the bandpass calibration efficiently corrects the bandpass shape. 
The spectrum shape slightly fluctuated at frequencies close to the RFI contamination, 
especially for the relatively faint sources. 
The flux is generally consistent with the spectrum model fitted using the flux measurements at a few 
frequency bands in the literature. The deviation between our measurement and the spectrum model,
e.g. source B2 1039+27, might be because of the intrinsic spectrum variation of the source that
can not be well-fitted by a low-order polynomial function. 

\begin{table*}
\begin{center}
\begin{minipage}{0.6\textwidth}
\begin{center}
\caption{Bright sources used for bandpass shape inspection. The flux values are all in 
the unit of mJy}\label{tab:b2sources}
\scriptsize
\begin{tabular}{lcccccc}\hline\hline
Source name    &  $74$ MHz $^a$   & $151$ MHz $^b$   & $365$ MHz $^c$   & $408$ MHz $^d$  & $1400$ MHz $^e$ & $4850$ MHz $^f$\\ \hline
B2 1010+27A    &  $1440\pm170$ & $910\pm66$  & $424\pm26$  & $470\pm72$  & $158\pm4.8$ & $40\pm6.0$ \\
B2 1016+26     &  --           & $700\pm63$  & $357\pm25$  & $304\pm70$  & $121\pm4.3$ & $39\pm5.9$ \\
B2 1018+26     &  $1660\pm190$ & $1400\pm79$ & $640\pm57$  & $735\pm80$  & $228\pm7.7$ & $76\pm11$  \\
%B2 1020+27     &  --           & $580\pm64$  & $460\pm48$  & $275\pm75$  & $165\pm5$   & $113\pm15$ \\
B2 1036+26     & $1180\pm140$  & $870\pm59$  & --          & $354\pm70$  & $145\pm4.7$ & $46\pm6.9$ \\
B2 1037+26     & $1840\pm260$  & $860\pm63$  & $408\pm32$  & $425\pm72$  & $124\pm4.4$ & $30\pm7.0$ \\
B2 1039+27     & $1880\pm200$  & $1200\pm66$ & $559\pm30$  & $603\pm72$  & $149\pm4.5$ & $31\pm4.7$ \\
B2 1054+26     & $4620\pm490$  & $2650\pm133$& $1520\pm53$ & $1260\pm100$&$436\pm15$   & $95\pm14$  \\
B2 1138+27     & $2150\pm230$  & $1380\pm97$ & $816\pm26$  & $873\pm80$  & $284\pm8.5$ & $92\pm14$  \\
B2 1202+26     & $2250\pm240$  & $2110\pm124$& $1250\pm172$& $969\pm80$  & $379\pm11$  & $141\pm21$ \\
B2 1203+26B    & $770\pm130$   & $670\pm98$  & $415\pm38$  & $420\pm72$  & $156\pm4.7$ & $63\pm9.5$ \\
\hline\hline
\end{tabular}
\end{center}
{\footnotesize
\noindent
$^{a}$ The VLA Low-Frequency Sky Survey \citep{2007AJ....134.1245C}.\\
$^{b}$ The 7C survey of radio sources at 151 MHz \citep{1996MNRAS.282..779W}.\\
$^{c}$ The Texas Survey of Radio Sources \citep{1996AJ....111.1945D}\\
$^{d}$ The B2 Catalogue of radio sources \citep{1972A&AS....7....1C}.\\
$^{e}$ The NRAO VLA Sky Survey \citep{1998AJ....115.1693C}.\\
$^{f}$ A New Catalog of 53522 4.85 GHz Sources \citep{1991ApJS...75....1B}.
}
%\tablecomments{
%}
\end{minipage}
\end{center}
\end{table*}

\subsection{Flux of detected sources}

\begin{figure}
\centering
\includegraphics[width=0.47\textwidth]{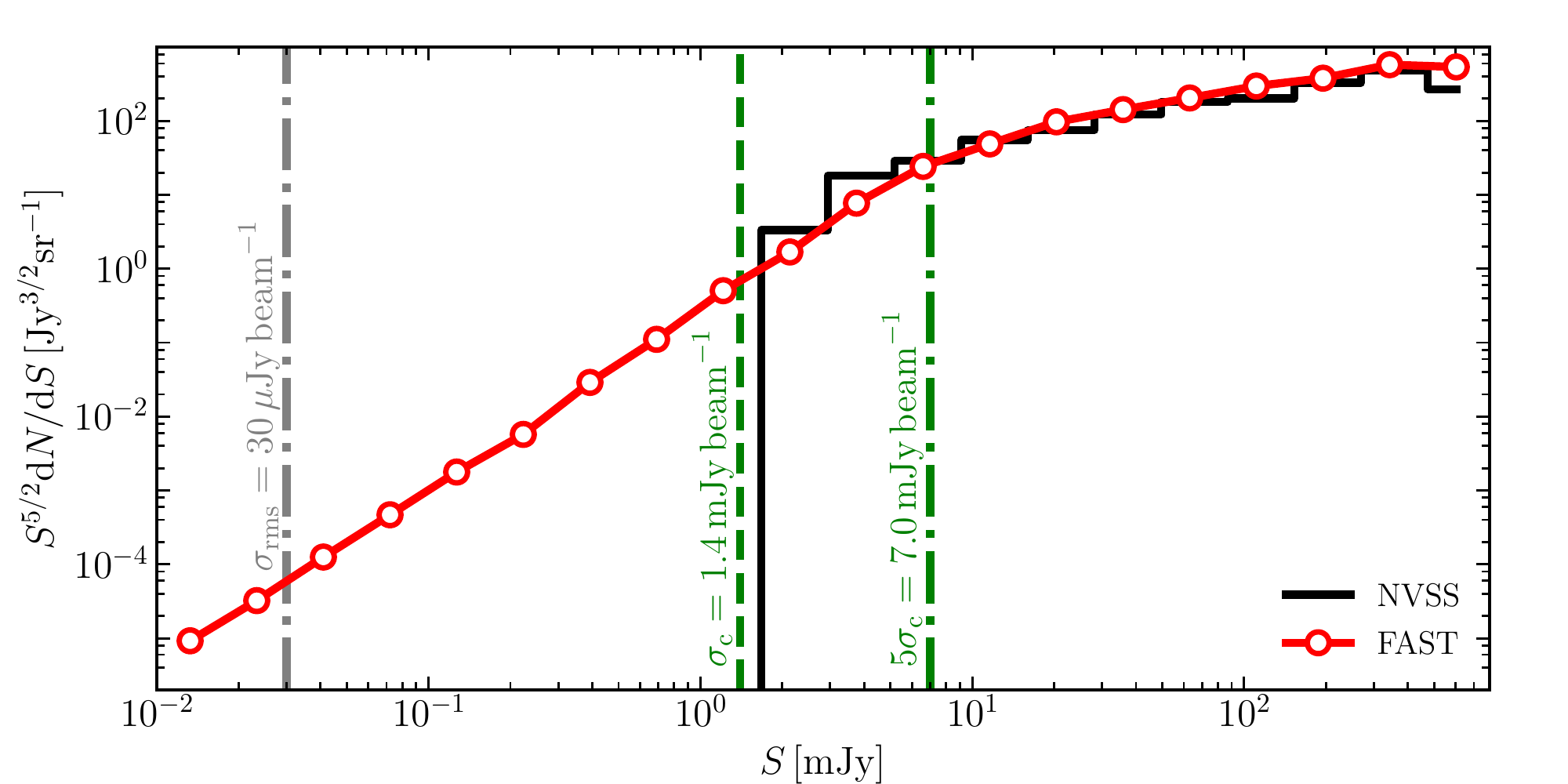}
\caption{
 The differential number count of the continuum sources in the survey area.
The black stepping curves display the differential number count of the NVSS catalog, 
while the red circle markers with solid lines display the results of our map.
The green vertical dash-dot line indicates the flux limit of 
$5\sigma_{\rm c} \approx 7\,{\rm mJy}\,{\rm beam}^{-1}$,
where $\sigma_{\rm c}$, shown with the green vertical dashed line, is the confusion limit 
due to the background unresolved sources.
The gray vertical dashed line indicates the $30\,{\rm \mu Jy}\,{\rm beam}^{-1}$ flux limit of the map.
}\label{fig:dNdS}
\end{figure}

According to the measurements in \refsc{sec:tsys}, in our survey
the pixel noise level is $\sim 25.9$ mK with the frequency resolution of $28$ kHz at 
high-frequency band. The corresponding flux limit at $1400$ MHz with a bandwidth of 
$50$ MHz should be $\sim 30\,{\rm \mu Jy}\,{\rm beam}^{-1}$. We apply a source
finding algorithm, i.e. the {\tt DAOStarFinder} 
\footnote{\url{https://photutils.readthedocs.io/en/stable/api/photutils.detection.DAOStarFinder.html}},
to our map with the threshold of $30\,{\rm \mu Jy}\,{\rm beam}^{-1}$ and aperture size of 
$3\,{\rm arcmin}$ and more than three thousand continuum sources are detected. 
The flux-weighted differential number count of the detected continuum sources is shown 
using the red circle markers with the solid curve in \reffg{fig:dNdS}.

However, we should also consider the confusion limit for the continuum sources \citep{1974ApJ...188..279C,2017PASA...34...13M},
\begin{equation}
\sigma_{\rm c} \approx 0.2 \left(\frac{\nu}{\rm GHz}\right)^{-0.7} \left(\frac{\theta_{\rm FWHM}}{\rm arcmin}\right)^2 \approx 1.4\,{\rm mJy}\,{\rm beam}^{-1}.
\end{equation}
This is much larger than the limit given above. Sources fainter than a few $\sigma_{\rm c}$ would be confused and not detected as individual sources. 

To check our survey results, we compare 
the continuum flux density of the sources in the observed field with those in the NRAO-VLA Sky Survey (NVSS) catalog \citep{2008AJ....136..684K}. We use the integrated flux density of NVSS sources from a combined radio objects catalog with flux and position corrections\footnote{\url{http://www.aoc.nrao.edu/~akimball/radiocat.shtml}}. The flux limit of the NVSS catalog is given as $2\,{\rm mJy}\,{\rm beam}^{-1}$. There are $3161$ NVSS sources in the survey area, i.e. 
$9\overset{{\rm h}}{} \, < {\rm R.A.}< \,13\overset{{\rm h}}{}$
and $+25\overset{\circ}{.}83\,<{\rm Dec}<\,+27\overset{\circ}{.}08$.
The flux-weighted differential number count for sources in the NVSS catalog is also shown in
\reffg{fig:dNdS} with the black stepping curves. 
The detected continuum sources using our map is consistent with the NVSS down to 
$\sim 7\,{\rm mJy}\,{\rm beam}^{-1} \approx 5\sigma_{\rm c}$. 
At the faint end  below 7 mJy (marked in the figure by the vertical dash-dot line), the number of sources detected by our survey begins to fall below that of the NVSS, which is unsurprising because the NVSS has much higher angular resolution and therefore lower the confusion limit.  

In order to make source-by-source flux measurement comparison, we select 
isolated bright sources from the full NVSS sample according to the following criteria:
\begin{enumerate}[i)]
\item We reject the sources that have neighbors' flux over 10\% of the
centra source within $9\,{\rm arcmin}$, 
i.e. about three times of the beam width \citep{1996ApJS..103..427G}. 
Such selection criteria reject more than $90\%$ of the NVSS sources in the field.
\item Then we remove the sources with flux less than $14\,{\rm mJy}\,{\rm beam}^{-1}$.
We adopt such an aggressive flux limit to avoid confusion from the background noise.
Another $3\%$ source is rejected according to this criteria. 
\item In the end, we pick the closely scanned sources that are 
$0.5\theta_{\rm FWHM}\sim1.5\,{\rm arcmin}$ or less from the center of at least one FAST beam. 
\end{enumerate}

A total of $81$ isolated sources meet these selection criteria, making up the isolated sample. This sample is listed in \reftb{tab:sources}. 
This isolated sample of sources is used for the following source-by-source 
flux measurement comparison.

\subsubsection{Flux comparison with time-ordered data}\label{sec:fluxtod}

We first make a comparison of flux from the TOD. We use the mean flux density across the frequency range $1.375\sim1.425\,{\rm GHz}$, which is the same frequency range of the NVSS catalog \citep{1998AJ....115.1693C}.
Because the sky coverage partially overlaps between different days, 
the same source may be observed by different beams on different days. 
The number of sources used for each feed is listed below,
\begin{center}
\begin{small}
\begin{tabular}{ccccccccccc} \hline\hline
Feed $\#$ & 01 & 02 & 03 & 04 & 05 & 06 & 07 & 08 & 09 & 10 \\ 
N      & 19 & 24 & 17 & 23 &  9 & 15 & 18 & 18 & 12 &  5 \\ \hline 
Feed $\#$ & 11 & 12 & 13 & 14 & 15 & 16 & 17 & 18 & 19 & \\
N      & 19 &  8 & 20 & 19 & 23 & 15 & 10 & 23 & 10 & \\\hline\hline
\end{tabular}
\end{small}
\end{center}

\begin{figure*}
\begin{minipage}[t]{0.47\textwidth}
\centering
\includegraphics[width=\textwidth]{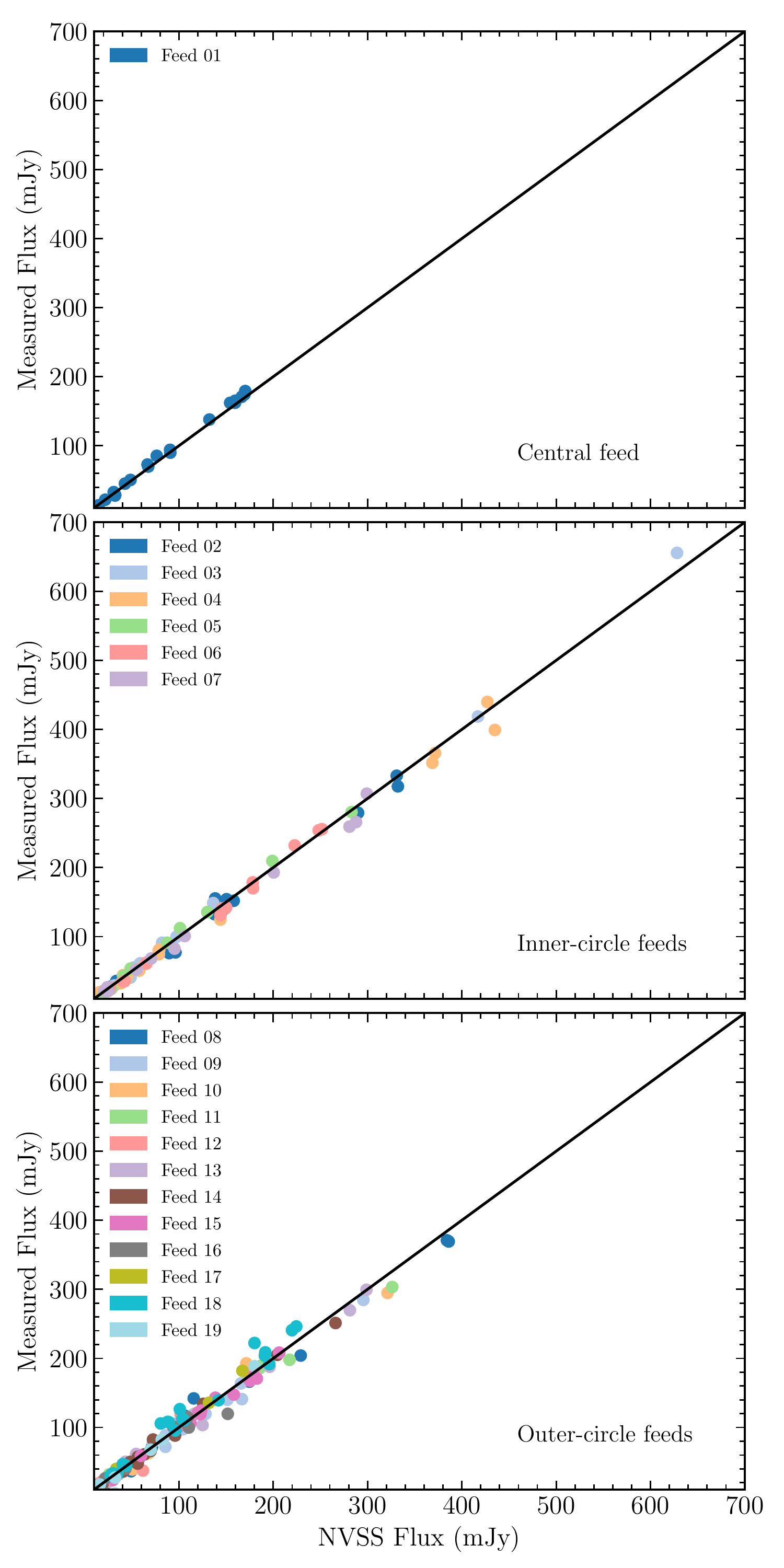}
\caption{
The flux-flux diagram compares the measurements using the TOD with those in the NVSS catalog. 
}\label{fig:fluxfluxtod}                             
\end{minipage} \hfill
\begin{minipage}[t]{0.47\textwidth}
\centering
\includegraphics[width=\textwidth]{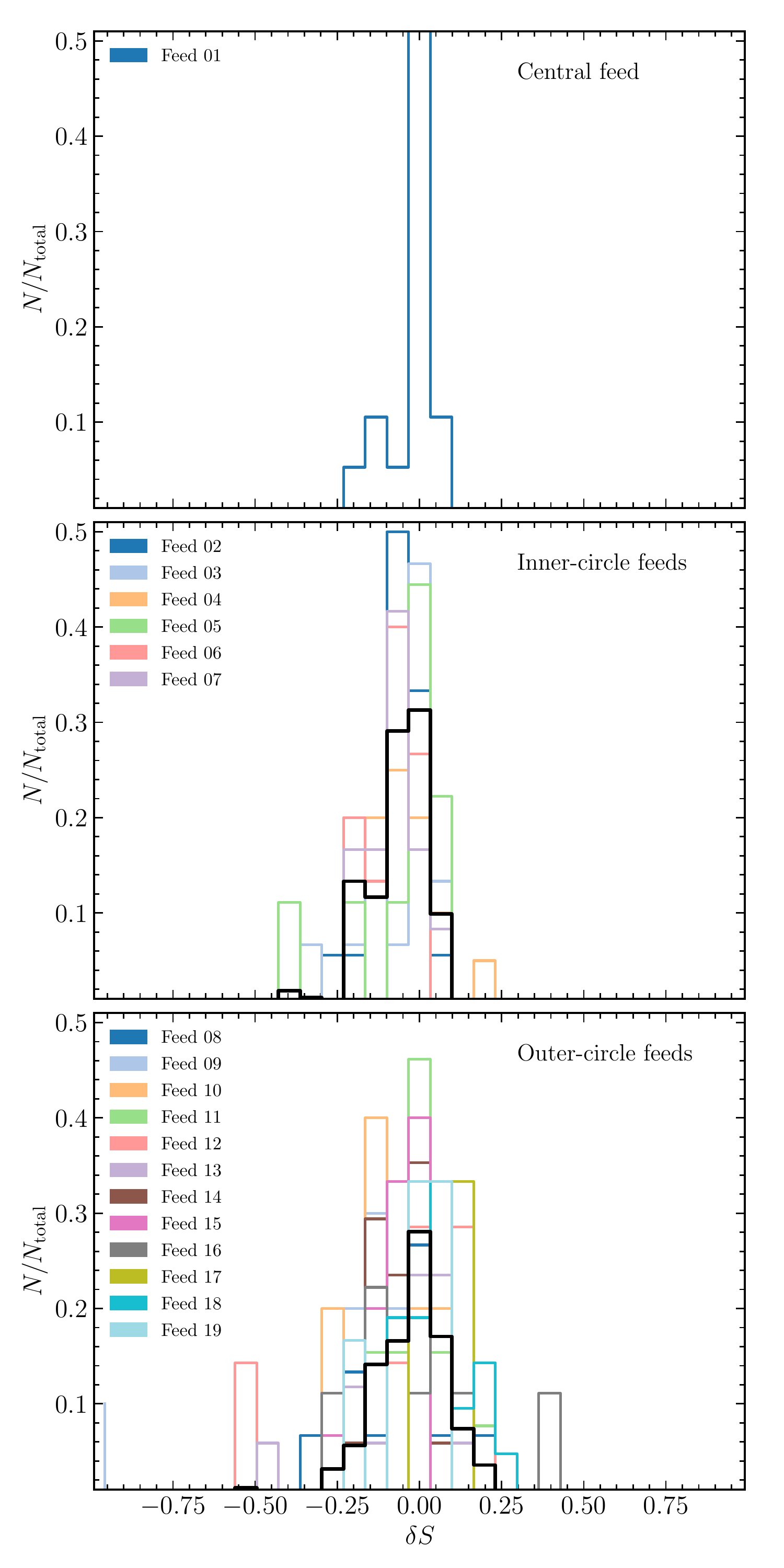}
\caption{The histogram of relative flux residual between our measurements 
using the TOD and the NVSS catalog, i.e. \refeq{eq:rdiff}.
}\label{fig:fluxhisttod}                             
\end{minipage}
\end{figure*}

\begin{figure*}
\begin{minipage}[t]{0.47\textwidth}
\centering
\includegraphics[width=\textwidth]{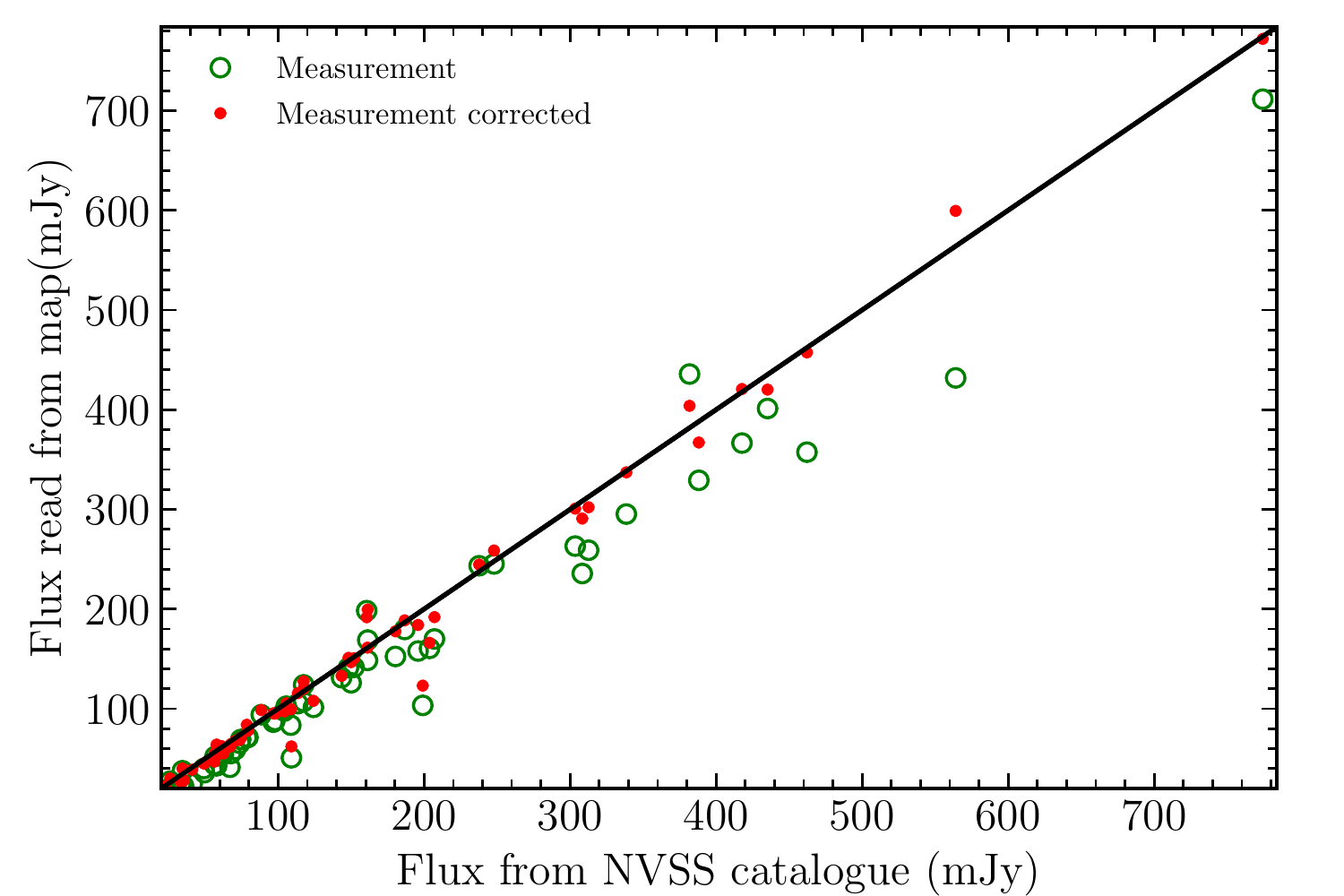}
\includegraphics[width=\textwidth]{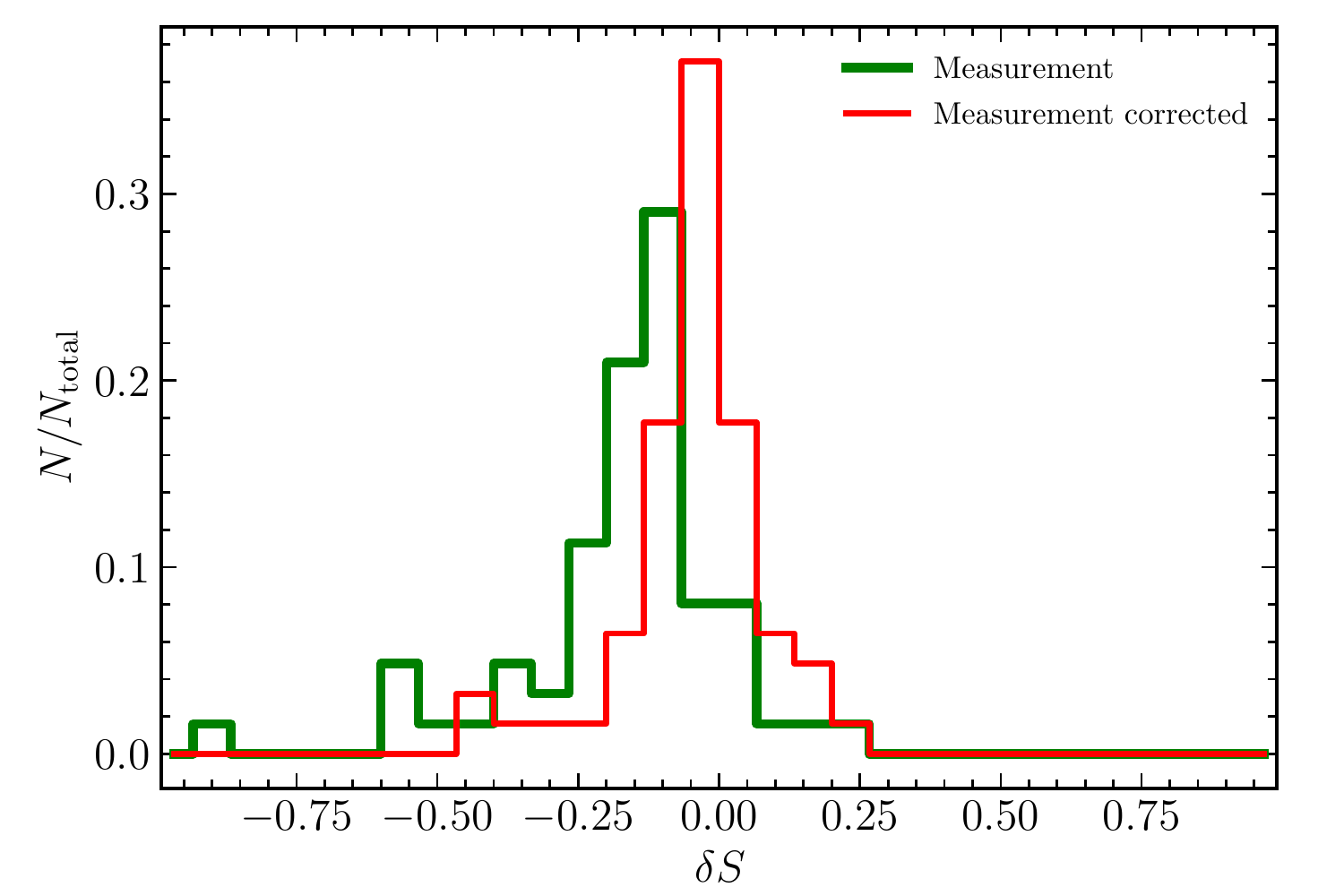}
\caption{Map-domian flux measurements comparing. 
The flux measurements before and after correction with simulation are shown in 
green and red, respectively.
Top panel: the x-axis is the flux values from the NVSS catalog and the y-axis is the flux
measurements from the map; Bottom-panel: the histogram statistic of the relative flux residual, 
i.e. \refeq{eq:rdiff}. All the measurements use an aperture radius size of $1.5$ arcmin.
}\label{fig:fluxcorrection}                             
\end{minipage} \hfill
\begin{minipage}[t]{0.47\textwidth}
\centering
\includegraphics[width=\textwidth]{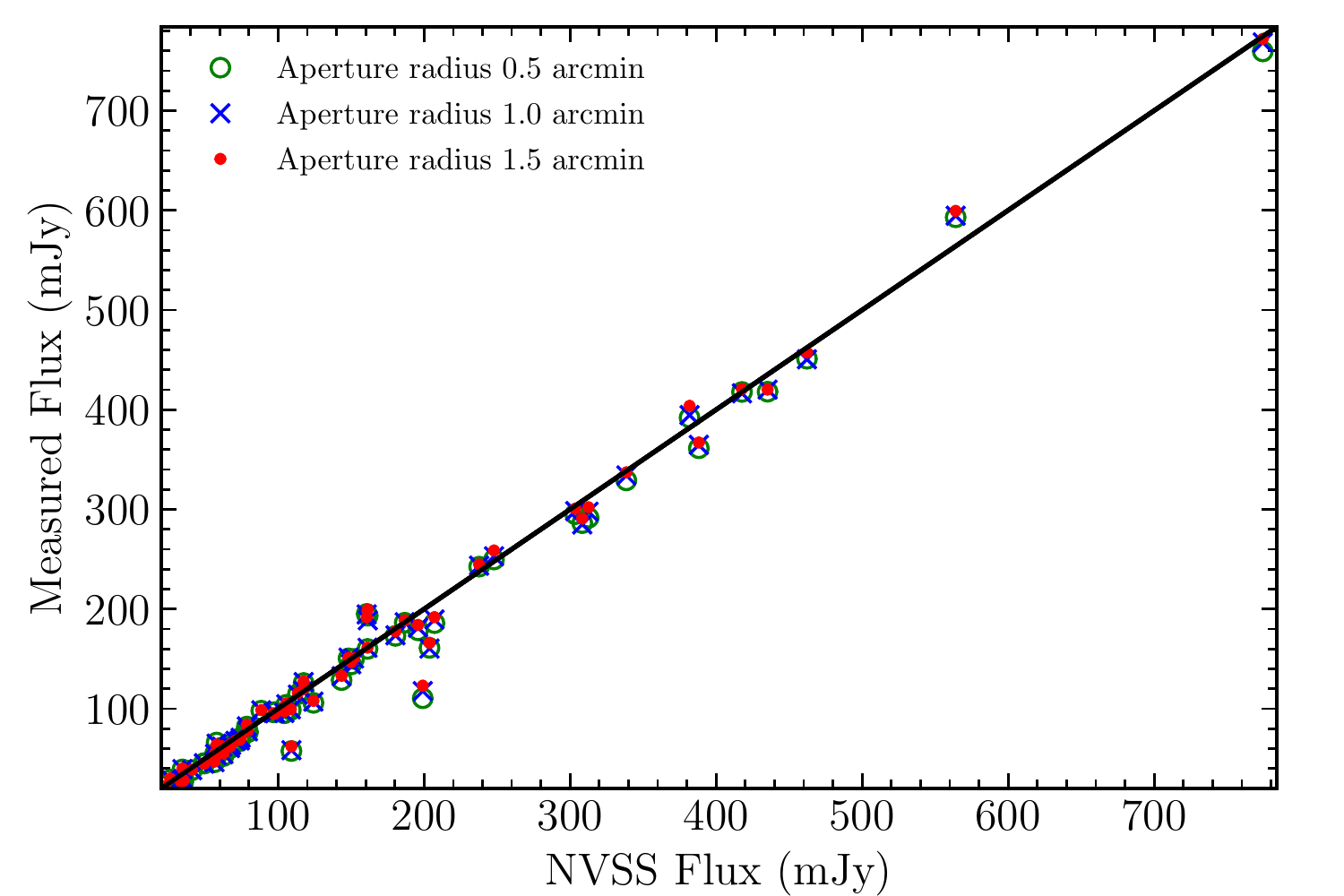}
\includegraphics[width=\textwidth]{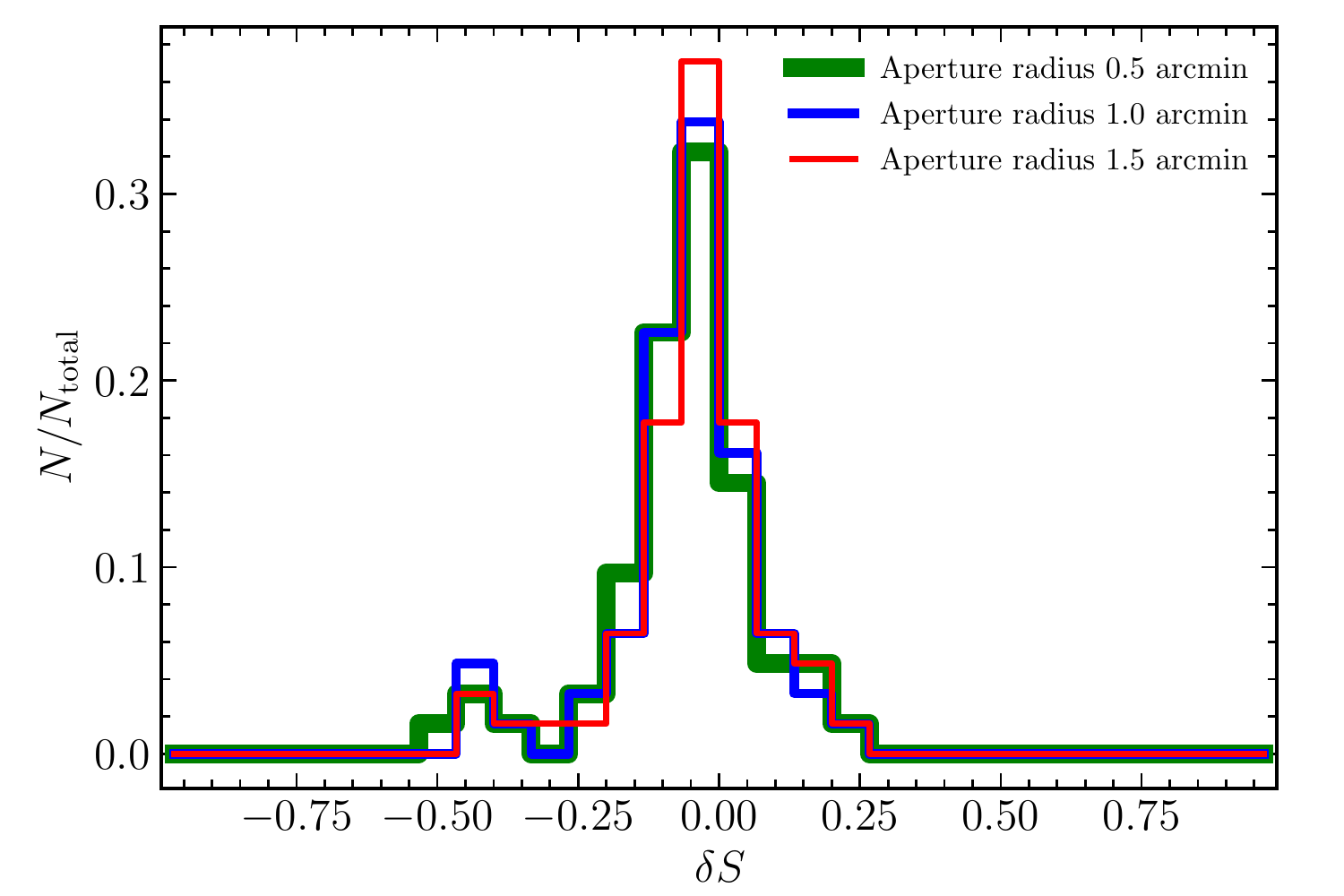}
\caption{Map-domain flux measurements with different aperture radius sizes.
Top panel: the x-axis is the flux values from the NVSS catalog and the y-axis is the flux
measurements from the map; Bottom-panel: the histogram statistic of the relative flux residual, 
i.e. \refeq{eq:rdiff}.
}\label{fig:fluxmap}                             
\end{minipage}
\end{figure*}

The measured flux density is extracted by taking the spectrum density 
at the time when the source has the minimal angular distance to the feed center, and is compared with the expected flux, which is obtained by multiplying The NVSS flux density with a Gaussian beam profile according to the angular distance to the beam center,
\begin{align}
\hat{S}_{\rm NVSS} = S_{\rm NVSS} \exp\left[ -\frac{1}{2} 
\frac{\theta^2_{\rm min}}{\left(\theta_{\rm FHWM}/(2\sqrt{2\ln2})\right)^2} \right].
\end{align}

The flux-flux comparison is plotted in \reffg{fig:fluxfluxtod}. The NVSS sources scanned by different feeds are shown with different colors and those 
sources scanned by the central, inner circle, and outer circle of the 
FAST feed arrays are shown in the top, middle, and bottom sub-panels, respectively.
The measured flux densities are shown to be consistent with the NVSS catalog.

We quantified the scattering of the measurements using the relative flux residual 
with respect to the total flux, 
\begin{equation}\label{eq:rdiff}
\delta S= \left(S - \hat{S}_{\rm NVSS}\right) / \sqrt{S \hat{S}_{\rm NVSS}},
\end{equation}
where $S$ represents the extracted fluxes from our TOD. 
The histogram statistic of the flux residuals is illustrated in \reffg{fig:fluxhisttod}.
The sources scanned by different feeds are also shown with different colors and
those sources scanned with the feeds in the central, the inner circle and 
the outer circle of the feed array is shown in the top, middle, and bottom
sub-panels, respectively.
The black solid curves show the 
averaged histogram of the feed categories. 
All the histograms are normalized with the total number of measurements.
The rms of the relative flux residual, i.e. 
$\sqrt{\sum \delta S^2/N_{\rm total}}\times100\%$, of the three feed categories are 
$4.2\%$, $5.3\%$ and $6.9\%$, respectively.
The source flux measured with the central feed has less scattering than
measurements with the rest of the feeds. 
The increasing residue for feeds in the inner and outer circle of the feed array is probably a result of an error in the beam  model, as the beams are more distorted as we move out from the center. 

We also check the flux measurements uncertainty between the observation on different days. The results show that the flux residual rms of different days are generally consistent. We can also take advantage of the repeated observation of the same strip on 
March 9th, 2021, and March 14th, 2021.  
We estimate the relative residual rms using the flux differences of the 
same sources between these two observations. 
$$\sqrt{\sum \delta S_{\rm ab}^2/N_{\rm total}/2}\times 100\% = 3.7\%,$$
where $\delta S_{\rm ab} = (S_{\rm a} - S_{\rm b})/\sqrt{S_{\rm a} S_{\rm b}}$ is the 
flux difference.
As the observations on such two days have the same pointing direction, 
the systematic effect, such as the beam effect, is canceled. 
If the source flux variation between the short period is negligible, such 
relative residual rms indicates the calibration uncertainties in 
our point source flux measurements. 
The residual between these two observations is significantly less than the residual between our measurements and the NVSS catalog.
The additional discrepancies between our results and the NVSS database could result from a number of different factors. For instance, a less accurate beam model or the flux variation of the NVSS source. 
Thus, the $\sim 6.3 \%$ flux dispersion on average indicates an upper bound on the 
residual gain variations after the calibration process.

\subsubsection{Flux comparison with the combined map}\label{sec:fluxmap}

Next, we make the flux density comparison using a map with a fine angular resolution created using the map-making process, as described in \citet{2018ApJ...861...49H}.
We average the flux density of pixels within an aperture radius of
$0.5\theta_{\rm FWHM} \sim 1.5\,{\rm arcmin} $ via \citep{2011MNRAS.411..993F}
\begin{equation}
S = \frac{\sum_i S(\theta_i)}{\sum_i K'(\theta_i)}, \,\,
\sigma^2_{S} = \frac{\sum_i K'^2(\theta_i)}{\left(\sum_i K'(\theta_i)\right)^2}
\frac{\sum_i\left(S(\theta_i) - K' S\right)^2}{\sum_i K'^2(\theta_i)},
\end{equation}
where $\sigma_{S}$ is the measurements error, 
$\theta_i$ is the angular separation to the center pixel 
and $K'(\theta_i)$ represents the kernel function used in the map domain,
\begin{equation}
K'(\theta_i) = \exp\left[ - \frac{1}{2} \left(\frac{\theta_i}{\sigma_{K'}}\right)^2 \right],
\end{equation}
where $\sigma_{K'} = \sigma_K + \theta_{\rm FWHM} / \left(2\sqrt{2\ln 2}\right)$, 
$\sigma_K$ is the kernel size of \refeq{eq:todkernel} that applied during the map-making
and $\theta_{\rm FWHM}$ is the beam size.
We use the same NVSS sources selected in \refsc{sec:fluxtod} for comparison.

However, a direct comparison of bright pixels in the map with the NVSS source flux would show a large bias. This is because the sources do not always transit across the beam center, but in the map-making process no correction has been made, as we can not presume that we know the sources and their positions. With a sufficiently large number of scans, the sources would be completely sampled, and the flux 
measurements taken after constructing the maps would be unbiased.
However, due to the limited number of scans, and also RFI flagging, noise diode injection, and abandoning of data from bad beams, the sources are far from completely sampled. In order to recover the bias raises from the incomplete sampling of the sources,
we simulate the TOD using the NVSS catalog. The simulated TOD has the same sky coordinates and mask as the real data and is projected to the map domain using the same map-making procedure as observations.
We discover that the flux from most sources is pretty biased.
The map-domain flux values are then corrected using the difference between the flux from the simulation and the NVSS catalog. The comparison of the flux before and after correction is shown in \reffg{fig:fluxcorrection}.
The top panel shows the flux-flux comparison between map-domain measurements and the
NVSS catalog and the bottom panel show the histogram statistic of the relative flux residual defined in \refeq{eq:rdiff}.
It is obvious that the measurements are significantly biased in the absence of flux correction. 
The flux correction makes a significant improvement, i.e. the measurement's relative uncertainty
is improved from $17.1\%$ to $6.3\%$ after the flux correction.

We also investigate how aperture size affects flux measurements. We vary the aperture radius size
between $0.5$ arcmin, $1.0$ arcmin and $1.5$ arcmin and show the comparison results in
\reffg{fig:fluxmap}. 
All the measurements are corrected using the simulation with the corresponding aperture size. 
With varying aperture radius sizes, the flux values only slightly varied without a clear systematic trend.

The map-domain flux measurements result in about $6.3\%$ relative uncertainty, which is consistent 
with the uncertainty of the TOD flux measurements. It shows that our map-making procedure
is accurate enough for continuum flux measurements. 
For the flux check in this work, we only selected a small number of bright, isolated point sources.
We leave the work of identifying weak and diffuse sources to future studies. 

\section{Summary}\label{sec:summary}

The neutral hydrogen (\hi) is known to trace the galaxies in the 
post-reionization era. 
A comprehensive wide-field of the extragalactic \hi survey could provide valuable
information for both cosmology and astrophysics research. 
In this work, we report the time-ordered data (TOD) analysis pipeline designed for
drift-scan observation using the Five-hundred-meter Aperture 
Spherical Telescope (FAST).

The data analyzed in this work were collected over a few nights spanning in 
2019, 2020, and 2021. During the $4$ hours drift scan of each night, 
the FAST telescope points at a fixed altitude angle and the
observation covers right ascension (R. A.) range from 9 hr to 13 hr,
which overlaps with the Northern Galactic Cap (NGP) area of the Sloan
Digital Sky Survey (SDSS). 
The FAST L-band 19-feed receiver is used
in our observation. The feed array is rotated by $23\overset{\circ}{.}4$ to
optimize the coverage. The pointing Dec. shift by $10.835\,{\rm arcmin}$
between different days to enlarge the survey area. 

The noise diode signal, as the relative calibrator, is injected for $1\,{\rm s}$ 
in every $8\,{\rm s}$. The noise diode signal is used for calibrating the bandpass
gain and temporal drift of the gain. Our analysis indicates that the observation data
in 2019 have significant bandpass shape variation during the $4$ hours drift-scan
observation. The bandpass shape of the data in 2021 becomes much more stable.
The major data analysis focuses on the $7$ nights observations in 2021.

We applied the {\tt SumThreshold} and {\tt SIR} radio frequency
interference (RFI) flagging program to the bandpass calibrated data.
In order to enhance the RFI signal and protect the potentially existing \hi emission lines, 
the RFI flagging is applied to the feed averaged data. 
Due to the contamination of the Global Navigation Satellite Systems, the
data between the frequency range of $1135$-$1310$ MHz are mostly flagged.
Besides, about $40\%$ and $14\%$ data are flagged in the frequency range of
$1050$-$1135$ MHz (the low-frequency band) and $1310$-$1450$ MHz (the high-frequency band),
respectively.

We develop the temporal drift calibration strategy that estimates the
gain variation across the drift-scan observation by applying a
wiener filter on the gain variation measurements. The Wiener filter
is designed according to the 1/f noise temporal power spectrum model,
which is constrained using the observation data. With our calibration 
strategy, a temporal oscillation of the gain is observed in the $4$ hours drift scan
and such oscillation can be well calibrated.

The absolute flux calibration is done by calibrating the noise diode spectrum 
against the celestial source 3C286. The calibration observation is made 
also in drift scan mode. The noise diode spectrum shape is stable during the 
$7$ nights observations in 2021.
Besides, using the drift scan observation of bright source 3C286, 
we check the beam profile for each of the feeds. The beam profiles of all the 
$19$ feeds are significantly asymmetric. 

Due to the systematic background noise level variation during the observation time,
a significant temporal baseline variation is observed with the gain-calibrated data. 
Especially, some sharp variations are shown in one night of the observations.
Such baseline variations are subtracted by fitting with a baseline template, 
which is constructed using the baseline averaged across different feeds.
After baseline subtraction, the calibrated data are zero-centered and transferred to the
standard map-making procedure.

We check the standing-wave ripples across the frequency axis by estimating  
the delay spectrum using well-calibrated data.
The standing-wave signature is more prominent in the bandpass measurements than in the 
sky observation, probably because this standing wave is induced by the noise diode. 
We use the low-pass filter with the size of $\tau_{\rm c}=0.7\,{\mu s}$ to suppress
both the noise and standing-wave signature during the bandpass determination. 
The standing-wave peak is negligible during our observations. 

We check the measurement noise level using the TOD following the method 
introduced in \cite{2021MNRAS.505.3698W}. The noise level of the 
low- and high-frequency bands are $116.39$ mK and $99.47$ mK, respectively. 
The noise level also varies between different feeds. 
The feed in the outer circle of the feed array has a noise level
increasing more than $10\%$ than the central feed. 
The noise level reduces slowly after integrating the measurements via map-making,
due to weak RFI contamination, residual sky emission, or correlated noise.

We also study the systematic uncertainties by comparing the continuum flux measurements
with the NVSS catalog. 
By applying the source-finding algorithm with the threshold 
of $30\,{\rm \mu Jy}\,{\rm beam}^{-1}$, i.e. the flux limit due to the map rms, 
more than three thousand continuum sources are detected within our survey field.
However, most of them are confused sources due to the angular resolution limit of the FAST beam.
The flux-weighted differential number counts for the detected sources are consistent with 
the NVSS catalog down to $\sim 7\,{\rm mJy}\,{\rm beam}^{-1}$, which is about $5$ times of
the confusion limit.
Finally, we chose $81$ isolated NVSS sources with flux over 
$14\,{\rm mJy}\,{\rm beam}^{-1}$ in our survey
field and find that the calibrated data shows about $4.2\%$, $5.3\%$ and
$6.9\%$ measurement uncertainties for the central feed, inner circle feeds and 
outer circle feeds, respectively. 
Such uncertainty varies between the measurements of central feed, inner-circle feeds, 
and outer-circle feeds. Finally, there is about $6.3\%$ uncertainty on average, which
is consistent with the map-domain flux measurements. 

\section*{Acknowledgements}

This work made use of the data from FAST (Five-hundred-meter Aperture Spherical radio Telescope). 
FAST is a Chinese national mega-science facility, operated by 
National Astronomical Observatories, Chinese Academy of Sciences.
We acknowledge the support of the National SKA Program of China (Nos.~2022SKA0110100, 2022SKA0110200, 2022SKA0110203), the National Natural Science Foundation of China (Nos.~11975072, 11835009), 
the CAS Interdisciplinary Innovation Team (JCTD-2019-05), and the science research grants from the China Manned Space Project with No.~CMS-CSST-2021-B01.
LW is a UK Research and Innovation Future Leaders Fellow [grant MR/V026437/1].
%%%%%%%%%%%%%%%%%%%%%%%%%%%%%%%%%%%%%%%%%%%%%%%%%%%
\section*{Data Availability}

The data underlying this article will be shared on reasonable request to the corresponding author.

%%%%%%%%%%%%%%%%%%%%% REFERENCES %%%%%%%%%%%%%%%%%%


\bibliography{main}
\bibliographystyle{aasjournal}



%%%%%%%%%%%%%%%%%% APPENDICES %%%%%%%%%%%%%%%%%%%%%
%
\appendix
\section{Continuum source catalogue}
\begin{table*}[h!]
\begin{center}
\caption{The flux density of the 81 isolated continuum sources over 
$14\,{\rm mJy}\,{\rm beam}^{-1}$ in the survey field. 
The sources' sky coordinates at the J2000 epoch are presented in the 
`RA' and `Dec' columns; The flux density measurements read from the 
combined radio objects catalogue \citep{2008AJ....136..684K} are 
presented in the `$S_{\rm NVSS}$' column; the flux density measurements from 
this work are presented in the `$S_{\rm FAST}$' column;
and the flux correction fraction with respected to `$S_{\rm NVSS}$' are 
presented in `corr.' column.
The flux measurements use $1.5$ arcmin aperture size and are corrected with the simulation. 
}\label{tab:sources}
\scriptsize
\hspace*{-1.5cm}
\begin{tabular}{ccrrr | ccrrr} \hline\hline
 RA (J2000) & Dec (J2000) & $S_{\rm NVSS}$ & $S_{\rm FAST}$ & Correction   & RA (J2000) & Dec (J2000) & $S_{\rm NVSS}$ & $S_{\rm FAST}$ & Correction \\
            &             & $[{\rm mJy}] $ & $[{\rm mJy}] $ & [$\%$] &            &             & $[{\rm mJy}] $ & $[{\rm mJy}] $ & [$\%$] \\\hline
$09\overset{\rm h}{} 03\overset{\rm m}{} 11\overset{{\rm s}}{.}67$ & $+26\overset{\circ}{} 09\overset{\prime}{} 40\overset{{\prime\prime}}{.}28$ & $ 195.87$ & $ 184.04 \pm  3.88$ &$  13.35$ &$09\overset{\rm h}{} 42\overset{\rm m}{} 36\overset{{\rm s}}{.}39$ & $+26\overset{\circ}{} 10\overset{\prime}{} 15\overset{{\prime\prime}}{.}27$ & $  35.09$ & $  27.41 \pm  0.90$ &$  13.08$ \\
$10\overset{\rm h}{} 18\overset{\rm m}{} 07\overset{{\rm s}}{.}13$ & $+26\overset{\circ}{} 09\overset{\prime}{} 42\overset{{\prime\prime}}{.}51$ & $ 180.38$ & $ 177.66 \pm  3.80$ &$  13.99$ &$09\overset{\rm h}{} 26\overset{\rm m}{} 00\overset{{\rm s}}{.}53$ & $+25\overset{\circ}{} 59\overset{\prime}{} 53\overset{{\prime\prime}}{.}30$ & $ 113.56$ & $ 115.60 \pm  2.97$ &$   9.24$ \\
$10\overset{\rm h}{} 57\overset{\rm m}{} 23\overset{{\rm s}}{.}15$ & $+26\overset{\circ}{} 01\overset{\prime}{} 30\overset{{\prime\prime}}{.}18$ & $ 186.74$ & $ 188.67 \pm  4.89$ &$   4.92$ &$11\overset{\rm h}{} 34\overset{\rm m}{} 02\overset{{\rm s}}{.}19$ & $+26\overset{\circ}{} 32\overset{\prime}{} 24\overset{{\prime\prime}}{.}54$ & $  15.62$ & $  13.75 \pm  0.39$ &$  13.94$ \\
$12\overset{\rm h}{} 38\overset{\rm m}{} 23\overset{{\rm s}}{.}54$ & $+26\overset{\circ}{} 33\overset{\prime}{} 46\overset{{\prime\prime}}{.}43$ & $ 104.30$ & $  97.35 \pm  3.34$ &$  -0.27$ &$12\overset{\rm h}{} 39\overset{\rm m}{} 52\overset{{\rm s}}{.}76$ & $+26\overset{\circ}{} 32\overset{\prime}{} 46\overset{{\prime\prime}}{.}57$ & $  49.14$ & $  45.11 \pm  1.52$ &$  10.42$ \\
$12\overset{\rm h}{} 48\overset{\rm m}{} 41\overset{{\rm s}}{.}35$ & $+26\overset{\circ}{} 33\overset{\prime}{} 56\overset{{\prime\prime}}{.}48$ & $  74.42$ & $  70.56 \pm  2.54$ &$   1.94$ &$10\overset{\rm h}{} 13\overset{\rm m}{} 36\overset{{\rm s}}{.}86$ & $+26\overset{\circ}{} 52\overset{\prime}{} 06\overset{{\prime\prime}}{.}67$ & $ 157.57$ & $ 139.77 \pm  3.66$ &$   9.56$ \\
$10\overset{\rm h}{} 45\overset{\rm m}{} 57\overset{{\rm s}}{.}36$ & $+26\overset{\circ}{} 52\overset{\prime}{} 39\overset{{\prime\prime}}{.}93$ & $  56.87$ & $  46.13 \pm  1.41$ &$  12.77$ &$10\overset{\rm h}{} 46\overset{\rm m}{} 40\overset{{\rm s}}{.}38$ & $+26\overset{\circ}{} 54\overset{\prime}{} 15\overset{{\prime\prime}}{.}73$ & $ 163.09$ & $ 155.16 \pm  3.46$ &$  13.96$ \\
$09\overset{\rm h}{} 21\overset{\rm m}{} 58\overset{{\rm s}}{.}57$ & $+26\overset{\circ}{} 21\overset{\prime}{} 22\overset{{\prime\prime}}{.}60$ & $ 206.52$ & $ 211.16 \pm  5.83$ &$   1.16$ &$10\overset{\rm h}{} 16\overset{\rm m}{} 00\overset{{\rm s}}{.}82$ & $+26\overset{\circ}{} 22\overset{\prime}{} 00\overset{{\prime\prime}}{.}12$ & $  27.87$ & $  23.13 \pm  0.81$ &$   2.94$ \\
$10\overset{\rm h}{} 48\overset{\rm m}{} 28\overset{{\rm s}}{.}59$ & $+26\overset{\circ}{} 22\overset{\prime}{} 34\overset{{\prime\prime}}{.}89$ & $ 237.69$ & $ 244.67 \pm  6.13$ &$   0.52$ &$10\overset{\rm h}{} 52\overset{\rm m}{} 52\overset{{\rm s}}{.}26$ & $+26\overset{\circ}{} 22\overset{\prime}{} 16\overset{{\prime\prime}}{.}24$ & $ 105.61$ & $ 105.37 \pm  2.62$ &$   2.66$ \\
$09\overset{\rm h}{} 27\overset{\rm m}{} 45\overset{{\rm s}}{.}09$ & $+26\overset{\circ}{} 13\overset{\prime}{} 00\overset{{\prime\prime}}{.}51$ & $ 150.04$ & $ 146.74 \pm  3.36$ &$  13.82$ &$10\overset{\rm h}{} 40\overset{\rm m}{} 37\overset{{\rm s}}{.}59$ & $+26\overset{\circ}{} 13\overset{\prime}{} 24\overset{{\prime\prime}}{.}81$ & $  29.96$ & $  25.01 \pm  1.14$ &$  13.45$ \\
$10\overset{\rm h}{} 54\overset{\rm m}{} 56\overset{{\rm s}}{.}64$ & $+26\overset{\circ}{} 12\overset{\prime}{} 47\overset{{\prime\prime}}{.}26$ & $ 338.51$ & $ 337.11 \pm  7.16$ &$  12.35$ &$12\overset{\rm h}{} 11\overset{\rm m}{} 29\overset{{\rm s}}{.}70$ & $+26\overset{\circ}{} 14\overset{\prime}{} 50\overset{{\prime\prime}}{.}20$ & $  67.80$ & $  64.70 \pm  1.25$ &$  14.18$ \\
$09\overset{\rm h}{} 49\overset{\rm m}{} 06\overset{{\rm s}}{.}44$ & $+26\overset{\circ}{} 01\overset{\prime}{} 57\overset{{\prime\prime}}{.}71$ & $  20.78$ & $  11.63 \pm  0.87$ &$  15.45$ &$12\overset{\rm h}{} 57\overset{\rm m}{} 12\overset{{\rm s}}{.}52$ & $+26\overset{\circ}{} 01\overset{\prime}{} 40\overset{{\prime\prime}}{.}62$ & $  62.57$ & $  55.22 \pm  1.23$ &$   1.79$ \\
$09\overset{\rm h}{} 18\overset{\rm m}{} 56\overset{{\rm s}}{.}17$ & $+26\overset{\circ}{} 33\overset{\prime}{} 30\overset{{\prime\prime}}{.}99$ & $  20.90$ & $  18.22 \pm  0.82$ &$  22.58$ &$10\overset{\rm h}{} 39\overset{\rm m}{} 30\overset{{\rm s}}{.}97$ & $+26\overset{\circ}{} 33\overset{\prime}{} 03\overset{{\prime\prime}}{.}88$ & $ 143.48$ & $ 133.02 \pm  3.53$ &$   1.31$ \\
$10\overset{\rm h}{} 59\overset{\rm m}{} 07\overset{{\rm s}}{.}98$ & $+26\overset{\circ}{} 34\overset{\prime}{} 07\overset{{\prime\prime}}{.}50$ & $  56.83$ & $  55.56 \pm  1.47$ &$   6.01$ &$12\overset{\rm h}{} 36\overset{\rm m}{} 30\overset{{\rm s}}{.}59$ & $+26\overset{\circ}{} 35\overset{\prime}{} 19\overset{{\prime\prime}}{.}78$ & $ 564.07$ & $ 599.41 \pm 20.89$ &$  29.70$ \\
$11\overset{\rm h}{} 22\overset{\rm m}{} 04\overset{{\rm s}}{.}96$ & $+26\overset{\circ}{} 13\overset{\prime}{} 36\overset{{\prime\prime}}{.}73$ & $ 303.49$ & $ 300.70 \pm  5.84$ &$  12.37$ &$10\overset{\rm h}{} 22\overset{\rm m}{} 09\overset{{\rm s}}{.}14$ & $+26\overset{\circ}{} 23\overset{\prime}{} 13\overset{{\prime\prime}}{.}70$ & $ 161.25$ & $ 161.32 \pm  4.29$ &$   7.96$ \\
$10\overset{\rm h}{} 55\overset{\rm m}{} 47\overset{{\rm s}}{.}35$ & $+26\overset{\circ}{} 23\overset{\prime}{} 37\overset{{\prime\prime}}{.}82$ & $ 152.02$ & $ 150.63 \pm  3.38$ &$   6.00$ &$12\overset{\rm h}{} 05\overset{\rm m}{} 32\overset{{\rm s}}{.}86$ & $+26\overset{\circ}{} 25\overset{\prime}{} 40\overset{{\prime\prime}}{.}65$ & $ 278.72$ & $ 268.41 \pm  4.18$ &$  27.54$ \\
$11\overset{\rm h}{} 25\overset{\rm m}{} 00\overset{{\rm s}}{.}85$ & $+26\overset{\circ}{} 06\overset{\prime}{} 09\overset{{\prime\prime}}{.}32$ & $  56.43$ & $  46.94 \pm  1.60$ &$   5.32$ &$12\overset{\rm h}{} 19\overset{\rm m}{} 56\overset{{\rm s}}{.}37$ & $+26\overset{\circ}{} 07\overset{\prime}{} 04\overset{{\prime\prime}}{.}07$ & $ 109.12$ & $  62.21 \pm  2.72$ &$  10.48$ \\
$12\overset{\rm h}{} 21\overset{\rm m}{} 02\overset{{\rm s}}{.}00$ & $+26\overset{\circ}{} 08\overset{\prime}{} 45\overset{{\prime\prime}}{.}74$ & $ 199.08$ & $ 123.17 \pm  2.70$ &$   9.95$ &$09\overset{\rm h}{} 02\overset{\rm m}{} 54\overset{{\rm s}}{.}70$ & $+25\overset{\circ}{} 55\overset{\prime}{} 44\overset{{\prime\prime}}{.}83$ & $  19.08$ & $  21.38 \pm  1.02$ &$  19.49$ \\
$09\overset{\rm h}{} 16\overset{\rm m}{} 47\overset{{\rm s}}{.}38$ & $+25\overset{\circ}{} 54\overset{\prime}{} 16\overset{{\prime\prime}}{.}05$ & $  73.79$ & $  68.79 \pm  2.34$ &$   4.71$ &$09\overset{\rm h}{} 28\overset{\rm m}{} 53\overset{{\rm s}}{.}21$ & $+25\overset{\circ}{} 54\overset{\prime}{} 37\overset{{\prime\prime}}{.}90$ & $  69.09$ & $  68.43 \pm  2.45$ &$   9.84$ \\
$09\overset{\rm h}{} 33\overset{\rm m}{} 15\overset{{\rm s}}{.}67$ & $+25\overset{\circ}{} 55\overset{\prime}{} 00\overset{{\prime\prime}}{.}69$ & $  61.57$ & $  62.32 \pm  2.06$ &$  12.20$ &$09\overset{\rm h}{} 44\overset{\rm m}{} 42\overset{{\rm s}}{.}33$ & $+25\overset{\circ}{} 54\overset{\prime}{} 43\overset{{\prime\prime}}{.}55$ & $ 774.39$ & $ 771.98 \pm 19.53$ &$   7.81$ \\
$09\overset{\rm h}{} 45\overset{\rm m}{} 53\overset{{\rm s}}{.}28$ & $+25\overset{\circ}{} 56\overset{\prime}{} 08\overset{{\prime\prime}}{.}37$ & $  54.64$ & $  55.52 \pm  1.17$ &$  22.19$ &$11\overset{\rm h}{} 08\overset{\rm m}{} 38\overset{{\rm s}}{.}99$ & $+25\overset{\circ}{} 56\overset{\prime}{} 13\overset{{\prime\prime}}{.}45$ & $  88.49$ & $  98.64 \pm  2.68$ &$   5.22$ \\
$09\overset{\rm h}{} 40\overset{\rm m}{} 13\overset{{\rm s}}{.}47$ & $+26\overset{\circ}{} 26\overset{\prime}{} 56\overset{{\prime\prime}}{.}68$ & $ 124.22$ & $ 107.93 \pm  2.82$ &$   5.26$ &$12\overset{\rm h}{} 53\overset{\rm m}{} 02\overset{{\rm s}}{.}00$ & $+26\overset{\circ}{} 27\overset{\prime}{} 45\overset{{\prime\prime}}{.}97$ & $ 203.70$ & $ 166.13 \pm  7.02$ &$   2.76$ \\
$09\overset{\rm h}{} 18\overset{\rm m}{} 37\overset{{\rm s}}{.}49$ & $+26\overset{\circ}{} 50\overset{\prime}{} 41\overset{{\prime\prime}}{.}10$ & $ 105.95$ & $ 103.85 \pm  3.05$ &$   8.58$ &$10\overset{\rm h}{} 42\overset{\rm m}{} 43\overset{{\rm s}}{.}59$ & $+26\overset{\circ}{} 50\overset{\prime}{} 02\overset{{\prime\prime}}{.}25$ & $ 148.24$ & $ 151.05 \pm  3.48$ &$   6.64$ \\
$12\overset{\rm h}{} 04\overset{\rm m}{} 12\overset{{\rm s}}{.}95$ & $+26\overset{\circ}{} 50\overset{\prime}{} 58\overset{{\prime\prime}}{.}95$ & $ 417.72$ & $ 420.74 \pm  9.65$ &$  12.98$ &$12\overset{\rm h}{} 19\overset{\rm m}{} 46\overset{{\rm s}}{.}58$ & $+26\overset{\circ}{} 17\overset{\prime}{} 34\overset{{\prime\prime}}{.}40$ & $  29.46$ & $  17.75 \pm  0.92$ &$  17.36$ \\
$12\overset{\rm h}{} 20\overset{\rm m}{} 27\overset{{\rm s}}{.}95$ & $+26\overset{\circ}{} 19\overset{\prime}{} 03\overset{{\prime\prime}}{.}54$ & $  32.41$ & $  17.92 \pm  0.28$ &$  12.38$ &$09\overset{\rm h}{} 13\overset{\rm m}{} 27\overset{{\rm s}}{.}12$ & $+26\overset{\circ}{} 02\overset{\prime}{} 23\overset{{\prime\prime}}{.}13$ & $  67.04$ & $  61.94 \pm  1.10$ &$  30.81$ \\
$09\overset{\rm h}{} 17\overset{\rm m}{} 15\overset{{\rm s}}{.}57$ & $+26\overset{\circ}{} 03\overset{\prime}{} 40\overset{{\prime\prime}}{.}53$ & $  58.42$ & $  49.17 \pm  1.41$ &$   7.95$ &$09\overset{\rm h}{} 40\overset{\rm m}{} 14\overset{{\rm s}}{.}71$ & $+26\overset{\circ}{} 03\overset{\prime}{} 29\overset{{\prime\prime}}{.}91$ & $ 462.22$ & $ 457.43 \pm 13.37$ &$  21.65$ \\
$09\overset{\rm h}{} 41\overset{\rm m}{} 28\overset{{\rm s}}{.}66$ & $+26\overset{\circ}{} 04\overset{\prime}{} 02\overset{{\prime\prime}}{.}56$ & $  79.37$ & $  77.97 \pm  2.18$ &$   8.46$ &$10\overset{\rm h}{} 05\overset{\rm m}{} 15\overset{{\rm s}}{.}06$ & $+26\overset{\circ}{} 03\overset{\prime}{} 35\overset{{\prime\prime}}{.}99$ & $ 180.10$ & $ 165.31 \pm  3.55$ &$  23.18$ \\
$10\overset{\rm h}{} 24\overset{\rm m}{} 19\overset{{\rm s}}{.}44$ & $+26\overset{\circ}{} 05\overset{\prime}{} 10\overset{{\prime\prime}}{.}71$ & $  20.86$ & $  18.40 \pm  0.50$ &$   5.51$ &$12\overset{\rm h}{} 22\overset{\rm m}{} 01\overset{{\rm s}}{.}84$ & $+26\overset{\circ}{} 04\overset{\prime}{} 15\overset{{\prime\prime}}{.}09$ & $  60.17$ & $  32.34 \pm  0.98$ &$  28.94$ \\
$09\overset{\rm h}{} 30\overset{\rm m}{} 27\overset{{\rm s}}{.}63$ & $+25\overset{\circ}{} 52\overset{\prime}{} 54\overset{{\prime\prime}}{.}26$ & $  78.58$ & $  83.92 \pm  2.25$ &$  16.53$ &$09\overset{\rm h}{} 17\overset{\rm m}{} 17\overset{{\rm s}}{.}69$ & $+26\overset{\circ}{} 45\overset{\prime}{} 51\overset{{\prime\prime}}{.}73$ & $  22.46$ & $  21.96 \pm  0.73$ &$  14.87$ \\
$10\overset{\rm h}{} 35\overset{\rm m}{} 16\overset{{\rm s}}{.}54$ & $+26\overset{\circ}{} 15\overset{\prime}{} 16\overset{{\prime\prime}}{.}45$ & $ 388.16$ & $ 367.07 \pm  7.47$ &$   9.77$ &$12\overset{\rm h}{} 02\overset{\rm m}{} 55\overset{{\rm s}}{.}33$ & $+26\overset{\circ}{} 15\overset{\prime}{} 18\overset{{\prime\prime}}{.}79$ & $  51.15$ & $  39.83 \pm  1.00$ &$  13.88$ \\
$11\overset{\rm h}{} 34\overset{\rm m}{} 34\overset{{\rm s}}{.}12$ & $+26\overset{\circ}{} 52\overset{\prime}{} 50\overset{{\prime\prime}}{.}95$ & $ 108.72$ & $  99.49 \pm  2.50$ &$  14.70$ &$09\overset{\rm h}{} 05\overset{\rm m}{} 05\overset{{\rm s}}{.}60$ & $+26\overset{\circ}{} 11\overset{\prime}{} 23\overset{{\prime\prime}}{.}24$ & $ 207.10$ & $ 191.89 \pm  4.47$ &$  10.57$ \\
$10\overset{\rm h}{} 11\overset{\rm m}{} 06\overset{{\rm s}}{.}75$ & $+26\overset{\circ}{} 38\overset{\prime}{} 21\overset{{\prime\prime}}{.}01$ & $  33.34$ & $  27.02 \pm  1.47$ &$   9.53$ &$10\overset{\rm h}{} 19\overset{\rm m}{} 26\overset{{\rm s}}{.}26$ & $+26\overset{\circ}{} 38\overset{\prime}{} 58\overset{{\prime\prime}}{.}41$ & $ 117.41$ & $ 119.92 \pm  3.15$ &$  10.87$ \\
$10\overset{\rm h}{} 25\overset{\rm m}{} 08\overset{{\rm s}}{.}32$ & $+26\overset{\circ}{} 36\overset{\prime}{} 39\overset{{\prime\prime}}{.}24$ & $  14.53$ & $   8.41 \pm  1.04$ &$  17.35$ &$10\overset{\rm h}{} 31\overset{\rm m}{} 56\overset{{\rm s}}{.}02$ & $+26\overset{\circ}{} 38\overset{\prime}{} 20\overset{{\prime\prime}}{.}94$ & $  96.73$ & $  95.41 \pm  3.55$ &$   9.19$ \\
$10\overset{\rm h}{} 57\overset{\rm m}{} 24\overset{{\rm s}}{.}93$ & $+26\overset{\circ}{} 38\overset{\prime}{} 00\overset{{\prime\prime}}{.}02$ & $ 435.24$ & $ 420.13 \pm 14.35$ &$   4.35$ &$11\overset{\rm h}{} 25\overset{\rm m}{} 20\overset{{\rm s}}{.}26$ & $+26\overset{\circ}{} 38\overset{\prime}{} 02\overset{{\prime\prime}}{.}07$ & $  49.37$ & $  40.30 \pm  1.67$ &$   5.63$ \\
$09\overset{\rm h}{} 23\overset{\rm m}{} 43\overset{{\rm s}}{.}10$ & $+26\overset{\circ}{} 34\overset{\prime}{} 49\overset{{\prime\prime}}{.}94$ & $  41.15$ & $  38.81 \pm  1.30$ &$  34.81$ &$12\overset{\rm h}{} 42\overset{\rm m}{} 48\overset{{\rm s}}{.}21$ & $+25\overset{\circ}{} 58\overset{\prime}{} 40\overset{{\prime\prime}}{.}76$ & $ 308.33$ & $ 290.77 \pm  6.32$ &$  17.88$ \\
$11\overset{\rm h}{} 06\overset{\rm m}{} 08\overset{{\rm s}}{.}35$ & $+25\overset{\circ}{} 52\overset{\prime}{} 13\overset{{\prime\prime}}{.}04$ & $  38.26$ & $  39.91 \pm  1.97$ &$  10.67$ &$11\overset{\rm h}{} 27\overset{\rm m}{} 05\overset{{\rm s}}{.}77$ & $+25\overset{\circ}{} 50\overset{\prime}{} 55\overset{{\prime\prime}}{.}17$ & $ 117.51$ & $ 127.43 \pm  3.32$ &$   2.96$ \\
$12\overset{\rm h}{} 00\overset{\rm m}{} 05\overset{{\rm s}}{.}07$ & $+26\overset{\circ}{} 44\overset{\prime}{} 18\overset{{\prime\prime}}{.}34$ & $  34.49$ & $  40.04 \pm  1.21$ &$   6.54$ &$10\overset{\rm h}{} 33\overset{\rm m}{} 40\overset{{\rm s}}{.}04$ & $+26\overset{\circ}{} 42\overset{\prime}{} 57\overset{{\prime\prime}}{.}77$ & $  16.66$ & $  16.56 \pm  0.57$ &$  13.50$ \\
$10\overset{\rm h}{} 54\overset{\rm m}{} 40\overset{{\rm s}}{.}25$ & $+26\overset{\circ}{} 41\overset{\prime}{} 25\overset{{\prime\prime}}{.}69$ & $  63.90$ & $  59.84 \pm  1.59$ &$   7.20$ &$10\overset{\rm h}{} 14\overset{\rm m}{} 14\overset{{\rm s}}{.}85$ & $+25\overset{\circ}{} 52\overset{\prime}{} 00\overset{{\prime\prime}}{.}01$ & $  37.40$ & $  30.04 \pm  1.06$ &$  21.30$ \\
$09\overset{\rm h}{} 54\overset{\rm m}{} 39\overset{{\rm s}}{.}79$ & $+26\overset{\circ}{} 39\overset{\prime}{} 24\overset{{\prime\prime}}{.}55$ & $ 312.67$ & $ 302.08 \pm  5.82$ &$  13.77$ &$10\overset{\rm h}{} 39\overset{\rm m}{} 50\overset{{\rm s}}{.}71$ & $+27\overset{\circ}{} 01\overset{\prime}{} 48\overset{{\prime\prime}}{.}14$ & $  14.86$ & $  14.63 \pm  0.73$ &$   5.01$ \\
$12\overset{\rm h}{} 24\overset{\rm m}{} 21\overset{{\rm s}}{.}36$ & $+27\overset{\circ}{} 01\overset{\prime}{} 42\overset{{\prime\prime}}{.}20$ & $  97.89$ & $  95.13 \pm  2.15$ &$   7.08$ &$09\overset{\rm h}{} 31\overset{\rm m}{} 32\overset{{\rm s}}{.}80$ & $+26\overset{\circ}{} 40\overset{\prime}{} 48\overset{{\prime\prime}}{.}18$ & $  70.83$ & $  68.01 \pm  2.04$ &$  12.48$ \\
$11\overset{\rm h}{} 02\overset{\rm m}{} 53\overset{{\rm s}}{.}30$ & $+27\overset{\circ}{} 03\overset{\prime}{} 39\overset{{\prime\prime}}{.}38$ & $  35.19$ & $  28.59 \pm  1.19$ &$  18.43$ &$12\overset{\rm h}{} 50\overset{\rm m}{} 21\overset{{\rm s}}{.}97$ & $+27\overset{\circ}{} 04\overset{\prime}{} 32\overset{{\prime\prime}}{.}95$ & $  49.53$ & $  44.96 \pm  1.43$ &$  18.40$ \\
$09\overset{\rm h}{} 15\overset{\rm m}{} 01\overset{{\rm s}}{.}99$ & $+27\overset{\circ}{} 01\overset{\prime}{} 45\overset{{\prime\prime}}{.}69$ & $  26.15$ & $  30.06 \pm  0.49$ &$  10.34$ & & & & & \\
\hline\hline
\end{tabular}
\end{center}
\end{table*}

\end{document}